\documentclass[sigconf]{acmart}

\usepackage{tabularx}
\usepackage{multirow}
\usepackage{color}
\usepackage{colortbl}
\usepackage{natbib} 
\usepackage{amsmath}
\usepackage{enumitem}
\usepackage{stfloats}
\newcommand{\eg}{e.g.,}
\newcommand{\ie}{i.e.,}
\newcommand{\etal}{\textit{et al.}}
\newcommand{\bpstart}[1]{\vspace{1mm}\noindent{\textbf{#1.}}}

\AtBeginDocument{%
  }

\copyrightyear{2025}
\acmYear{2025}
\setcopyright{cc}
\setcctype{by-sa}
\acmConference[CHI '25]{CHI Conference on Human Factors in Computing Systems}{April 26-May 1, 2025}{Yokohama, Japan}
\acmBooktitle{CHI Conference on Human Factors in Computing Systems (CHI '25), April 26-May 1, 2025, Yokohama, Japan}\acmDOI{10.1145/3706598.3713158}
\acmISBN{979-8-4007-1394-1/25/04}

\begin{document}

\title["It Brought the Model to Life"]{"It Brought the Model to Life": Exploring the Embodiment of Multimodal I3Ms for People who are Blind or have Low Vision}

\author{Samuel Reinders}
\email{samuel.reinders@monash.edu}
\orcid{0000-0001-5627-413X}
\affiliation{
  \institution{Monash University}
  \city{Melbourne}
  \state{Victoria}
  \country{Australia}
}

\author{Matthew Butler}
\email{matthew.butler@monash.edu}
\orcid{0000-0002-7950-5495}
\affiliation{
  \institution{Monash University}
  \city{Melbourne}
  \state{Victoria}
  \country{Australia}
}

\author{Kim Marriott}
\email{kim.marriott@monash.edu}
\orcid{0000-0002-9813-0377}
\affiliation{
  \institution{Monash University}
  \city{Melbourne}
  \state{Victoria}
  \country{Australia}
}

\renewcommand{\shortauthors}{Reinders et al.}

\begin{abstract}
  3D-printed models are increasingly used to provide people who are blind or have low vision (BLV) with access to maps, educational materials, and museum exhibits. Recent research has explored interactive 3D-printed models (I3Ms) that integrate touch gestures, conversational dialogue, and haptic vibratory feedback to create more engaging interfaces. Prior research with sighted people has found that imbuing machines with human-like behaviours, \ie\, embodying them, can make them appear more lifelike, increasing social perception and presence. Such embodiment can increase engagement and trust. This work presents the first exploration into the design of embodied I3Ms and their impact on BLV engagement and trust. In a controlled study with 12 BLV participants, we found that I3Ms using specific embodiment design factors, such as haptic vibratory and embodied personified voices, led to an increased sense of liveliness and embodiment, as well as engagement, but had mixed impact on trust.
\end{abstract}




\begin{teaserfigure}
  \includegraphics[width=1.00\columnwidth]{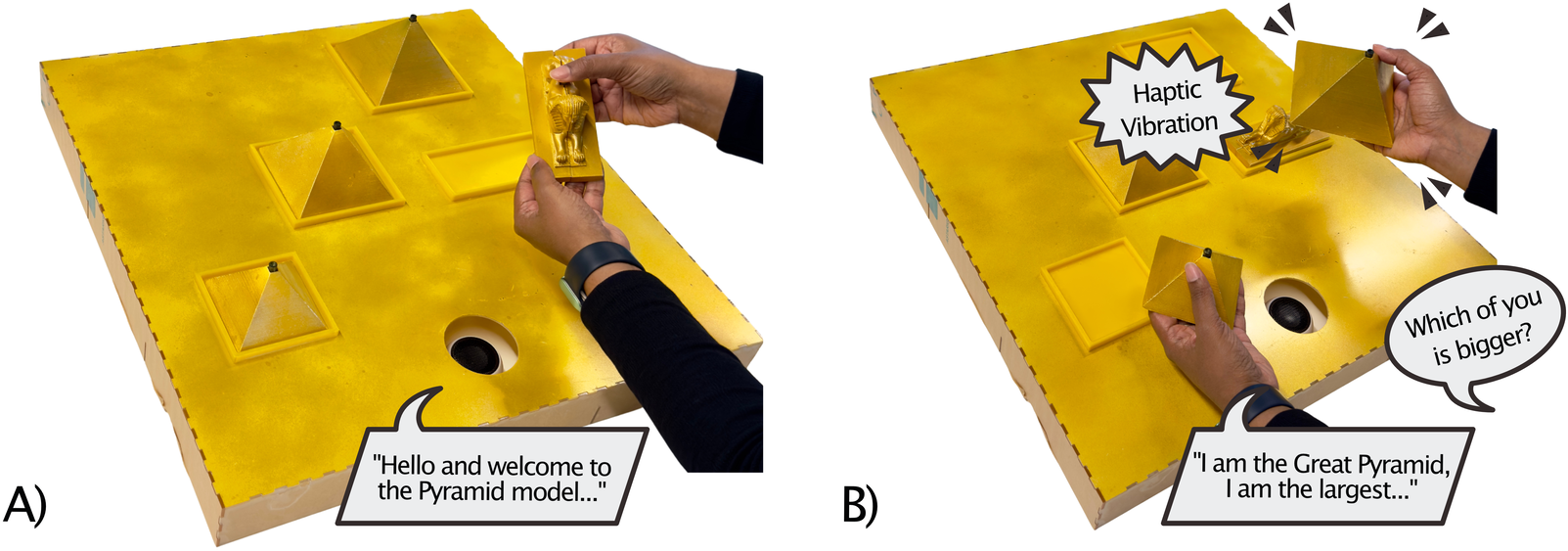}
  %TC:ignore
  \caption{We designed two interactive 3D-printed models (I3Ms) that could operate using various embodied design factors. Shown here is the Egyptian Pyramid I3M operating in High Embodied Mode (HEM). In A), the model introduces itself as a user picks up the Sphinx; In B), the user holds the Great Pyramid and Pyramid of Menkaure and asks which is larger. The Great Pyramid then emits localised embodied haptics and responds using an embodied personified voice and first-person narration.}~\label{fig:teaser}
  \Description{This figure includes two images showcasing one of the interactive 3D-printed models (I3Ms) we designed. Seen is the Egyptian Pyramid I3M operating in High Embodied Mode (HEM). It includes a base made of acrylic that is painted gold. It includes four 3D-printed objects -- the Sphinx, and the Pyramids of Khufu, Menkaure, and Khafre. They are also painted gold. In A), the model introduces itself as a user picks up the Sphinx; In B), the user holds the Great Pyramid and Pyramid of Menkaure and asks which is larger, to which the Great Pyramid emits localised haptics and responds using an embodied personified voice and first-person narration.}
  %TC:endignore  
\end{teaserfigure}

\maketitle

\section{Introduction}
The lack of equitable access to graphical information, such as educational materials and navigation maps, can significantly reduce opportunities and overall quality of life for people who are blind or have low vision (BLV)~\cite{butler2017understanding,Sheffield2016}. In recent years, 3D printing has been used to create accessible graphics~\cite{Stangl2015,Buehler2016,Holloway2018,Hu2015}. Unlike traditional accessible graphics, 3D printing enables the fabrication of tangible models that directly represent three-dimensional objects and concepts. This allows a broader range of content to be effectively conveyed non-visually. 3D-printed models have demonstrated improved tactual understanding and mental model development compared to tactile graphics, as well as increased engagement~\cite{Holloway2018}. 

It is now becoming common to add button or touch-triggered audio labels to 3D-printed models~\cite{Ghodke2019,Holloway2018,Shi2016,Davis2020}, creating \textbf{interactive 3D-printed models (I3Ms)}. Audio labels support independent exploration and reduce the need for braille labels, which many blind people cannot read~\cite{NFIB2009}, may not fit on the model~\cite{Holloway2018}, and can distort the model surface~\cite{Holloway2018,Shi2016}. However, while useful, audio labels can only provide limited, predetermined information. With advances in intelligent agents, recent work has begun integrating conversational agents into I3Ms~\cite{Shi2017b,Quero2019,Reinders2020,Reinders2023}. These models afford greater agency and independence to BLV users, allowing them to generate their own queries and potentially access unlimited information about the model. One study involving the co-design of an I3M that combined touch gestures, conversational dialogue, and haptic vibratory feedback, found that participants desired an experience that felt more personal and `\textit{alive}'~\cite{Reinders2023}.

Previous research has shown that imbuing machines with human-like behaviours and characteristics, that is -- embodying them -- can make them appear more lifelike and alive, increasing social perception and presence~\cite{Cassell2001,Lankton2015,Lester1997,Nowak2003,Shamekhi2018}. Such embodiment has been found to increase subjective user engagement~\cite{Cassell2001,Shamekhi2018} and trust~\cite{Bickmore2001,Bickmore2013,Rheu2021,Shamekhi2018}. A variety of \textbf{conversational embodiment} factors have been identified to increase embodiment, including speech that mimics human voices~\cite{Cassell2001}, greeting users~\cite{Luria2019,Cassell2001,Shamekhi2018,Lester1997,Nowak2003}, conversational turn-taking~\cite{Cassell2001,Kontogiorgos2020}, small talk~\cite{Liao2018,Pradhan2019,Shamekhi2018,Cassell2001,Bickmore2001}, and exhibiting personality~\cite{Lester1997}. \textbf{Visual embodiment} attributes, such as giving the agent a face~\cite{Shamekhi2018,Bickmore2001,Bickmore2013,Kontogiorgos2020}, gestures~\cite{Cassell2001,Bickmore2001}, and employing gaze~\cite{Kontogiorgos2020,Shamekhi2018}, have also been found to increase embodiment, as has \textbf{physical embodiment}~\cite{Luria2017,Kidd2004}.

However, virtually all prior research has been conducted with sighted users and has not considered BLV users, who may not be able to fully discern or perceive visual characteristics or physical embodiment. Since the usefulness of an accessible graphic or interface depends on users' willingness to engage with and accept the information it provides~\cite{Betsy1993,Wu2017,Abdolrahmani2018}, engagement and trust are critical for BLV users. We believe that embodiment, and its links with engagement and trust, may hold significant potential for I3Ms, which are designed to be spoken to, picked up, and touched. Embodied I3Ms could enable BLV students or self-learners to engage in deeper, more meaningful experiences with content, and therefore serve as a catalyst for their broader adoption. Here, we present the first exploration into the design of more embodied I3Ms and the impact of embodiment on the engagement and trust of BLV users. We believe our study is the first to explore embodiment in the context of both BLV users and I3Ms. 

We selected five non-visual design factors and created two I3Ms -- the \textbf{Saturn V Rocket} and \textbf{Egyptian Pyramids} -- that could be configured in two states -- \textit{High Embodied Mode (HEM)} and \textit{Low Embodied Mode (LEM)}. These states differed based on the embodiment design factors. Those factors relating to conversational embodiment (\textit{introductions and small talk, embodied personified voices} and \textit{embodied narration style}) have previously been found to increase embodiment. Physical embodiment factors (\textit{embodied haptic vibratory feedback} and \textit{location of speech output}) were more novel and were motivated by feedback from BLV users interacting with I3Ms~\cite{Reinders2023}. We conducted a within-subject user study with 12 BLV participants, using established questionnaires and subjective ratings to examine how \textit{lively}, \textit{engaging} and \textit{trustworthy} participants perceived each model to be. The main findings of our study include:

\begin{itemize}
    \item Participants perceived the HEM I3Ms as having a greater sense of liveliness, appearing more embodied compared to LEM I3Ms; 
    \item HEM I3Ms appeared to be more engaging. This adds to research showing that more embodied conversational agents and social robots increase subjective user engagement, establishing that this relationship extends to I3Ms in the context of BLV users;
    \item Differences in trust between LEM and HEM I3Ms were inconclusive, suggesting the impact of embodiment on trust may be more limited.
\end{itemize}

Our findings, which represent the first exploration into the embodiment and social perception of embodied I3Ms, provide initial design recommendations for creating I3Ms that BLV users find engaging. These recommendations will be of critical interest to the accessibility research community and to practitioners designing I3Ms for accessible exhibits in public spaces, such as museums and galleries, or as accessible materials for classroom use. 

\section{Related Work}
This work builds on research on: accessible graphics and interactive 3D-printed models for BLV users; conversational agents; and embodied agents.

\subsection{Accessible Graphics}
BLV people face challenges accessing graphical information, which impacts education opportunities~\cite{butler2017understanding}, makes independent travel difficult~\cite{Sheffield2016}, and causes disengagement with culture and the creative arts~\cite{Bartlett2019}. These barriers can lead to reductions in confidence and overall quality of life~\cite{Keeffe2005}.

Graphical information can be made available in formats that improve non-visual access. Traditionally, accessible graphics -- known as raised line drawings or \textbf{tactile graphics} -- have been used to assist BLV people in accessing information. Tactile graphics are frequently used to facilitate classroom learning~\cite{Aldrich2001,Rosenblum2015} and orientation and mobility (O\&M) training~\cite{Blades1999,Rowell2005}. Work has been conducted on the development of \textbf{interactive tactile graphics}, including the NOMAD~\cite{NOMAD}, IVEO~\cite{IVEO}, and Talking Tactile Tablet~\cite{Miele2006,TTT}. These devices combine printed tactile overlays with touch-sensitive surfaces, enabling BLV users to explore the graphics tactually and access audio labels by interacting with predefined touch areas. However, as these systems rely on printed tactile graphics, their scope is limited to two-dimensional content.

\textbf{3D-printed models} are an increasingly common alternative to tactile graphics. They enable a broader range of material to be produced, particularly for concepts that are inherently three-dimensional in nature. In recent years, the production cost and effort of 3D-printed models have fallen to levels comparable to tactile graphic production. 3D-printed models are increasingly being applied in various accessible graphic areas: mapping and navigation~\cite{gual2012visual,Holloway2018,Holloway2019b,Holloway2022,Nagassa2023}; special education~\cite{Buehler2016}; art galleries~\cite{karaduman2022beyond,Butler2023}; books~\cite{kim2015,Stangl2015}; mathematics~\cite{Brown2012,Hu2015}; graphic design~\cite{McDonald2014}; science~\cite{wedler2012applied,Hasper2015}; and programming~\cite{kane2014}. Compared to tactile graphics, 3D-printed models have been shown to improve tactual understanding and mental model development among BLV people~\cite{Holloway2018}. However, as with traditional tactile graphics, the provision of written descriptions or braille labelling presents challenges. The limited space on models and the low-fidelity of 3D-printed braille can significantly impact the utility and readability of labels~\cite{Brown2012,Taylor2015,Shi2016}.

\subsection{Interactive 3D-Printed Models}
To address labelling challenges, limitations on the type of content that can be produced, and to create more engaging and interactive experiences, there is growing interest in the development of \textbf{interactive 3D-printed models (I3Ms)}. By combining 3D-printed models with low-cost electronics and/or smart devices, many I3Ms now include button or touch-triggered audio labels that provide verbal descriptions of the printed model~\cite{Landau2009,Shi2016,Reichinger2016,Giraud2017,Gotzelmann2017,Holloway2018,Ghodke2019,Davis2020}. Such I3Ms have been applied across various BLV-accessible graphic areas, including: art~\cite{Holloway2019,Bartolome2019,Butler2023}; education~\cite{Ghodke2019,Shi2019,Reinders2020}; and mapping and navigation~\cite{Gotzelmann2017,Holloway2018,Shi2020}. I3Ms with audio labels are especially useful for BLV users who are not fluent braille readers. Stored as text and synthesised in real-time, audio labels are easier to update compared to labels on non-interactive models. Many I3Ms also support multiple levels of audio labelling~\cite{Holloway2018,Shi2019,Reinders2020}, extracted through unique button presses or touch gestures, enabling them to convey far more information than the written descriptions supplied alongside tactile graphics and non-interactive models.

I3Ms are inherently multimodal. Multimodality can improve the adaptability of a system~\cite{Reeves2004}, and when modalities are combined, they can increase the resolution of information the system conveys~\cite{Edwards2015} and enable more natural interactions~\cite{Bolt1980}. For BLV users, combining modalities has been shown to improve confidence and independence~\cite{Quero2021}. Modality adaptability allows BLV users to choose interaction methods based on context, ability, or effort. For example, a user may be uncomfortable engaging in speech interaction in public due to privacy concerns~\cite{Abdolrahmani2018}, opting instead to use button or gesture-based inputs. Richer resolutions of information can be achieved when combining modalities, \eg\, the tactile features of a 3D-printed model with haptic vibratory and auditory outputs. Combining modalities is critical to overcoming the \textit{`bandwidth problem'}, in which BLV users' non-visual senses cannot match the capacity of vision, necessitating their combined use~\cite{Edwards2015}.

While early I3Ms primarily relied on button or gesture-based triggered audio labels, recent research has explored integrating speech interfaces and conversational agents. This shift is driven by research finding that BLV people find voice interaction convenient~\cite{Azenkot2013}, along with widespread adoption~\cite{Pradhan2018} and high usage~\cite{Abdolrahmani2018} of conversational agents among BLV users. For instance, Quero \etal~\cite{Quero2019} combined a tactile graphic of a floor plan with a conversational agent that focused on indoor navigation; however, voice interaction was performed through a connected smartphone rather than the graphic itself. Other works have developed voice-controlled agents to guide BLV users in exploring 3D-printed representations of gallery pieces~\cite{Bartolome2019,Quero2018}. These systems, however, have primarily focused on basic command-driven interactions more analogous to voice menus rather than conversational dialogue. Shi \etal~\cite{Shi2017b,Shi2019} proposed incorporating conversational agents into I3Ms to allow BLV users to expand their understanding of the modelled content.

Recent research into I3Ms has begun to explore modalities beyond audio and touch. Quero \etal~\cite{Quero2018} designed an I3M representing an art piece that integrated localised audio, wind, and heat output. However, participants faced challenges in interpreting the semantic mapping of modalities, \eg\, whether heat represented the morning sun or the shine of starlight. In our previous work, we found that BLV users desired I3Ms that combined touch, haptic vibratory feedback, and conversational dialogue~\cite{Reinders2020}. Additionally, we co-designed an I3M with BLV co-designers to explore how these modalities could create natural interactions~\cite{Reinders2023}, inspired by the \textit{`Put-That-There'} paradigm~\cite{Bolt1980}. This work led to five I3M design recommendations, including: support interruption-free tactile exploration; leverage prior interaction experience with personal technology; support customisation and personalisation; support more natural dialogue; and tightly coupled haptic feedback. These studies motivated our current work, with participants finding the I3M engaging, and several beginning to embody it.

\subsection{The Embodiment of Agents}
Dourish~\cite{Dourish2001} presents a seminal view that embodiment \textit{``denotes a form of participative status''}, where embodied interaction in natural forms of communication is influenced both by physical presence and context. They posit that this perspective applies to \textit{``spoken conversations just as much as to apples or bookshelves''}. Within the design of agent-based systems, embodiment is largely understood as the use of different modalities -- \eg\ voice, visual output, gestures, gaze -- to imbue machines with human-like behaviours and characteristics, making them appear more \textit{`alive'}; with an enhanced perception of social presence that is capable of approximating human-human social interaction~\cite{cassell2000more,Lester1997,Lankton2015,Biocca1999}. Such agents are often described as being more \textit{`lifelike'} or possessing \textit{`lifelikeness'}~\cite{Lester1997,Cassell1999,Cassell1999b,Cassell1999c,Lester1999}.

Research into embodiment has predominantly focused on sighted users. As systems become more embodied, users' perceptions of social presence can increase, motivating users to treat them more favourably~\cite{Reeves2004}. Lankton \etal~\cite{Lankton2015} described how social presence can make systems appear more sociable, warm, and personal, while Cassell~\cite{Cassell2001} noted that embodying technology allows users to locate intelligence, illuminating what would otherwise be an \textit{`invisible computer'}. Importantly, embodied agents exhibiting higher levels of social presence and perception have been shown to enhance user perception of engagement~\cite{Shamekhi2018,Luger2016,Heuwinkel2013,Cassell2001} and trust~\cite{Bickmore2001,Shamekhi2018,Bickmore2013,Rheu2021}. 

In the HCI community, efforts to embody intelligent agents have focused on enhancing social perception through conversational, visual, and physical attributes. \textbf{Conversational embodiment} includes mimicking human voices~\cite{Cassell2001}, small talk~\cite{Liao2018,Pradhan2019,Cassell2001,Shamekhi2018,Bickmore2001}, greetings~\cite{Cassell2001,Luria2019,Shamekhi2018,Lester1997}, and conversational turn-taking~\cite{Cassell2001,Kontogiorgos2020}. Lester \etal~\cite{Lester1997} described the \textit{persona effect}, proposing that social presence can increase when agents exhibit personality, making them appear more lifelike. Many conversational agents, like Siri, incorporate attributes of conversational embodiment.

\textbf{Visually embodied} agents, which are often also conversationally embodied, associate systems with virtual avatars or characters, many of which have faces~\cite{Cassell2001,Kontogiorgos2020,Bickmore2013}, and are capable of gesturing~\cite{Cassell2001,Bickmore2001} and gaze~\cite{Kontogiorgos2020,Shamekhi2018,Bickmore2001}. Visual embodiment  extends beyond the visual feedback that mainstream conversational agents emit, \eg\ rings of light or animations used to indicate that agents are processing or `thinking'. Depending on the task, visually and conversationally embodied agents can improve social perceptions~\cite{Shamekhi2018,Luria2019,Cassell2001,Nowak2003}, trust~\cite{Bickmore2001,Bickmore2013,Rheu2021,Shamekhi2018}, and engagement~\cite{Shamekhi2018,Cassell2001}. 

Embodied agents can extend beyond virtual embodiment and include physical bodies~\cite{Kontogiorgos2020,Luria2017}. Lura \etal~\cite{Luria2017} found that users' situational awareness was higher when using a physically embodied robot to perform smart-home tasks compared to an unembodied voice agent. Robots capable of emitting human-like warmth have been associated with increased perceptions of friendship and presence~\cite{Nie2012}. Recent research has explored the use of haptic vibratory feedback to create lifelike cues, such as heartbeats~\cite {Borgstedt2023} and handshaking~\cite{Bevan2015}. \textbf{Physically embodied} agents can be perceived as having higher social presence than non-physically embodied agents~\cite{Kidd2004}. Additionally, they have been found to be more forgivable during unsuccessful interactions; however, depending on their realism, thi can become distracting in high-stakes scenarios~\cite{Kontogiorgos2020}.

\subsection{The Embodiment of I3Ms}
The design of embodied agents has traditionally focused on the perception of sighted users. However, in the last decade, work has begun exploring how human-human conversational cues can be converted into non-visual formats for BLV users. Many of these efforts utilise haptic belts/headsets~\cite{rader2014,McDaniel2018} or AR glasses~\cite{Qiu2016,Qiu2020} to convey body movements like head shaking, nodding, and gaze. Despite this, the impact of agent embodiment, and specifically embodied I3Ms, on BLV users' perceptions remains unstudied.

In our previous work, we found that a number of participants desired I3Ms that felt more lively and human~\cite{Reinders2023}. Participants suggested integrating a conversational agent with a personality, incorporating haptic vibratory feedback to make the I3M feel alive, and enabling speech to originate directly from the model itself. This desire for more human-like interactions aligns with findings with conversational agents. Choi \etal~\cite{Choi2020} reported that many BLV users valued human-like conversation with conversational agents as critical in relationship building, while Abdolrahmani \etal~\cite{Abdolrahmani2018} observed that BLV users preferred agents they could talk with as if they were other people rather than pieces of technology. Karim \etal~\cite{Karim2023} recommended that agents have customisable personalities, recognising that such features may not be relevant in all scenarios, such as group settings. Collins \etal~\cite{Collins2023}, however, found hesitance among BLV users towards embodied AI agents in VR applications. Whether these perspectives and desires extend to I3Ms is unknown.

Further impetus for studying I3M embodiment comes from the links between embodiment, engagement, and trust with embodied agents that have been previously identified for sighted users. Trust and engagement are especially critical for BLV users, as the usefulness of accessible graphics, aids, or tools depends on users' willingness to engage with and accept/rely on the information they provide. This directly determines the extent to which users rely on these tools~\cite{Betsy1993,Wu2017,Abdolrahmani2018}. Therefore, it is crucial to explore whether I3Ms can be effectively embodied and whether embodiment fosters greater trust and engagement between users and their I3Ms. 

\section{Embodied Design Factors}
\label{sec:Design}
Our work was influenced by interpretations of embodiment and embodied interaction proposed by Dourish~\cite{Dourish2001} and Cassell~\cite{Cassell2001}. Interfaces can be embodied using a range of design characteristics, including visual embodiment, conversational embodiment, and physical embodiment. However, as BLV users, particularly those who are totally blind, may not be able to fully discern visual characteristics, we approached I3M embodiment from a purely non-visual perspective. Traditional visual embodiment design factors, such as virtual avatars or gaze, were not considered. We drew from existing literature to identify a range of design factors shown to enhance perceived levels of social perception and embodiment. These were split between \textbf{conversational} and \textbf{physical embodiment}.

\subsection{Model Selection \& Design}
\begin{figure*}[ht!]
\centering
    \includegraphics[width=1\textwidth]{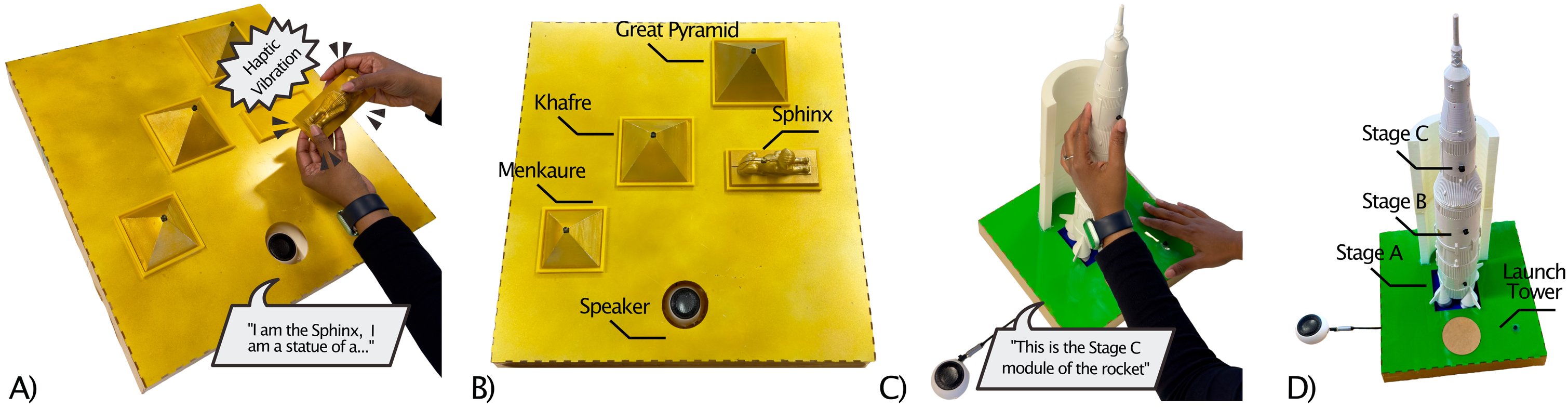}
    %TC:ignore
    \caption{\small In A), the Egyptian Pyramids I3M is configured in HEM, showing a user pressing the Sphinx touchpoint, triggering localised haptics and a personified voice response via a speaker contained within the I3M's enclosure; B) shows the Pyramid I3M's components -- the Great Pyramid, Sphinx, and Khafre/Menkaure. In C), the Saturn V Rocket I3M is in LEM, with a user pressing the Stage C touchpoint, triggering speech output via an external speaker; D) shows the Rocket I3M's components -- detachable Stage A/B/C rocket modules, and the Launch Tower.}~\label{fig:RocketPyramidI3M}
    \Description{This figure includes four images showcasing our two I3Ms. In A), the Egyptian Pyramids I3M is configured in HEM. The user is pressing the touchpoint on the Sphinx, which emits localised haptics, and, using an embodied personified voice and first-person narration, responds through a speaker contained within the I3M's enclosure. In B), the I3M consists of four 3D-printed components -- the Great Pyramid, Sphinx, and Pyramids of Khafre/Menkaure -- each of these is labelled, as is the speaker that is embedded inside the I3M enclosure/case in HEM. In C), the Saturn V Rocket I3M is configured in LEM. It includes a base made of acrylic that is painted green, representing the ground. Shown is a user pressing the Stage C module touchpoint. Speech output is played back through a speaker that is housed externally from the I3Ms enclosure — ``This is the Stage C module of the rocket''; In D), the Saturn V Rocket I3M consists of four 3D-printed components, the Stage A, Stage B and Stage C rocket modules, which can be detached, and the rockets Launch Tower. The rocket components can be stacked on top of one another as they have magnets attached.}
    %TC:endignore
\end{figure*}

To investigate conversational and physical embodiment, we created two I3Ms -- (1) the \textbf{Egyptian Pyramids} and (2) the \textbf{Saturn V Rocket} [Figure~\ref{fig:RocketPyramidI3M}]. These subjects were selected because they represent the types of materials commonly found in museums, galleries, or in science or history classes. Additionally, they facilitated the design of models with multiple components that could be individually picked up, detached, and manipulated, which has been shown to increase engagement~\cite{Reinders2020}. Each I3M can be configured in two states -- \textit{High Embodied Mode (HEM)} or \textit{Low Embodied Mode (LEM)} -- based on five non-visual design factors (Sections~\ref{sec:ConversationalEmbodiment} \& ~\ref{sec:PhysicalEmbodiment}).

\begin{figure}[hb!]
\centering
  \includegraphics[width=1.00\columnwidth]{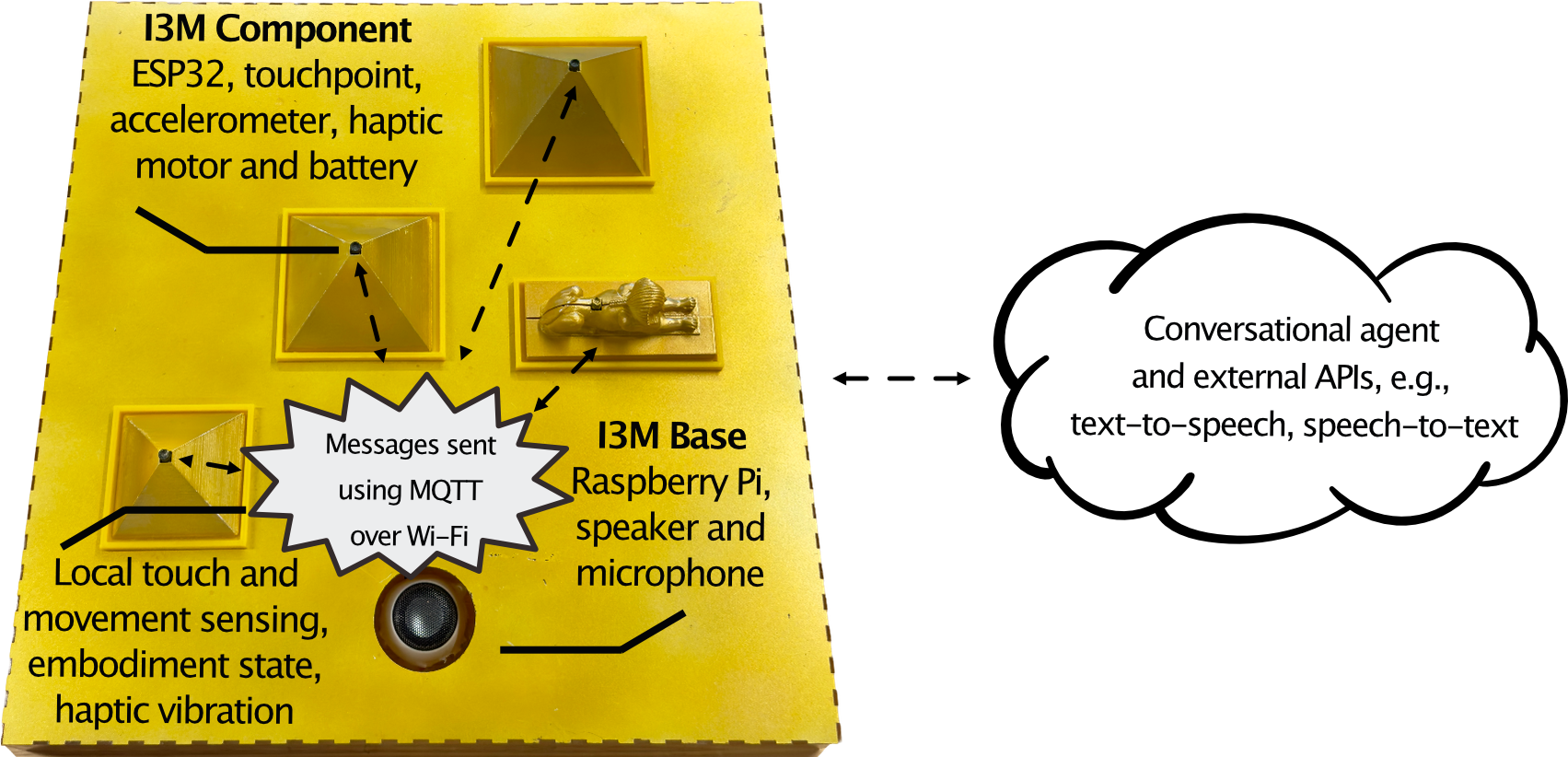}
  %TC:ignore
  \caption{I3M Architecture. Shown is the Pyramid I3M in HEM. The base houses a Raspberry Pi, speaker, and mic. Components enable local sensing, each equipped with a microcontroller, touchpoint, accelerometer, haptic motor, and battery. The Pi manages embodiment state, controls speech input and output, and connects to the conversational agent.}~\label{fig:Architecture}
  \Description{This figure includes an image outlining the architecture of the HEM I3Ms. Shown is a picture of the Egyptian Pyramid I3M, with annotations on top pointing to the various components and hardware it includes. The I3M Base includes a Raspberry Pi, speaker, and microphone. Four I3M Components are highlighted, each including an ESP32, touchpoint, accelerometer, haptic motor and battery. A bubble is depicted in between the Base and Components, highlighting the communication of messages using MQTT over Wi-Fi. These messages include local touch and movement sensing, embodiment state, and haptic vibrations. To the right is a bubble encompassing all the external APIs and libraries that the I3Ms use. This includes Dialogflow for the conversational agent, as well as text-to-speech and speech-to-text libraries.}
  %TC:endignore
\end{figure}

The base of each I3M was constructed from laser-cut acrylic, serving as a stand to hold the constituent components of the I3M and to house a Raspberry Pi, speaker, and microphone (Figure~\ref{fig:Architecture}). The Pi powered each I3M and handled the following responsibilities:

\begin{itemize}
    \item Maintaining a Wi-Fi connection with each I3M component and operating as a message broker (MQTT) to facilitate messaging between the Raspberry Pi and each component.
    \item Managing the embodied state of the I3M by setting the design factors to operate in either HEM or LEM mode.
    \item Controlling speech and auditory output through the connected speaker and microphone, using the Picovoice Porcupine wake word library and Google Cloud Speech-to-Text and Text-to-Speech for speech input and synthesis.
    \item Connecting to the I3M's conversational agent, built using Google Dialogflow, and integrating ChatGPT to perform external searches triggered by Dialogflow's fallback intent.
\end{itemize} 

Each I3M had four 3D-printed components. For the Saturn V Rocket, these were the \textit{Stage A}, \textit{Stage B}, and \textit{Stage C} modules, and the \textit{Launch Tower}. For the Egyptian Pyramid, these included the \textit{Sphinx}, \textit{Great Pyramid}, \textit{Pyramid of Menkaure}, and \textit{Pyramid of Khafre}. Each component had an ESP32 microcontroller embedded in the print, providing localised touch and movement sensing and haptic vibratory output. The microcontroller had integrated Wi-Fi, capacitive touch-sensing GPIO pins, and was connected to a touchpoint 3D-printed with conductive filament, a 3.7V 400mAh lithium polymer battery, an MPU6050 accelerometer and gyroscope, a DRV2605L haptic motor controller, and a vibrating haptic disc.

\subsubsection{Touch Gestures.}
Each I3M component, such as the Sphinx, had a touchpoint that protruded from its surface and was printed using a touch capacitive filament. Touch sensors underwent an automatic calibration process upon I3M startup. Touch gestures were implemented based on findings from our previous co-design work~\cite{Reinders2023}. To enable independent tactile exploration, touchpoints needed to be activated before gestures could be used. Users could perform an \textit{Activate Press} gesture by holding the touchpoint down for one second. This would activate the sub-component and play an audio label identifying it. Once activated, users could perform a \textit{Double Press} gesture to cycle through 12 audio labels that provided different facts and information about the sub-component. For example, for the Sphinx, this included details about its location, the material it was made from, its construction date, purpose, size, and physical appearance. The final gesture supported was a \textit{Long Press}, performed by holding down the touchpoint for two seconds. This gesture invoked the I3M's conversational agent.

\subsubsection{Conversational Agent.}
Users could perform a \textit{Long Press} or use the wake word -- \textit{`Hey Model'} -- to invoke the I3M's conversational agent. The attached microphone recorded user queries, which were processed by the agent built using Google Dialogflow. The agent was trained to answer questions related to each component using a corpus containing all the information extractable via touch gestures. This dataset contained an average of 500 words of facts per component. For example, for the Sphinx, this included details about its missing nose, potential astronomical significance, historical restorations, and symbolic meaning. Like audio labels, speech responses were output through the I3M's speaker, synthesised using the same high-quality voice. If the agent could not answer a query directly, it offered to perform an external search. Upon approval, the agent would send the query to ChatGPT, configured with a context to ensure consistency with the other agent outputs. For example, \textit{``You are an intelligent assistant that answers questions in 25 words or less about ancient Egypt and the Pyramids, including the Pyramid of Menkaure, Khafre, Khufu, and the Sphinx of Giza''}.

\subsubsection{Haptic Vibratory Feedback.}
\label{sec:Haptics}
Each model component contained a haptic disc capable of delivering localised vibratory feedback, generated using the DRV2605 Waveform library. Haptic vibrations were emitted as system feedback when touch gestures were performed, following established recommendations~\cite{Reinders2023}. These corresponded to the type of gesture executed: Activate (\textit{Strong Buzz, 150ms)}, Double (\textit{Strong Short Double Click}), or Long Press (\textit{Strong Buzz, 500ms}).

\subsection{Conversational Embodiment Design Factors}
\label{sec:ConversationalEmbodiment}
In earlier work, we observed that I3M conversational agents should speak with high-quality, human-like voices~\cite{Reinders2023}. When designing our I3Ms, the researchers made the decision that voice quality should be independent of embodiment state, to prevent preferences for high-quality voices from overpowering other design factors.

\subsubsection{DF\#1: Introductions and Small Talk.}
\label{sec:Smalltalk}
Choi \etal~\cite{Choi2020} found that BLV people prefer agents that engage in human-like conversation, as this can help in relationship building. Embodied agents have been designed to introduce themselves to users~\cite{Shamekhi2018,Cassell2001,Luria2019,Lester1997} and engage in small talk~\cite{Liao2018,Pradhan2019,Cassell2001,Shamekhi2018}. In our I3Ms, when configured in HEM, the conversational agent introduces itself to the user, and could detect and respond to small talk during interactions, using a version of Dialogflow's small talk module. To avoid interrupting independent tactile exploration, HEM I3Ms only engage in user-initiated small talk, adhering to an established design recommendation~\cite{Reinders2023}. This design factor can be configured as follows:

\begin{itemize}
    \item \textbf{HEM}: When turned on, the I3M is introduced, \eg\, \textit{``Hello and welcome to the Pyramid model. Let's learn about ancient Egyptian history together!''}. Additionally, when a user initiates small talk, the I3M responds appropriately, \eg\, when greeted with \textit{``Hello Sphinx''}, it replies, \textit{``Hi, how are you?''}.
    
    \item \textbf{LEM}: When an I3M is turned on, a loading message is played, \eg\, \textit{``Loading Pyramid model''}. When a user engages in small talk, the I3M does not respond.
\end{itemize}

\subsubsection{DF\#2: Embodied Personified Voices.}
Influenced by works where users personified and imbued characters into conversational agents~\cite{Purington2017,Pradhan2019,Lester1997}, we designed HEM so that model components could `speak with their own unique voice'. This was achieved by assigning distinct synthesised voices with each model component, emulating aspects of a \textit{one-for-one} social presence~\cite{Luria2019}. HEM components cannot converse among themselves, in line with Luria \etal~\cite{Luria2019}, who observed that users felt discomfort when two active social presences interacted with each other. In contrast, LEM I3Ms employ a singular voice, operating under a \textit{one-for-all} social presence, where all model components are inhabited as a group.

\begin{itemize}
        \item \textbf{HEM}: Each I3M component speaks with its own unique synthesised voice, using a one-for-one social presence.
        
        \item \textbf{LEM}: All I3M components speak with a unified synthesised voice, using a one-for-all social presence.
\end{itemize}

\subsubsection{DF\#3: Embodied Narration Style.}
To further explore the personification of the models, we scripted speech output to be narrated from either a first or third-person perspective. This was influenced by work showing that some users anthropomorphise agents by using first and second-person pronouns~\cite{Liao2018,Coeckelbergh2011}.

\begin{itemize}
        \item \textbf{HEM}: I3M components phrase verbal responses using first-person narration, \eg\, \textit{``I am the Great Sphinx of Egypt. I am a statue of a reclining sphinx, a mythical creature. I have the head of a human and the body of a lion. Many suggest that my nose was lost to erosion, vandalism, or damage''}. This also extended to external search responses fetched using ChatGPT, which had a modified context, \eg\, \textit{``You are the Great Sphinx of Egypt and serve as an intelligent assistant, you answer questions in 25 words or less from a first-person perspective about yourself, ancient Egypt, and the Pyramids''}.
        
        \item \textbf{LEM}: Verbal responses are generated using objective third-person narration, \eg\, \textit{``This is the Great Sphinx of Egypt. It is a statue of a reclining sphinx, a mythical creature with the head of a human and the body of a lion. Many suggest its nose was lost to erosion, vandalism, or damage''}.
\end{itemize}
\newpage
\subsection{Physical Embodiment Design Factors}
\label{sec:PhysicalEmbodiment}
I3Ms are inherently designed to be perceived physically, \ie\, picked up and tactually observed or manipulated. We sought to explore ways in which the tangible nature of I3Ms could be enhanced by physically embodying presence.

\subsubsection{DF\#4: Embodied Vibratory Feedback.}
In our previous research, a BLV user described an I3M component as \textit{`lifeless'} except when it was emitting haptic vibrations~\cite{Reinders2023}. This influenced our focus on richer haptic vibratory feedback, supported by other work exploring how haptics imbue lifelike cues~\cite{Nie2012,Bevan2015, Borgstedt2023}. We designed HEM I3Ms to emit localised haptic vibratory feedback, creating a sense of physical presence. In contrast, LEM I3Ms generate haptics, but only as system feedback confirming gesture inputs. This decision aligns with the recommendations in ~\cite{Reinders2023}, since model usability could be impacted if haptics were turned off entirely.

\begin{itemize}
    \item \textbf{HEM}: I3M components generate haptic vibratory feedback to embody a sense of physical presence. Components use haptics to highlight themselves during interactions, \eg\, when a component identifies itself it emits a localised vibration (\textit{Transition Ramp Up - 0 to 100\%}), or when it is referenced during an auditory response (\textit{Strong Buzz, 1000ms}).

    \item \textbf{LEM}: Haptics are used only to confirm when a touch gesture has been correctly performed. Components do not use localised haptics to embody physical presence. 
\end{itemize}

\subsubsection{DF\#5: Location of Speech Output.}
Pradhan \etal~\cite{Pradhan2019} identified that an agent's proximity can influence how human-like it is perceived. Agents closer to the user, or capable of operating across multiple devices to broadcast ubiquity, are often thought of as more present. Due to size constraints, we could not embed speakers directly into I3M components. However, the location of the auditory output can be configured to vary by the position of the speaker.

\begin{itemize}
    \item \textbf{HEM}: The speaker used by the I3M to output speech is housed within the enclosure.
    \item \textbf{LEM}: The speaker used by the I3Ms to output speech is positioned externally from the enclosure, approximately 30cm to the left side of the model, reducing proximity.
\end{itemize}

\section{User Study -- Methodology}
\subsection{Hypotheses}
\label{sec:Hypotheses}
We designed a controlled user study to explore and understand whether I3Ms configured with different embodied design factors influence BLV end-users' perceptions of model embodiment, engagement, and trustworthiness. The study utilised both the Saturn V Rocket I3M and Egyptian Pyramids I3M, each configurable into two states -- \textit{High Embodied Mode (HEM)} and \textit{Low Embodied Mode (LEM)} -- as described in Section~\ref{sec:Design}. We hypothesised that:

\begin{itemize}[leftmargin=2mm, rightmargin=-1mm]
    \item \textbf{H\#1:} HEM I3Ms are perceived as more \textit{embodied} than LEM I3Ms
    \item \textbf{H\#2:} HEM I3Ms are perceived as more \textit{engaging} than LEM I3Ms
    \item \textbf{H\#3:} HEM I3Ms are perceived as more \textit{trustworthy} than LEM I3Ms
\end{itemize}
\newpage
\subsection{Participants}
Twelve BLV participants were recruited from our lab's participant contact pool (Table~\ref{tab:participants}). This sample size ($n=12$) falls within the range commonly seen in BLV accessibility studies, which often involve anywhere between 6-12 participants~\cite{Holloway2022,Nagassa2023,Shi2017b,Shi2020} due to the low incidence of blindness in the general population and associated recruitment challenges~\cite{Butler2021}.

Participants ranged in age from 27 to 78 years (\textit{$\mu$} = 50, \textit{$\sigma$} = 16.6). Nine participants self-reported as totally blind, while three reported being legally blind with low levels of light perception. Participants also varied in their prior experience with tactile graphics: Many reported substantial exposure and confidence ($n=5$), some reported some use but lacked confidence ($n=4$), and others reported limited or no exposure ($n=1$ and $n=2$, respectively). 

Familiarity with 3D-printed models was slightly less common. All participants regularly used conversational agents, including Google Assistant ($n=11$), Siri ($n=10$), Alexa ($n=4$), and ChatGPT ($n=4$). These interfaces were accessed on various devices, such as smartphones ($n=12$), smart speakers/displays ($n=11$), smartwatches ($n=6$), computers ($n=7$) and tablets ($n=5$).

\begin{table}[!htbp]
\small
\caption{Participant demographic information, detailing level of vision, accessible formats used, familiarity with tactile graphics and 3D models, and use of conversational agents.\label{tab:participants}}
\Description{Participant demographic information, level of vision, accessible formats used, familiarity with tactile graphics and 3D models, and use of conversational agents.}
\label{tab:table1}
\centering
\begin{tabular}{|l|p{0.15cm}|p{0.15cm}|p{0.15cm}|p{0.15cm}|p{0.15cm}|p{0.15cm}|p{0.15cm}|p{0.15cm}|p{0.15cm}|p{0.15cm}|p{0.15cm}|p{0.15cm}|} 
\hline
\cellcolor[HTML]{EFEFEF}\textbf{Participant \#:} & \cellcolor[HTML]{EFEFEF}\textbf{1} & \cellcolor[HTML]{EFEFEF}\textbf{2} & \cellcolor[HTML]{EFEFEF}\textbf{3} & \cellcolor[HTML]{EFEFEF}\textbf{4} & \cellcolor[HTML]{EFEFEF}\textbf{5} & \cellcolor[HTML]{EFEFEF}\textbf{6} & \cellcolor[HTML]{EFEFEF}\textbf{7} & \cellcolor[HTML]{EFEFEF}\textbf{8} &
\cellcolor[HTML]{EFEFEF}\textbf{9} & \cellcolor[HTML]{EFEFEF}\textbf{10} & \cellcolor[HTML]{EFEFEF}\textbf{11} &
\cellcolor[HTML]{EFEFEF}\textbf{12}\\ 
\hline
\multicolumn{13}{|l|}{\cellcolor[HTML]{EFEFEF}\textbf{Level of Vision:}} \\ 
\hline
Legally Blind & \checkmark & - & \checkmark & - & - & - & - & - & - & - & \checkmark & - \\ 
\hline
Totally Blind & - & \checkmark & - & \checkmark & \checkmark & \checkmark & \checkmark & \checkmark & \checkmark & \checkmark & - & \checkmark \\ 
\hline
\multicolumn{13}{|l|}{\cellcolor[HTML]{EFEFEF}\textbf{Accessible Formats Used:}} \\ 
\hline
Braille & \checkmark & \checkmark & \checkmark & \checkmark & \checkmark & \checkmark & - & \checkmark & \checkmark & \checkmark & \checkmark & \checkmark \\ 
\hline
Audio & \checkmark & \checkmark & \checkmark & \checkmark & \checkmark & \checkmark & \checkmark & \checkmark & \checkmark & \checkmark & \checkmark & \checkmark \\ 
\hline
Tactile Graphics & \checkmark & \checkmark & \checkmark & \checkmark & \checkmark & \checkmark & - & \checkmark & \checkmark & \checkmark & \checkmark & \checkmark \\ 
\hline
3D Models  & - & \checkmark & \checkmark & \checkmark & \checkmark & \checkmark & - & - & - & - & \checkmark & \checkmark \\ 
\hline
\multicolumn{13}{|l|}{\cellcolor[HTML]{EFEFEF}\textbf{Familiarity (1: Not Familiar - 4: Very Familiar):}} \\ 
\hline
Tactile Graphics & 2 & 3 & 3 & 4 & 4 & 4 & 1 & 3 & 4 & 1 & 4 & 3 \\ 
\hline
3D Models & 1 & 2 & 3 & 4 & 4 & 4 & 1 & 2 & 3 & 1 & 4 & 3 \\ 
\hline
\multicolumn{13}{|l|}{\cellcolor[HTML]{EFEFEF}\textbf{Conversational Interfaces Used:}} \\ 
\hline
Alexa & - & \checkmark & - & - & - & \checkmark & - & \checkmark & - & \checkmark & - & - \\ 
\hline
ChatGPT & \checkmark & \checkmark & - & - & \checkmark & - & - & \checkmark & - & - & - & - \\ 
\hline
Google Assistant & \checkmark & \checkmark & \checkmark & \checkmark & \checkmark & \checkmark & \checkmark & \checkmark & \checkmark & - & \checkmark & \checkmark \\ 
\hline
Siri & - & \checkmark & \checkmark & \checkmark & \checkmark & \checkmark & \checkmark & \checkmark & \checkmark & - & \checkmark & \checkmark \\ 
\hline
%\multicolumn{13}{|l|}{\cellcolor[HTML]{EFEFEF}\textbf{Accessed on Devices:}} \\ 
%\hline
%Smartphone & \checkmark & \checkmark & \checkmark & \checkmark & \checkmark & \checkmark & \checkmark & \checkmark & \checkmark & \checkmark & \checkmark & \checkmark \\ 
%\hline
%Tablet & - & \checkmark & - & \checkmark & - & \checkmark & - & - & \checkmark & - & \checkmark & - \\ 
%\hline
%Smartwatch & - & \checkmark & \checkmark & \checkmark & \checkmark & \checkmark & - & \checkmark & - & - & - & - \\  
%\hline
%Speaker/Display & - & \checkmark & \checkmark & \checkmark & \checkmark & \checkmark & \checkmark & \checkmark & \checkmark & \checkmark & \checkmark & \checkmark \\  
%\hline
%Computer & \checkmark & - & - & \checkmark & \checkmark & \checkmark & \checkmark & \checkmark & \checkmark & - & - & - \\ 
%\hline
\end{tabular}
\end{table}

\subsection{Study Measures}
\label{sec:StudyMeasures}

We used a series of questionnaires and asked participants to subjectively rate the I3Ms in order to investigate our hypotheses.

For \textbf{Hypothesis H\#1}, to measure how \textit{embodied} participants perceived the I3Ms, we utilised the \textbf{Godspeed Questionnaire Series (GQS)}~\cite{Bartneck2009}. Originally devised to measure users' social perceptions of robot and agent-based systems, GQS subscales such as ~\textit{anthropomorphism} and \textit{intelligence} have recently been applied to measure aspects of embodiment and liveliness in conversational agents ~\cite{Shamekhi2018} and robots with human-like abilities~\cite{Kontogiorgos2020}. We selected four subscales -- \textit{anthropomorphism, animacy, likeability}, and \textit{intelligence} -- as we felt each provides insight into components of embodied sociability. For example, \textit{anthropomorphism} captures the attribution of human-like characteristics, \textit{animacy} reflects perceptions of liveliness, \textit{likeability} gauges formation of positive impressions, and \textit{intelligence} focuses on perception of ability. Additionally, we also explored model embodiment by asking participants for their subjective perceptions by expressing the concept of `embodiment/embodied' using the terms `lively/liveliness' to ensure clarity. We felt this terminology would have more meaning to participants, and aligns with previous works that have used the similar terms `lifelike/lifelikeness' to describe embodied agents~\cite{Lester1997,Cassell1999,Cassell1999b,Cassell1999c,Lester1999}.

To explore \textbf{Hypothesis H\#2}, we used two \textit{engagement} measures -- the \textbf{User Engagement Scale [Short Form] (UES-SF)} and the \textbf{Playful Experiences Questionnaire (PLEXQ)}. The UES-SF measures user engagement as the depth of a user's perceived investment with a system~\cite{OBrien2016}. It consists of 12 five-point Likert items across four subscales -- \textit{focused attention, perceived usability, aesthetic appeal}, and \textit{reward}~\cite{Obrien2018}. The UES-SF has been widely applied across HCI to measure engagement in contexts such as interactive media~\cite{carlton2019}, video games~\cite{wiebe2014}, and 3D-printed building plans for BLV people~\cite{Nagassa2023}. We used all four subscales. To complement the UES-SF, we also used PLEXQ, which measures playfulness, pleasurable experiences, and playful engagement~\cite{Boberg2015}. PLEXQ is commonly used to assess how engaging games and game-like experiences are~\cite{Bischof2016,Cho2024}. We used eight subscales that we felt were most relevant to I3Ms -- \textit {captivation, challenge, control, discovery, exploration, humor, relaxation}, and \textit{sensation}. 

To supplement these measures of perceived engagement, we also captured two behavioural metrics (time spent and interactions performed during tasks), as more time spent with the model and more interactions may indicate greater immersion and enjoyment~\cite{OBrien2013,Doherty2018}. Note that comparing these between the LEM and HEM conditions was meaningful as the length of responses in both conditions was similar and every interaction type was available in both modes.

To investigate \textbf{Hypothesis H\#3} and measure \textit{trust}, we utilised the \textbf{Human-Computer Trust Model (HCTM)}. The HCTM conceptualises trust as a multifaceted construct, encompassing users' perceptions of the \textit{perceived risk, benevolence, competence}, and \textit{reciprocity} during interactions. These perceptions can influence users' reliance on a system and their likelihood of continued use~\cite{Gulati2019}. Previous work has demonstrated the utility of HCTM in assessing trustworthiness in conversational agents like Siri~\cite{Gulati2018}, machine learning systems~\cite{Guo2022}, and other human-like technologies such as large language models~\cite{Salah2023} and chatbots~\cite{Degachi2023}.

Based on pilot study feedback, we made minor adjustments to specific subscale items in the UES-SF and PLEXQ measures. For instance, the \textit{aesthetic appeal} (UES-SF) and \textit{sensation} (PLEXQ) subscales were adjusted, as concepts of ``attractiveness'', ``aesthetics'', and ``visuals'' held little meaning to our pilot user in BLV contexts. They recommended adding the phrase ``to my senses'' to these items. Per ~\cite{Obrien2018}'s guidance on modifying the UES-SF, we did not report an overall UES score, instead focusing on individual components of engagement. These adjustments also motivated our decision to add our own questions gathering participants' subjective ratings of the I3Ms, supplementing the validated scales and providing additional nuance and insight. The modified UES-SF and PLEXQ, along with GSQ and HCTM, are provided in the Appendices.

\subsection{Experiment Conditions}
Our user study used a within-subject design. All participants were exposed to (one) LEM and (one) HEM-configured I3M. To control for bias related to model type (Rocket/Pyramids) and design factor state (LEM/HEM), the order in which the I3Ms were presented, along with their associated LEM/HEM configuration, was counterbalanced. The activities that participants completed with each I3M remained the same, regardless of its LEM/HEM state, and could be completed in either mode without significant difficulty.

\subsection{Procedure}
\label{sec:StudyProcedure}
Each user study session lasted approximately two hours and included at least one researcher being present. Sessions began with the researcher providing an overview of the research project, and were divided into the following stages:

\begin{enumerate}[leftmargin=5.5mm]
    \item \textbf{Training}. Participants were guided through a 10-minute training exercise, which allowed them to familiarise themselves with how I3Ms operate. We designed and built an I3M that represented a non-descript sphere for training. Participants were first asked to explore the I3M tactually when it was turned off, before being taught how to extract basic information from the I3M using touch gestures, and asking questions through the conversational interface. The training I3M did not operate in either LEM or HEM mode and would, using a low-quality synthesised voice, only respond by confirming when interactions had been successfully performed (\eg\, ``Double Press'', ``Recording query''). Operating outside of LEM/HEM states was a deliberate design decision to allow training of basic interaction functionality without biasing future LEM/HEM exposure.
    \newline
    \item \textbf{Exposure to LEM/HEM I3Ms.}
    \begin{enumerate}
        \item \textbf{Activities}. Participants were introduced to their first I3M, configured in one of the LEM/HEM states, and completed a walkthrough activity. The I3M identified and described each component it included, with participants given the opportunity to tactually explore the I3M throughout. These walkthroughs were carefully curated so that participants encountered the majority of system functionality specific to the HEM/LEM design factors\footnote{The only exception to this was the small talk component of the \textit{introductions and small talk} design factor, which, unlike all other design factors that were explicit, required user initiation (see Section~\ref{sec:Smalltalk}).}. After the walkthrough, participants were given up to five minutes to \textit{explore} the models, during which they could interact with the I3M in any way they wished. Participants were then asked to complete four \textit{researcher-directed} information-gathering tasks (\eg\, finding out how long it took to build the Great Pyramid, what happened to the nose of the Sphinx, or the significance of the Pyramids). Participants could access this information using their choice of either touch gesture interaction or using the model's conversational agent. Before concluding, participants were given up to an additional five minutes for a \textit{free play} task designed to mimic undirected, real-world use. They were instructed to discover something interesting about the modelled concept that they were not aware of prior.

        \item \textbf{Questionnaire Scales}. Participants were taken through the questionnaire scales -- GSQ, PLEXQ, UES-SF, and HCTM. In addition, participants were asked to rate how \textit{lively}, \textit{engaging}, and \textit{trustworthy} the I3M was, using 5-point Likert scales. On average, these took 15 minutes to complete.
        \newline
        \item \textbf{Remaining Model}. Participants would then complete 2(a) and 2(b) again with the second I3M, configured in the remaining LEM/HEM state. Participants spent an average of 30 minutes with each I3M.
        \newline
    \end{enumerate}
    \item \textbf{Semi-Structured Interview}. At the end of the session, participants were asked questions about specific interactions with the LEM/HEM I3Ms and were asked to rank how \textit{lively}, \textit{engaging} and \textit{trustworthy} each model was. We also asked about the role each design factor played, and whether they impacted perceptions of how \textit{lively}, \textit{engaging}, and \textit{trustworthy} the I3Ms were. On average, it took 20 minutes to answer these questions.
\end{enumerate}

\subsection{Data Collection \& Analysis}
All sessions were video-recorded and subsequently transcribed. Collected data included responses to scale questions, semi-structured interview questions, and participant comments made during task completion. The time participants spent completing tasks, as well as the number of interactions they performed, were also recorded. Descriptive statistics were calculated on all subscale responses -- GSQ, UES-SF, PLEXQ, and HCTM -- as well as for interview questions that ranked the I3Ms and individual embodiment design factors. 

For data that did not follow a normal distribution (\eg\ our scale data), we conducted non-parametric statistical tests, specifically Wilcoxon signed-rank tests. Binomial tests were performed on the rankings of the I3Ms and the impact ratings of individual embodiment design factors. Paired t-tests were conducted on data that was normally distributed (\eg\ time taken to complete tasks). We opted for one-tailed tests because our hypotheses were explicit in nature (described in Section~\ref{sec:Hypotheses}). This approach was deliberate, allowing us to dedicate more power to detecting effects in one direction.

The analysis should be interpreted in light of both the exploratory nature of our user study and our small sample size ($n$ $=$ 12). This influenced our approach in two ways. First, from the outset, due to our small sample size, we decided not to use $p$ $<$ 0.05 as the sole determinant of significance~\cite{Shamekhi2018,Cramer2004}, instead marking results $p$ $<$ 0.05 as \textit{significant} ($^\ast$) and 0.05 $<=$ $p$ $<=$ 0.1 as \textit{marginally significant} ($^\wedge$). Second, given the exploratory nature of our work, in order to reduce the risk of overlooking meaningful results (false negatives), we chose not to apply corrections for multiple comparisons (although this does increase the risk of false positives).

\section{Results}
\label{sec:Results}
Results are presented based on our hypotheses (Section~\ref{sec:Hypotheses}) and separated across the \textbf{embodiment} of the I3Ms, \textbf{engagement}, and \textbf{trust}. Each section presents quantitative results, including questionnaire scales, rankings of the models and HEM/LEM-configured design factors, and qualitative results, in the form of participant responses from the semi-structured interview.

\subsection{Embodiment of I3Ms}
\subsubsection{\textbf{GSQ Scales.}}
HEM I3Ms elicited higher mean scores compared to I3Ms with LEM across all GSQ subscales -- \textit{anthropomorphism ($\mu$ $=$ +0.67), animacy ($\mu$ $=$ +0.46), likeability ($\mu$ $=$ +0.27)}, and \textit{intelligence ($\mu$ $=$ +0.29)}. We conducted one-tailed Wilcoxon signed-rank tests for each GSQ subscale (Table~\ref{table:1}), using design configuration (HEM, LEM) as the independent variable to assess significance. The positive effect of HEM was statistically significant across all GSQ subscales -- \textit{anthropomorphism} ($p=0.011$), \textit{animacy} ($p=0.029$), \textit{likeability} ($p=0.008$), and \textit{intelligence} ($p=0.010$).

\begin{table*}[b!]
%TC:ignore
\caption{Impact of HEM and LEM conditions on I3M social perceptions and embodiment using four GSQ subscales. Participants also rated how lively the I3Ms felt. Descriptive stats, including median (M), mean ($\mu$), standard deviation (SD), and variance (V) are shown for both conditions. Wilcoxon signed-rank texts assess significance, with $p < 0.05$ considered significant ($^\ast$), $0.05 \leq p \leq 0.1$ marginally significant ($^\wedge$), and one-tailed $z > 1.28$ and $z > 1.645$ indicating 90\% and 95\% confidence, respectively.}
%TC:endignore
\label{table:1}
\centering
\begin{tabular}{@{}cccccccccccclll@{}}
\toprule
\multicolumn{15}{c}{\textit{\textbf{Godspeed Questionnaire}}} \\ \midrule
\multicolumn{1}{c|}{\multirow{2}{*}{}} &
  \multicolumn{8}{c|}{\textbf{Descriptive Statistics}} &
  \multicolumn{6}{c}{\textbf{Wilcoxon Test}} \\
\multicolumn{1}{c|}{} &
  \multicolumn{4}{c|}{\textbf{HEM I3Ms}} &
  \multicolumn{4}{c|}{\textbf{LEM I3Ms}} &
  \multicolumn{6}{c}{} \\
\multicolumn{1}{c|}{\textbf{Scale}} &
 \textit{M} &
  $\mu$ &
 \textit{SD} &
  \multicolumn{1}{c|}{V} &
 \textit{M} &
  $\mu$ &
 \textit{SD} &
  \multicolumn{1}{c|}{V} &
 \textit{z-score} &
 \textit{p-value} &
  \multicolumn{4}{c}{} \\ \midrule
\multicolumn{1}{l|}{\textbf{Anthropomorphism}} &
  4.00 &
  3.75 &
  1.36 &
  \multicolumn{1}{c|}{1.85} &
  3.00 &
  3.08 &
  1.35 &
  \multicolumn{1}{c|}{1.83} &
  2.288 &
  \multicolumn{1}{l}{0.011$^\ast$} &
  \multicolumn{4}{c}{} \\
\multicolumn{1}{l|}{\textbf{Animacy}} &
  5.00 &
  4.17 &
  1.14 &
  \multicolumn{1}{c|}{1.31} &
  4.00 &
  3.71 &
  1.37 &
  \multicolumn{1}{c|}{1.87} &
  1.895 &
  \multicolumn{1}{l}{0.029$^\ast$} &
  \multicolumn{4}{c}{} \\
\multicolumn{1}{l|}{\textbf{Likeability}} &
  5.00 &
  4.58 &
  0.68 &
  \multicolumn{1}{c|}{0.47} &
  4.50 &
  4.31 &
  0.78 &
  \multicolumn{1}{c|}{0.60} &
  2.428 &
  \multicolumn{1}{l}{0.008$^\ast$} &
  \multicolumn{4}{c}{} \\
\multicolumn{1}{l|}{\textbf{Intelligence}} &
  5.00 &
  4.46 &
  0.71 &
  \multicolumn{1}{c|}{0.50} &
  4.50 &
  4.17 &
  0.90 &
  \multicolumn{1}{c|}{0.81} &
  2.333 &
  \multicolumn{1}{l}{0.010$^\ast$} &
  \multicolumn{4}{c}{} \\ \midrule
\multicolumn{15}{c}{\textit{\textbf{Did you find that the I3M felt lively?}}} \\ \midrule
\multicolumn{1}{l|}{\textbf{Liveliness}} &
  5.00 &
  4.50 &
  0.65 &
  \multicolumn{1}{c|}{0.42} &
  4.00 &
  4.00 &
  0.91 &
  \multicolumn{1}{c|}{0.83} &
  2.121 &
  \multicolumn{1}{l}{0.017$^\ast$} &
  \multicolumn{4}{c}{} \\ \bottomrule
\end{tabular}
\end{table*}
\begin{table*}[b!]
%TC:ignore
\caption{Impact of design factors on the perceived liveliness of HEM I3Ms. Descriptive stats are provided. Significance is tested using Binomial tests, with $p < 0.05$ considered significant ($^\ast$), $0.05 \leq p \leq 0.1$ marginally significant ($^\wedge$). Likert responses were collapsed to a binary scale (N: Strongly Disagree, Disagree, Neutral, and Y: Agree, Strongly Agree). The number of successes (\textit{k}), trials (\textit{n}) and confidence intervals are included.}
%TC:endignore
\label{table:5}
\begin{tabular}{@{}cccccccc@{}}
\toprule
\multicolumn{8}{c}{\textit{\textbf{Did the design factors impact your perception of how lively the HEM I3M was?}}}               \\ \midrule
\multicolumn{1}{c|}{\multirow{3}{*}{\textbf{Design Factor}}} &
  \multicolumn{4}{c|}{\multirow{2}{*}{\textbf{Descriptive Statistics}}} &
  \multicolumn{3}{c}{\multirow{2}{*}{\textbf{Binomial Test}}} \\
\multicolumn{1}{c|}{}                                     & \multicolumn{4}{c|}{}                          & \multicolumn{3}{c}{}       \\
\multicolumn{1}{c|}{} &
  \textit{M} &
  \textit{$\mu$} &
  \textit{SD} &
  \multicolumn{1}{c|}{\textit{V}} &
  \textit{k, n} &
  \textit{p-value} &
  \textit{95th CI} \\ \midrule
\multicolumn{1}{l|}{\textbf{DF\#1: Introductions \& Small Talk}} & 4.00 & 4.17 & 0.55 & \multicolumn{1}{c|}{0.31} & 11, 12 & \multicolumn{1}{r}{<0.001$^\ast$} & 0.661, 1.000 \\
\multicolumn{1}{l|}{\textbf{DF\#2: Embodied Personified Voices}} & 4.00 & 4.25 & 0.72 & \multicolumn{1}{c|}{0.52} & 10, 12 & \multicolumn{1}{r}{0.003$^\ast$} & 0.562, 1.000 \\
\multicolumn{1}{l|}{\textbf{DF\#3: Embodied Narration Style}}             & 4.00 & 4.17 & 0.90 & \multicolumn{1}{c|}{0.81} & 10, 12 & \multicolumn{1}{r}{0.003$^\ast$} & 0.562, 1.000 \\
\multicolumn{1}{l|}{\textbf{DF\#4: Embodied Vibratory Feedback}}   & 4.50 & 4.33 & 0.85 & \multicolumn{1}{c|}{0.72} & 11, 12 & \multicolumn{1}{r}{<0.001$^\ast$} & 0.661, 1.000 \\
\multicolumn{1}{l|}{\textbf{DF\#5: Location of Speech Output}}                        & 4.00 & 4.00 & 0.82 & \multicolumn{1}{l|}{0.67} & 8, 12 & \multicolumn{1}{r}{0.057$^\wedge$} & 0.391, 1.000 \\ \bottomrule
\end{tabular}
\end{table*}

\subsubsection{\textbf{How Lively Were The I3Ms?}}
\label{sec:PerceptionLiveliness}
Participants rated how \textit{lively} each I3M was immediately after being exposed to it. HEM I3Ms were perceived as more \textit{lively} ($\mu$ $=$ +0.50) compared to the LEM configuration (Table~\ref{table:1}). This was statistically significant ($p=0.017$). In the post-activity interview, participants ranked the I3Ms based on perceived liveliness. Two-thirds of participants ($n=8$) indicated that the HEM I3Ms had higher liveliness compared to the LEM I3Ms, with the remainder split between no difference ($n=3$) and the LEM configuration ($n=1$). A binomial test revealed the difference between the number of participants who ranked the HEM I3Ms higher and those who either selected the LEM I3M or could tell no difference was statistically significant ($k=8$, $n=12$, $p=0.019$).

Most participants were emphatic in their selection. P2 described how the HEM I3M's design factors \textit{``brought it [the model] to life''}, continuing, \textit{``it [the HEM I3M] created a relationship, [it is] like dealing with something that is alive, it *is* talking to you''}. P8 referred to the HEM I3M as \textit{``more human-like and interactive, more natural''}, while P11 noted that the HEM I3M \textit{``was more like an entity... less of a computer program''}. Similarly, P4 mentioned that the HEM I3M \textit{``seemed to want to interact with me ... [whereas] the other one could have been talking to the moon''}. Six participants explicitly referred to the HEM I3M as being \textit{``more human-like''} in their explanations (P3, P4, P8, P10, P11 and P12), with P3 also stating that the LEM I3M was \textit{``too machine-like''}. However, P5 felt that the HEM and LEM configurations appeared just as lively as one another, stating that \textit{``they were both pretty active''}. One participant (P1) selected the LEM Rocket I3M as the most lively, citing specific elements of that model's design as the determining factor. P1 explained, \textit{``the rocket... [its] three sections made it more real''} (P1).
\newpage
\subsubsection{\textbf{Impact of Design Factors on Liveliness.}}
Participants rated how each HEM design factor influenced their perception of I3M \textit{liveliness} using a 5-point Likert scale (Table~\ref{table:5}). All factors appeared to influence perceptions of the liveliness of the HEM I3Ms. \textit{Embodied vibratory feedback} elicited the highest mean score, while \textit{location of speech output} scored the lowest. Binomial tests revealed that the difference between the number of participants who agreed or strongly agreed, and those who were neutral or below, was statistically significant for all factors, apart from \textit{location of speech output}, which was marginally significant ($k=8$, $n=12$, $p=0.057$).

\subsection{Engagement of I3Ms}
\subsubsection{\textbf{UES \& PLEXQ Scales.}}
HEM I3Ms elicited higher mean scores across three of the four UES-SF subscales (Table~\ref{table:4}) -- \textit{focused attention ($\mu$ $=$ +0.17), perceived usability ($\mu$ $=$ +0.45)}, and \textit{aesthetic appeal ($\mu$ $=$ +0.14)}. Wilcoxon results for these subscales were significant for both \textit{perceived usability} ($p=0.001$) and \textit{aesthetic appeal} ($p=0.029$), but marginally significant for \textit{focused attention} ($p=0.067$). The \textit{reward} subscale, however, showed a higher mean score for LEM I3Ms ($\mu$ $=$ +0.11), and was non-significant.

HEM I3Ms exhibited marginally higher mean scores across seven PLEXQ subscales. Significant results were found for \textit{control} ($\mu$ $=$ +0.22, $p=0.016$) and \textit{sensation} ($\mu$ $=$ +0.17, $p=0.029$), while \textit{humor} ($\mu$ $=$ +0.16) was marginally significant ($p=0.092$). One subscale, \textit{exploration}, showed a higher mean score for LEM I3Ms ($\mu$ $=$ +0.08).

\begin{table*}[b!]
\caption{Impact of HEM and LEM conditions on the engagement of the I3Ms, using four UES-SF and eight PLEXQ subscales. Participants were also asked to rate how engaging the I3Ms felt.}
\label{table:4}
\begin{tabular}{@{}cccccccccccclll@{}}
\toprule
\multicolumn{15}{c}{\textit{\textbf{User Engagement Scale}}} \\ \midrule
\multicolumn{1}{c|}{\multirow{2}{*}{}} &
  \multicolumn{8}{c|}{\textbf{Descriptive Statistics}} &
  \multicolumn{6}{c}{\textbf{Wilcoxon Test}} \\
\multicolumn{1}{c|}{} &
  \multicolumn{4}{c|}{\textbf{HEM I3Ms}} &
  \multicolumn{4}{c|}{\textbf{LEM I3Ms}} &
  \multicolumn{6}{c}{\textbf{}} \\
\multicolumn{1}{c|}{\textbf{Scale}} &
  \textit{M} &
  $\mu$ &
  \textit{SD} &
  \multicolumn{1}{c|}{V} &
  \textit{M} &
  $\mu$ &
  \textit{SD} &
  \multicolumn{1}{c|}{V} &
  \textit{z-score} &
  \textit{p-value} &
  \multicolumn{4}{c}{} \\ \midrule
\multicolumn{1}{l|}{\textbf{Focused Attention}} &
  4.00 &
  3.86 &
  0.95 &
  \multicolumn{1}{c|}{0.90} &
  4.00 &
  3.69 &
  1.00 &
  \multicolumn{1}{c|}{0.99} &
  1.500 &
  \multicolumn{1}{l}{0.067$^\wedge$} &
  \multicolumn{4}{c}{} \\
\multicolumn{1}{l|}{\textbf{Perceived Usability}} &
  5.00 &
  4.56 &
  0.55 &
  \multicolumn{1}{c|}{0.30} &
  4.00 &
  4.11 &
  0.84 &
  \multicolumn{1}{c|}{0.71} &
  3.025 &
  \multicolumn{1}{l}{0.001$^\ast$} &
  \multicolumn{4}{c}{} \\
\multicolumn{1}{l|}{\textbf{Aesthetic Appeal}} &
  5.00 &
  4.53 &
  0.60 &
  \multicolumn{1}{c|}{0.36} &
  4.00 &
  4.39 &
  0.64 &
  \multicolumn{1}{c|}{0.40} &
  1.889 &
  \multicolumn{1}{l}{0.029$^\ast$} &
  \multicolumn{4}{c}{} \\
\multicolumn{1}{l|}{\textbf{Reward}} &
  5.00 &
  4.64 &
  0.54 &
  \multicolumn{1}{c|}{0.29} &
  5.00 &
  4.75 &
  0.43 &
  \multicolumn{1}{c|}{0.19} &
  -2.000 &
  \multicolumn{1}{l}{0.977} &
  \multicolumn{4}{c}{} \\ \midrule
\multicolumn{15}{c}{\textit{\textbf{Playful Experiences Questionnaire}}} \\ \midrule
\multicolumn{1}{l|}{\textbf{Captivation}} &
  3.00 &
  3.03 &
  1.34 &
  \multicolumn{1}{c|}{1.80} &
  3.00 &
  3.00 &
  1.20 &
  \multicolumn{1}{c|}{1.44} &
  0.233 &
  \multicolumn{1}{l}{0.408} &
  \multicolumn{4}{c}{} \\
\multicolumn{1}{l|}{\textbf{Challenge}} &
  5.00 &
  4.61 &
  0.49 &
  \multicolumn{1}{c|}{0.24} &
  5.00 &
  4.58 &
  0.55 &
  \multicolumn{1}{c|}{0.30} &
  0.333 &
  \multicolumn{1}{l}{0.369} &
  \multicolumn{4}{c}{} \\
\multicolumn{1}{l|}{\textbf{Control}} &
  4.00 &
  4.25 &
  0.76 &
  \multicolumn{1}{c|}{0.58} &
  4.00 &
  4.03 &
  0.76 &
  \multicolumn{1}{c|}{0.58} &
  2.138 &
  \multicolumn{1}{l}{0.016$^\ast$} &
  \multicolumn{4}{c}{} \\
\multicolumn{1}{l|}{\textbf{Discovery}} &
  5.00 &
  4.50 &
  0.60 &
  \multicolumn{1}{c|}{0.36} &
  5.00 &
  4.42 &
  0.72 &
  \multicolumn{1}{c|}{0.52} &
  0.905 &
  \multicolumn{1}{l}{0.183} &
  \multicolumn{4}{c}{} \\
\multicolumn{1}{l|}{\textbf{Exploration}} &
  4.00 &
  4.42 &
  0.60 &
  \multicolumn{1}{c|}{0.35} &
  5.00 &
  4.50 &
  0.60 &
  \multicolumn{1}{c|}{0.36} &
  -1.732 &
  \multicolumn{1}{l}{0.954} &
  \multicolumn{4}{c}{} \\
\multicolumn{1}{l|}{\textbf{Humor}} &
  4.00 &
  4.22 &
  0.85 &
  \multicolumn{1}{c|}{0.73} &
  4.00 &
  4.06 &
  0.97 &
  \multicolumn{1}{c|}{0.94} &
  1.328 &
  \multicolumn{1}{l}{0.092$^\wedge$} &
  \multicolumn{4}{c}{} \\
\multicolumn{1}{l|}{\textbf{Relaxation}} &
  5.00 &
  4.31 &
  0.91 &
  \multicolumn{1}{c|}{0.82} &
  4.00 &
  4.22 &
  0.82 &
  \multicolumn{1}{c|}{0.67} &
  0.676 &
  \multicolumn{1}{l}{0.249} &
  \multicolumn{4}{c}{} \\
\multicolumn{1}{l|}{\textbf{Sensation}} &
  5.00 &
  4.53 &
  0.60 &
  \multicolumn{1}{c|}{0.36} &
  4.00 &
  4.36 &
  0.67 &
  \multicolumn{1}{c|}{0.45} &
  1.897 &
  \multicolumn{1}{l}{0.029$^\ast$} &
  \multicolumn{4}{c}{} \\ \midrule
\multicolumn{15}{c}{\textit{\textbf{Did you find that the I3M felt engaging?}}} \\ \midrule
\multicolumn{1}{l|}{\textbf{Engagement}} &
  5.00 &
  4.67 &
  0.47 &
  \multicolumn{1}{c|}{0.22} &
  4.00 &
  4.25 &
  0.60 &
  \multicolumn{1}{c|}{0.35} &
  2.236 &
  \multicolumn{1}{l}{0.013$^\ast$} &
  \multicolumn{4}{c}{} \\ \bottomrule
\end{tabular}
\end{table*}
\newpage
\subsubsection{\textbf{How Engaging Were The I3Ms?}}
\label{sec:PerceptionEngagement}
When asked to rate how \textit{engaging} each I3M was, results indicated that HEM I3Ms were more \textit{engaging} ($\mu$ $=$ +0.42) than LEMs (Table~\ref{table:4}). This difference was statistically significant ($p=0.013$). In the post-activity interview, the majority of participants ($n=10$) ranked the HEM I3Ms as more engaging than the LEM condition, while the remaining participants ($n=2$) found no difference. A binomial test revealed that these rankings were statistically significant ($k=10$, $n=12$, $p<0.001$).

Participants clearly articulated their reasons, with seven directly referencing the HEM I3M as being either \textit{``more engaging''} or \textit{``interactive''} in their explanations (P2, P3, P5, P8, P9, P10, and P12). P2, who felt the HEM I3M was more engaging, described the difference as \textit{``one is more [like] reading an encyclopedia and the other [the HEM I3M] is an experience''}, adding that they found the HEM I3M to be more `playful'. P7 expanded on this, \textit{``[I] wanted to ask [the HEM I3M] more questions, I wanted to get more information, whereas [I] just accepted [the LEM I3M] as fact''}. P8 went a step further, stating that the LEM I3M was \textit{``boring''} while the HEM I3M was \textit{``more enthusiastic''}. P10 compared the HEM I3M to their screen reader, \textit{``Jaws is neutral, it is not interactive, it gets hypnotic. Using different voices is way more engaging!''}. Despite being more interested in the subject matter of their LEM I3M, P8 ultimately found the HEM I3M more engaging, explaining, \textit{``I was interested in Egypt more than space ... but the way the rocket acted made it more engaging''}.

As a further indication of engagement, participants spent more time interacting with the HEMs ($\mu$ $=$ 99.5 seconds) compared to the LEMs ($\mu$ $=$ 68.8) during the free play exercise (Table~\ref{table:10}). A one-tailed paired t-test revealed this to be statistically significant ($t=2.899$, $p=0.007$). Participants also engaged in more interactions with the HEMs ($\mu$ $=$ 3.3) compared to the LEMs ($\mu$ $=$ 2.2). Wilcoxon results indicated this difference was marginally significant ($p=0.069$).

Participants also chose to spend more time using the HEM I3Ms during both the \textit{model exploration} ($\mu$ $=$ +29.8 seconds) and \textit{researcher-directed} ($\mu$ $=$ +54.4) tasks. Paired t-tests revealed that the effect of the HEMs was statistically significant for time spent during \textit{model exploration} ($t=2.025$, $p=0.035$) and marginally significant for time spent completing the \textit{researcher-directed} tasks ($t=1.595$, $p=0.070$). 

Regarding interactions, participants performed more interactions with the HEMs during these tasks. Paired t-tests indicated marginal significance for both \textit{model exploration} ($\mu$ $=$ +1.8, $t=1.583$, $p=0.071$) and \textit{researcher-directed} tasks ($\mu$ $=$ +0.5, $t=1.732$, $p=0.056$). Notably, during the \textit{researcher-directed} tasks, participants occasionally chose to continue interacting beyond what was required to complete a task, performing additional interactions. These instances favoured the HEMs ($n=10$) over the LEMs ($n=4$).

\begin{table*}[!ht]
\caption{Task completion for HEM and LEM conditions across three sets of tasks: Model Exploration, Researcher-Directed, and Free Play. Descriptive statistics are reported for time spent (seconds) and interactions undertaken. Values have been rounded to one decimal place. Statistical significance was tested using paired t-tests ($t$) and Wilcoxon tests ($z$), where appropriate.}
\label{table:10}
\begin{tabular}{@{}ccccccccccclll@{}}
\toprule
\multicolumn{13}{c}{\textit{\textbf{Task Completion (Time Spent)}}} \\ \midrule
\multicolumn{1}{c|}{\multirow{2}{*}{}} &
  \multicolumn{6}{c|}{\textbf{Descriptive Statistics}} &
  \multicolumn{6}{c}{\textbf{Statistical Test}} \\
\multicolumn{1}{c|}{} &
  \multicolumn{3}{c|}{\textbf{HEM I3Ms}} &
  \multicolumn{3}{c|}{\textbf{LEM I3Ms}} &
  \multicolumn{6}{c}{} \\
\multicolumn{1}{c|}{\textbf{Task}} &
 \textit{M} &
  $\mu$ &
  \multicolumn{1}{c|}{\textit{SD}} &
 \textit{M} &
  $\mu$ &
  \multicolumn{1}{c|}{\textit{SD}} &
 \textit{test-statistic} &
 \textit{p-value} &
  \multicolumn{4}{c}{} \\ \midrule
\multicolumn{1}{l|}{\textbf{Model Exploration}} &
  240.0 &
  228.9 &
  \multicolumn{1}{c|}{63.2} &
  180.0 &
  199.1 &
  \multicolumn{1}{c|}{40.8} &
  $t=$ 2.025 &
  \multicolumn{1}{l}{0.035$^\ast$} &
  \multicolumn{4}{c}{} \\
\multicolumn{1}{l|}{\textbf{Researcher-Directed}} &
  592.5 &
  592.3 &
  \multicolumn{1}{c|}{120.5} &
  557.5 &
  537.9 &
  \multicolumn{1}{c|}{150.7} &
  $t=$ 1.595 &
  \multicolumn{1}{l}{0.070$^\wedge$} &
  \multicolumn{4}{c}{} \\
\multicolumn{1}{l|}{\textbf{Free Play}} &
  92.5 &
  99.5 &
  \multicolumn{1}{c|}{39.9} &
  60.0 &
  68.8 &
  \multicolumn{1}{c|}{17.0} &
  $t=$ 2.899 &
  \multicolumn{1}{l}{0.007$^\ast$} &
  \multicolumn{4}{c}{} \\ \midrule
\multicolumn{13}{c}{\textit{\textbf{Task Completion (Interactions Undertaken)}}} \\ \midrule
\multicolumn{1}{l|}{\textbf{Model Exploration}} &
  13.0 &
  11.9 &
  \multicolumn{1}{c|}{2.6} &
  10.0 &
  10.1 &
  \multicolumn{1}{c|}{3.0} &
  $t=$ 1.583 &
  \multicolumn{1}{l}{0.071$^\wedge$} &
  \multicolumn{4}{c}{} \\
\multicolumn{1}{l|}{\textbf{Researcher-Directed}} &
  4.5 &
  4.8 &
  \multicolumn{1}{c|}{0.9} &
  4.0 &
  4.3 &
  \multicolumn{1}{c|}{0.5} &
  $t=$ 1.732 &
  \multicolumn{1}{l}{0.056$^\wedge$} &
  \multicolumn{4}{c}{} \\
\multicolumn{1}{l|}{\textbf{Free Play}} &
  2.0 &
  3.3 &
  \multicolumn{1}{c|}{2.1} &
  2.0 &
  2.2 &
  \multicolumn{1}{c|}{0.7} &
  $z=$ 1.483 &
  \multicolumn{1}{l}{0.069$^\wedge$} &
  \multicolumn{4}{c}{} \\ \bottomrule
\end{tabular}
\end{table*}

\subsubsection{\textbf{Impact of Design Factors on Engagement.}}
Participants rated how each design factor, presented in the HEM state, impacted their perception of I3M \textit{engagement} (Table~\ref{table:6}). All five design factors appeared to influence how engaging the HEM I3Ms were perceived and were statistically significant -- \textit{introductions \& small talk} and \textit{location of speech output} (both had $k=11$, $n=12$, $p<0.001$), and the remaining three factors (all $k=10$, $n=12$, $p=0.003$).

\begin{table*}[!ht]
\caption{Impact of the design factors on the perception of how engaging the HEM I3Ms appeared.}
\label{table:6}
\begin{tabular}{@{}cccccccc@{}}
\toprule
\multicolumn{8}{c}{\textit{\textbf{Did the design factors impact your perception of how engaging the HEM I3M was?}}}               \\ \midrule
\multicolumn{1}{c|}{\multirow{3}{*}{\textbf{Design Factor}}} &
  \multicolumn{4}{c|}{\multirow{2}{*}{\textbf{Descriptive Statistics}}} &
  \multicolumn{3}{c}{\multirow{2}{*}{\textbf{Binomial Test}}} \\
\multicolumn{1}{c|}{}                                     & \multicolumn{4}{c|}{}                          & \multicolumn{3}{c}{}       \\
\multicolumn{1}{c|}{} &
  \textit{M} &
  \textit{$\mu$} &
  \textit{SD} &
  \multicolumn{1}{c|}{\textit{V}} &
  \textit{k, n} &
  \textit{p-value} &
  \textit{95th CI} \\ \midrule
\multicolumn{1}{l|}{\textbf{DF\#1: Introductions \& Small Talk}} & 4.00 & 4.08 & 0.76 & \multicolumn{1}{c|}{0.58} & 11, 12 & \multicolumn{1}{r}{<0.001$^\ast$} & 0.661, 1.000 \\
\multicolumn{1}{l|}{\textbf{DF\#2: Embodied Personified Voices}} & 4.00 & 4.25 & 0.72 & \multicolumn{1}{c|}{0.52} & 10, 12 & \multicolumn{1}{r}{0.003$^\ast$} & 0.562, 1.000 \\
\multicolumn{1}{l|}{\textbf{DF\#3: Embodied Narration Style}}             & 4.50 & 4.25 & 0.92 & \multicolumn{1}{c|}{0.85} & 10, 12 & \multicolumn{1}{r}{0.003$^\ast$} & 0.562, 1.000 \\
\multicolumn{1}{l|}{\textbf{DF\#4: Embodied Vibratory Feedback}}   & 4.00 & 4.25 & 0.72 & \multicolumn{1}{c|}{0.52} & 10, 12 & \multicolumn{1}{r}{0.003$^\ast$} & 0.562, 1.000 \\
\multicolumn{1}{l|}{\textbf{DF\#5: Location of Speech Output}}                        & 4.00 & 4.25 & 0.60 & \multicolumn{1}{c|}{0.35}                      & 11, 12 & \multicolumn{1}{r}{<0.001$^\ast$} & 0.661, 1.000 \\ \bottomrule
\end{tabular}
\end{table*}
\newpage
\subsection{Trustworthiness of I3Ms}
\subsubsection{\textbf{HCTM Scales.}}
\begin{table*}[!ht]
\caption{Impact of HEM and LEM conditions on perceptions of how trustworthy the I3Ms were, using four HCTM subscales. Participants were also asked to rate how trustworthy the I3Ms felt.}
\label{table:3}
\begin{tabular}{@{}cccccccccccclll@{}}
\toprule
\multicolumn{15}{c}{\textit{\textbf{Human Computer Trust Model}}} \\ \midrule
\multicolumn{1}{c|}{\multirow{2}{*}{}} &
  \multicolumn{8}{c|}{\textbf{Descriptive Statistics}} &
  \multicolumn{6}{c}{\textbf{Wilcoxon Test}} \\
\multicolumn{1}{c|}{} &
  \multicolumn{4}{c|}{\textbf{HEM I3Ms}} &
  \multicolumn{4}{c|}{\textbf{LEM I3Ms}} &
  \multicolumn{6}{c}{\textbf{}} \\
\multicolumn{1}{c|}{\textbf{Scale}} &
  \textit{M} &
  $\mu$ &
  \textit{SD} &
  \multicolumn{1}{c|}{V} &
  \textit{M} &
  $\mu$ &
  \textit{SD} &
  \multicolumn{1}{c|}{V} &
  \textit{z-score} &
  \textit{p-value} &
  \multicolumn{4}{c}{} \\ \midrule
\multicolumn{1}{l|}{\textbf{Perceived Risk}} &
  1.00 &
  1.58 &
  0.72 &
  \multicolumn{1}{c|}{0.52} &
  2.00 &
  1.86 &
  0.98 &
  \multicolumn{1}{c|}{0.95} &
  2.428 &
  \multicolumn{1}{l}{0.008$^\ast$} &
  \multicolumn{4}{c}{} \\
\multicolumn{1}{l|}{\textbf{Benevolence}} &
  4.00 &
  4.14 &
  0.95 &
  \multicolumn{1}{c|}{0.90} &
  4.00 &
  4.03 &
  0.90 &
  \multicolumn{1}{c|}{0.80} &
  1.633 &
  \multicolumn{1}{l}{0.051$^\wedge$} &
  \multicolumn{4}{c}{} \\
\multicolumn{1}{l|}{\textbf{Competence}} &
  5.00 &
  4.50 &
  0.55 &
  \multicolumn{1}{c|}{0.31} &
  4.00 &
  4.31 &
  0.66 &
  \multicolumn{1}{c|}{0.43} &
  2.646 &
  \multicolumn{1}{l}{0.004$^\ast$} &
  \multicolumn{4}{c}{} \\
\multicolumn{1}{l|}{\textbf{Reciprocity}} &
  4.00 &
  4.31 &
  0.70 &
  \multicolumn{1}{c|}{0.49} &
  4.00 &
  4.08 &
  0.80 &
  \multicolumn{1}{c|}{0.63} &
  1.929 &
  \multicolumn{1}{l}{0.027$^\ast$} &
  \multicolumn{4}{c}{} \\ \midrule
\multicolumn{15}{c}{\textit{\textbf{Did you find that the I3M felt trustworthy?}}} \\ \midrule
\multicolumn{1}{l|}{\textbf{Trustworthiness}}  &
  5.00 &
  4.58 &
  0.49 &
  \multicolumn{1}{c|}{0.24} &
  5.00 &
  4.50 &
  0.65 &
  \multicolumn{1}{c|}{0.42} &
  0.577 &
  \multicolumn{1}{l}{0.282} &
  \multicolumn{4}{c}{} \\ \bottomrule
\end{tabular}
\end{table*}
HEM I3Ms outperformed LEM I3Ms across all four HCTM subscales (Table~\ref{table:3}) -- \textit{perceived risk}\footnote{The \textit{perceived risk} subscale relates to the willingness of a user to engage with a system despite possible risks. A lower perceived risk score is desired, signifying that the user is more willing to interact with the system.} \textit{($\mu$ $=$ -0.28), benevolence ($\mu$ $=$ +0.11), competence ($\mu$ $=$ +0.19)}, and \textit{reciprocity ($\mu$ $=$ +0.23)}. Wilcoxon results were significant for three HCTM subscales -- \textit{perceived risk} ($p=0.008$), \textit{competence} ($p=0.004$), and \textit{reciprocity} ($p=0.027$). The remaining subscale, \textit{benevolence}, was marginally significant ($p=0.051$).

\subsubsection{\textbf{How Trustworthy Were The I3Ms?}}
\label{sec:PerceptionTrust}
When rating the \textit{trustworthiness} of each I3M (Table~\ref{table:3}), results were less clear. A minor increase in mean score for HEM I3Ms ($\mu$ $=$ +0.08) over LEM I3Ms was observed; however, this difference was nonsignificant ($p=0.282$). In the post-activity interview ranking, the majority of the participants ($n=9$) indicated that there was no major discernible difference in trust between the HEM and LEM models. The remaining participants ($n=3$) ranked the HEM I3Ms higher. A binomial test showed that this result was nonsignificant ($k=3$, $n=12$, $p=0.819$).

Participants described feeling indecisive, often basing their interpretation of trust on the believability of the information provided by the models. P2 noted that both models were \textit{``just giving [them] facts''} and that their manner of acting or speaking did not matter. P4 explained that as both HEM and LEM I3Ms were \textit{``accessing information from the internet ... that [they] trusted it to only access certain [appropriate] things''}. Despite reporting no major differences in their ratings, P11 suggest that \textit{``incorrect facts, [when] talking in first person reduces trust''}, while P5, who also rated no difference, focused on the salient physical design of the I3Ms, expressing concern about \textit{`the height of the rocket ... knocking it over''}.

\subsubsection{\textbf{Impact of Design Factors on Trust.}}
Participants were divided on how the individual HEM design factors influenced their perceptions of the \textit{trustworthiness} of the I3Ms (Table~\ref{table:9}). \textit{Embodied vibratory feedback} elicited the highest mean score, while \textit{embodied narration style} scored the lowest. Apart from \textit{embodied vibratory feedback}, which was marginally significant ($k=8$, $n=12$, $p=0.057$), all other design factors were nonsignificant.

\begin{table*}[!ht]
\caption{Impact of the design factors on the perception of how trustworthy the HEM I3Ms appeared.}
\label{table:9}
\begin{tabular}{@{}cccccccc@{}}
\toprule
\multicolumn{8}{c}{\textit{\textbf{Did the design factors impact your perception of how trustworthy the HEM I3M was?}}}               \\ \midrule
\multicolumn{1}{c|}{\multirow{3}{*}{\textbf{Design Factor}}} &
  \multicolumn{4}{c|}{\multirow{2}{*}{\textbf{Descriptive Statistics}}} &
  \multicolumn{3}{c}{\multirow{2}{*}{\textbf{Binomial Test}}} \\
\multicolumn{1}{c|}{}                                     & \multicolumn{4}{c|}{}                          & \multicolumn{3}{c}{}       \\
\multicolumn{1}{c|}{} &
  \textit{M} &
  \textit{$\mu$} &
  \textit{SD} &
  \multicolumn{1}{c|}{\textit{V}} &
  \textit{k, n} &
  \textit{p-value} &
  \textit{95th CI} \\ \midrule
\multicolumn{1}{l|}{\textbf{DF\#1: Introductions \& Small Talk}} & 4.00 & 3.67 & 0.62 & \multicolumn{1}{c|}{0.39} & 7, 12 & \multicolumn{1}{l}{0.158} & 0.315, 1.000 \\
\multicolumn{1}{l|}{\textbf{DF\#2: Embodied Personified Voices}} & 3.50 & 3.67 & 0.75 & \multicolumn{1}{c|}{0.56} & 6, 12 & \multicolumn{1}{l}{0.335} & 0.245, 1.000 \\
\multicolumn{1}{l|}{\textbf{DF\#3: Embodied Narration Style}}             & 3.00 & 3.42 & 0.95 & \multicolumn{1}{c|}{0.91} & 5, 12 & \multicolumn{1}{l}{0.562} & 0.181, 1.000 \\
\multicolumn{1}{l|}{\textbf{DF\#4: Embodied Vibratory Feedback}}   & 4.00 & 3.83 & 0.69 & \multicolumn{1}{c|}{0.47} & 8, 12 & \multicolumn{1}{l}{0.057$^\wedge$} & 0.391, 1.000 \\
\multicolumn{1}{l|}{\textbf{DF\#5: Location of Speech Output}}                        & 3.50 & 3.67 & 0.75 & \multicolumn{1}{c|}{0.56}                      & 6, 12 & \multicolumn{1}{l}{0.335} & 0.245, 1.000 \\ \bottomrule
\end{tabular}
\end{table*}

\section{Discussion}
\subsection{Hypothesis \#1: HEM I3Ms are perceived as more \textit{embodied} than LEM I3Ms}
Participants felt that HEM I3Ms were more lively, with an increased perception of embodiment compared to the LEM I3Ms. When ranking the I3Ms on liveliness, 8/12 selected HEM I3Ms (Section~\ref{sec:PerceptionLiveliness}). This was also statistically significant, with most participants providing emphatic explanations for their selections, supporting H\#1.

Results from the Godspeed Questionnaire Series also supported H\#1 and allowed us to explore different dimensions and key aspects of embodiment, adding nuance to our understanding of how the I3Ms appeared embodied (Table~\ref{table:1}). All GQS subscales were statistically significant. It is particularly noteworthy that the \textit{anthropomorphism} and \textit{animacy} subscales were rated more positively for HEM I3Ms, as these subscales deal with the attribution of human-like behaviours and perception of life~\cite{Bartneck2009}, which are key components of embodiment. The \textit{likeability} and \textit{intelligence} subscales were also perceived more favourably for HEM I3Ms.

Overall, our findings and participant comments help to support H\#1 that BLV users do perceive I3Ms configured with HEM design factors as more embodied, \textit{`present'}, and \textit{`lively'}. These findings align with work on the embodiment of other interfaces in non-BLV contexts, including robots~\cite{Kontogiorgos2020} and conversational agents~\cite{Luria2017,Shamekhi2018,Luria2019}, which has found that imbuing machines with human-like behaviours can enhance their perceived embodiment and sense of presence, creating a sense of \textit{`being there'}~\cite{Shamekhi2018}. 

\subsection{Hypothesis \#2: HEM I3Ms are perceived as more \textit{engaging} than LEM I3Ms}
Participants rated higher levels of engagement with HEM I3Ms compared to LEM I3Ms. When ranking which I3M was more engaging, the overwhelming majority (10/12) selected HEM (Section~\ref{sec:PerceptionEngagement}). These differences were statistically significant, supporting H\#2.

We used the User Engagement Scale and Playful Experiences Questionnaire (Table~\ref{table:4}) to explore different dimensions of engagement to add to and help contextualise our understanding of how engaging participants found the I3Ms. The UES-SF indicated that HEM I3Ms had higher average scores than LEM I3Ms across 3/4 subscales. Of these subscales, two were statistically significant, and one was marginally significant. These subscales focus on the extent to which users feel absorbed in an interaction (\textit{focused attention}) and the usability/negative affect experienced during interactions (\textit{perceived usability}), both of which are critical dimensions of engagement~\cite{Obrien2018}. The \textit{aesthetic appeal} subscale, which the researchers adjusted to be more meaningful in BLV contexts -- encompassing aesthetics beyond those purely visual -- was also statistically significant. This is particularly noteworthy, with HEM-configured I3Ms providing more engaging sensory experiences. The \textit{reward} subscale, centred on valued experiential outcomes, was not significant with respect to our hypothesis, and was the only subscale where LEM I3Ms received higher mean scores. This may indicate that the additional presence and feedback of HEM I3Ms could, in some circumstances, reduce initial curiosity.

It is our belief that several PLEXQ subscales may have had reduced meaning to participants, potentially as a result of the controlled nature and limited time exposure of the study. For example, subscales that focus on finding something hidden (\textit{discovery}) or unwinding through playful experiences (\textit{relaxation}) may have been less relevant. Interestingly, one subscale, \textit{exploration}, elicited higher mean values for LEM I3Ms, possibly for similar reasons to the UES-SF \textit{reward} subscale. Despite this, several PLEXQ results added support to H\#2, with subscales related to excitement (\textit{sensation}), enjoyment/amusement (\textit{humor}), and power (\textit{control}) being more favourably observed with HEM I3Ms.

Across all tasks, participants spent more time interacting with the HEM I3Ms and performed a greater number of interactions compared to the LEM I3Ms. We feel this was particularly noteworthy during the undirected free play task, which was designed to mimic real-world use. Time spent and the number of interactions were statistically significant and marginally significant, respectively. These metrics have previously been used as behavioural measures of engagement in HCI~\cite{OBrien2013,Doherty2018}, and provide further evidence that participants were more engaged with the HEM I3Ms. They also complement the UES-SF, which ties engagement
to the depth of a user's investment with a system~\cite{OBrien2016}.

Based on emphatic participant discussions and clear preferences when directly asked, our study provides evidence supporting H\#2 that HEM I3Ms are perceived as more engaging. Our scale data adds important nuance to this understanding, with key UES-SF and PLEXQ subscales showing that HEM I3Ms were more favourably perceived. Although ratings for both HEM and LEM I3Ms were generally high, which aligns with works showing that BLV users find 3D-printed models and I3Ms engaging~\cite{Nagassa2023,Reinders2020,Shi2019}, our results suggest that HEM I3Ms are perceived as even more engaging. More broadly, our findings align with prior work demonstrating that interfaces imbued with more human-like behaviours can positively impact end-user perceptions of engagement~\cite{Lester1997,Cassell1999,Shamekhi2018}.

\subsection{Hypothesis \#3: HEM I3Ms are perceived as more \textit{trustworthy} than LEM I3Ms}
Our participants were mixed on whether there were any discernible differences in the trustworthiness of HEM and LEM I3Ms. Hesitance was observed when participants were asked to rank which I3M was the more trustworthy model, with 9/12 indicating no major discernible difference between them (Section~\ref{sec:PerceptionTrust}). Participants expressed indecisiveness, often basing their interpretation of trust solely on the believability of the information output from the models, rather than their interactions with the I3Ms. 

On the other hand, results from the Human-Computer Trust Model (Table~\ref{table:3}) suggest that HEM I3Ms may lead to greater trust, as all four HCTM subscales were either significant ($=3$) or marginally significant ($=1$). Results did, however, reveal only very minor differences in mean scores between HEM I3Ms and LEM I3Ms. Despite this, these scales relate to the willingness of users to engage with a system despite the possible risks (\textit{perceived risk}), whether a system possesses the functionalities needed to depend on it (\textit{competence}), and a willingness to spend more time using it when support situations arise (\textit{reciprocity}). The \textit{benevolence} subscale, which was marginally significant, focuses on whether users believe that a system has the abilities required to help them achieve their goals.

Despite the statistical significance of the HCTM results, we believe that, overall, our results provide only mixed support for Hypothesis H\#3. It is the researchers' view that the HCTM questionnaire and subjective ratings may have been interpreted differently by participants. While the scale data appears to have successfully captured different dimensions of trust based on interactions with the HEM/LEM models, many participants, when asked about the concept of trust subjectively
(Section~\ref{sec:PerceptionTrust}), focused solely on the believability of the information provided by the models, independent of their HEM/LEM state. They tended to prioritise believability of information over considerations as to whether their interactions with the models and their behaviours influenced perceptions of reliance or competence. Our findings also suggest that the HEM design factors tested do not appear to play a strong role in influencing trust. As visually embodied conversational agents have been shown to be subjectively more trustworthy in non-BLV contexts~\cite{Heuwinkel2013,Bickmore2013,Sidner2018,Shamekhi2018,Gulati2018}, it is clear that more research is needed in order to better understand the impact of embodiment on trustworthiness of I3Ms for BLV users, particularly in real-world situations.

\subsection{How Did The Design Factors Impact Embodiment, Engagement and Trust?}
Broadly speaking, our embodiment design factors and their implementations were well-received, contributing to how embodied and engaging the HEM I3Ms were perceived. However, connections to trust were mixed. These are discussed below in order of importance.
\newline
\newline
\vspace{-1mm}\bpstart{DF\#4: Embodied Vibratory Feedback} Participants shared their enthusiasm regarding haptics. Referring to the liveliness of the HEM I3Ms, P3 described how the use of haptics made them feel \textit{``more three-dimensional... realistic''}. Participants also discussed how haptics made the HEM I3Ms more engaging, including P5, \textit{``[haptics] added an extra sense of interaction''}. These emphatic reactions align with other work, where haptics added life to I3Ms~\cite{Reinders2023}, and support the growing body of research exploring how haptics can create physically embodied, lifelike cues~\cite{Nie2012,Bevan2015, Borgstedt2023}.

\bpstart{DF\#2: Embodied Personified Voices} Participants highlighted how unique voices shaped their perception of embodiment. P9 was emphatic about how it made the HEM model feel more alive, \textit{``[it] made [the components] seem like they were their own things''}. P6 detailed how the one-for-one social presence of the HEM model gave components their own `character'. Regarding engagement, several participants found the voices helpful for tracking which I3M component was active and talking. P2 explained, \textit{``different voices ... broke the [HEM] model up into separate parts, it defined the objects better''}. P8 reflected, \textit{``[a] single voice does not stack up''}. This aligns with work by Choi \etal~\cite{Choi2020}'s finding that many BLV users value human-like conversation with agents. 

\bpstart{DF\#3: Embodied Narration Style} Participants frequently linked first-person narration to engagement and embodiment, creating a transformative experience. P2 described it as making interactions feel \textit{``like you were talking to someone, vs it talking to you''}. Similarly, P4 remarked that it made conversations seem \textit{``more human-like ... on a one-on-one basis''}. While P9 noted that it enhanced HEM presence, they initially felt that \textit{``... it was a bit strange anthropomorphizing [a system]... like when Microsoft talks like a human''}. The positive reception of first-person narration aligns with other research that has found that users anthropomorphise agents by using first and second-person pronouns~\cite{Liao2018,Coeckelbergh2011}.

\bpstart{DF\#5: Location of Speech Output} Participants noted that closer coupling with the I3M and its speaker impacted engagement. P4 found the HEM I3M's audio output \textit{``easier to take in''}, while P8 appreciated being able to \textit{``focus more directly on [the HEM I3M]''}. P3 described HEM as \textit{``more intimate and engaging''}. The researchers believe that embedding speakers inside each printed component could further enhance embodiment and engagement. P5 suggested this would make the I3Ms \textit{``feel more alive''}, echoing feedback from previous work~\cite{Reinders2023}. One possibility may be to use ultrasonics to project speech output~\cite{Iravantchi2020}, acting as digital ventriloquism. 

\bpstart{DF\#1: Introductions \& Small Talk} This factor had less impact on participants, who were less likely to reference it. The researchers believe this may be because it was not as explicit as other design factors. Many participants forgot that the HEM I3Ms introduced themselves, and small talk, being reliant on user initiation, was minimal. Only two participants (P2 and P4) engaged in small talk. Their responses revealed no meaningful deviation from the wider participant group regarding embodiment, engagement, and trust, suggesting that small talk played a minimal role. Despite this, P2 found it made the HEM I3M seem more playful, while P4 said introductions felt like an \textit{``invitation''} to interact. Conversely, P6 felt HEM introductions diminished their independence, stating that \textit{``I did not like it [introducing itself]; I like finding out what things are by [looking at/touching] them''}. These mixed reactions align with other works which
have found that BLV users have varying perspectives on how proactive I3Ms should be~\cite{Reinders2020}, suggesting that conversational strategies effective in embodied agents for sighted users~\cite{Liao2018,Pradhan2019,Cassell2001,Shamekhi2018} may hold less meaning in BLV contexts.

\subsection{Can I3Ms Be Embodied \& What Is Their Impact?}
Our study supports the larger idea that I3Ms are able to be perceived as \textit{more embodied} \textbf{and are} \textit{more engaging}. This follows from H\#1, where our findings support that BLV people perceive HEM I3Ms as more embodied and lively, and H\#2, where HEM I3Ms were found to be more engaging. In the context of I3Ms, it appears that model embodiment increases end-user engagement, specifically with I3Ms using the set of embodied design factors we implemented. However, the impact of embodiment on trust (H\#3) remains less clear, as BLV people generally appear to trust I3Ms regardless of embodiment.

Prior research has shown that the embodiment of an interface can increase both engagement~\cite{Shamekhi2018,Luger2016,Heuwinkel2013,Cassell2001} and trust~\cite{Bickmore2001,Shamekhi2018,Bickmore2013,Rheu2021}. Our findings support the relationship between embodiment and engagement, but stop short of any impacts on trust. With the bulk of this previous research focusing on conversational agents and robots, our work extends this relationship to include contexts relevant to both I3Ms and with people who are BLV.

The relationship between embodied I3Ms and their impact on engagement is particularly significant in learning contexts, allowing BLV students or self-learners to engage in deeper, more meaningful experiences where the I3M can assist in teaching and testing their knowledge. Embodied agents have been used in learning environments, and have been found to enrich the learning experiences of students~\cite{Schroeder2013}. Pedagogic agents often rely on embodied designs, including combinations of visual and conversational embodiment~\cite{Lester1997,Moreno2001,Schroeder2013}. While prior research on non-embodied I3Ms in learning contexts has highlighted positive feedback from both teachers and BLV students~\cite{Shi2019}, we believe that understanding how I3Ms can be effectively embodied could further enhance engagement in educational settings, and facilitate broader adoption.

Based on the positive reception amongst participants of our embodied I3Ms, we propose extending our existing I3M design recommendations~\cite{Reinders2023} to include embodiment:

\begin{itemize}
   \item \textbf{\textit{Support more embodied experiences}}: I3Ms that support human-like behaviours and characteristics in their design can make them appear more human, lively, and engaging to use. This may involve combining aspects of physical and conversational embodiment, \eg\, introductions and small talk, embodied voices, embodied narration, embodied haptic vibratory feedback, and location of speech output.
\end{itemize}

\section{Limitations \& Future Work}
There are a number of limitations regarding this study. While some are lenses through which the results should be interpreted, others present exciting avenues for future investigation. 

Our investigation focused on five specific design factors, and our findings should be interpreted within that context. Future research should explore additional embodiment design factors. This could include models that are more autonomous, or incorporate visual embodiment elements, which may benefit low vision users with residual vision, \eg\, virtual avatars or the emittance of light. Physical embodiment could also be extended with additional design dimensions not covered in this study. For example, haptic perception could extend beyond vibratory feedback to include model texturing, scale, orientation, or the impact of detachable components.

Furthermore, our design factors were implemented in specific ways, presenting an opportunity to further explore how the design space of embodied I3Ms can be expanded through alternative implementations. Often, our implementations were shaped by technical limitations or model choice. For instance, with the \textit{location of speech output} design factor, embedding speakers within individual model components or using headphones could produce different results. Regarding model choice, while our models neatly segmented \textit{embodied personified voices} based on individual components, a different type of model, \eg\ a globe of Earth, could instead segment voices by country, continent, or hemisphere, depending on its purpose.

As our study looked at participants' subjective ratings to assess the impact of the design factors, future research could formally isolate and test each factor to understand its individual effect on embodiment. The \textit{small talk} component of DF\#1 was also rarely used, likely because it required user initiation, unlike the other design factors. While this was an intentional design decision (see Section~\ref{sec:Smalltalk}), future work should revisit small talk to determine if the design factor would be better utilised if it were agent-initiated.

We used validated scales to measure aspects of embodiment, engagement, and trust. Based on pilot feedback, we adjusted a small number of UES-SF and PLEXQ items to have more meaning in BLV contexts. We supplemented these scales with our own questions, providing additional nuance and insight. However, this raises a need for future work to explore how these instruments can be further modified and validated for BLV users, or whether entirely new tools should be developed to improve relevance and meaning.

Our study was conducted with 12 participants, consistent with similar studies involving BLV participants~\cite{Holloway2022,Nagassa2023,Shi2017b,Shi2020}. Nonetheless, we would like to run further sessions to confirm our findings. This presents multiple opportunities, such as using a broader selection of models, and conducting `in-the-wild' and longitudinal studies to examine how the relationship between end-users and I3Ms evolves over time. Additionally, as our work represents the first investigation into embodied I3Ms, and its exploratory nature influenced the experimental design and analysis, future research should focus on targeted testing and conservative analysis.

Regarding trust, given the lack of discernible differences between the trustworthiness of HEM and LEM I3Ms, future work should focus on re-examining the concept of trust. Future research should explore real-world trust scenarios where confidence in HEM/LEM I3Ms may hold greater significance and contexts where model behaviours could influence perception.

In our previous research, we identified the importance of I3Ms supporting customisation~\cite{Reinders2023}. In this study, participants had limited opportunities to personalise each model in order to better isolate the effects of each HEM design factor. Given the new insights into embodiment presented in this work, future work could seek to combine the HEMs described with greater user customisation.

This work utilised conversational agents specifically trained on information relevant to the modelled concepts. Rapid advancements in generative AI provide opportunities to expand the conversational capabilities of I3Ms. Future research should investigate how generative AI systems could be leveraged in place of bespoke conversational agents in order to enhance their flexibility.
\newpage
Finally, future work should examine embodiment and engagement with sighted users. Given the potential for I3Ms, such as those presented in this work, to be used in contexts like museums and classrooms, it is important that they be highly engaging for all users, contributing to more inclusive social experiences.

\section{Conclusion}
Increasing the engagement and trust that exists between BLV users and their accessible materials is of critical importance, as the ultimate utility of these resources is diminished if users are unwilling to engage with or trust them. This is particularly relevant in educational and cultural contexts, such as classrooms, museums, and galleries, where I3Ms are beginning to be deployed. 

Our work presents the first investigation into the relationship between embodiment, engagement, and trust in the context of I3Ms and BLV users. We created two I3Ms that supported a series of design factors aimed at making the models appear more embodied. Our findings revealed that our participants perceived many of the design factors as contributing to the embodiment of the I3Ms. Participants also believed that these design factors made the I3Ms more engaging. The impact on trust, however, was less clear. 

While many of the subscales we used revealed statistical significance in favour of HEM I3Ms, ratings for both HEM and LEM I3Ms were generally high. In the case of embodiment and engagement, rankings and participant comments added important context to these results, revealing clearer preferences. For example, while participants found I3Ms engaging overall, HEM I3Ms were perceived as even more engaging. This was further supported by our behavioural metrics, which showed that participants spent more time and performed more interactions with the HEMs.

Based on the positive reception of the embodied I3Ms, we established a new I3M design recommendation to complement those previously proposed to the I3M community~\cite{Reinders2023} -- \textit{support more embodied experiences}. We advocate that, due to the connections between embodiment and engagement, I3Ms should incorporate human-like behaviours and characteristics into their design.

We hope that this initial exploration into embodiment represents the first step toward a growing interest in embodied I3Ms, helping to facilitate their widespread adoption. One day, we envision similar models being found in schools and public spaces, such as museums and galleries, with configurable embodiment that allows BLV people to engage with content in ways that are more meaningful to them.

We believe our work is relevant in the broader context of research into embodiment, and its relationship to engagement and trust. Virtually all prior research has involved sighted users; as such, our study is the first to explore this relationship in the context of blind users who cannot discern visual embodiment cues. We hope our findings will inspire further exploration in this space. 

\begin{acks}
This research was supported by an Australian Government Research Training Program (RTP) Scholarship. We want to thank our participants for their time and expertise.  
\end{acks}
\newpage
\balance
\label{Bibliography}
\bibliographystyle{ACM-Reference-Format}
\bibliography{references}


\begin{thebibliography}{111}

%%% the Web address.

\ifx \showCODEN    \undefined \def \showCODEN     #1{\unskip}     \fi
\ifx \showISBNx    \undefined \def \showISBNx     #1{\unskip}     \fi
\ifx \showISBNxiii \undefined \def \showISBNxiii  #1{\unskip}     \fi
\ifx \showISSN     \undefined \def \showISSN      #1{\unskip}     \fi
\ifx \showLCCN     \undefined \def \showLCCN      #1{\unskip}     \fi
\ifx \shownote     \undefined \def \shownote      #1{#1}          \fi
\ifx \showarticletitle \undefined \def \showarticletitle #1{#1}   \fi
\ifx \showURL      \undefined \def \showURL       {\relax}        \fi
% invisible to TeX
\providecommand\bibfield[2]{#2}
\providecommand\bibinfo[2]{#2}
\providecommand\natexlab[1]{#1}
\providecommand\showeprint[2][]{arXiv:#2}

\bibitem[Abdolrahmani et~al\mbox{.}(2018)]%
        {Abdolrahmani2018}
\bibfield{author}{\bibinfo{person}{Ali Abdolrahmani}, \bibinfo{person}{Ravi
  Kuber}, {and} \bibinfo{person}{Stacy~M. Branham}.}
  \bibinfo{year}{2018}\natexlab{}.
\newblock \showarticletitle{"Siri Talks at You": An Empirical Investigation of
  Voice-Activated Personal Assistant (VAPA) Usage by Individuals Who Are
  Blind}. In \bibinfo{booktitle}{\emph{Proc. ACM SIGACCESS Conference on
  Computers \& Accessibility}} (Galway, Ireland) \emph{(\bibinfo{series}{ASSETS
  '18})}. \bibinfo{publisher}{ACM}, \bibinfo{address}{New York, NY, USA},
  \bibinfo{pages}{249--258}.
\newblock
\showISBNx{978-1-4503-5650-3}
\href{https://doi.org/10.1145/3234695.3236344}{doi:\nolinkurl{10.1145/3234695.3236344}}


\bibitem[Aldrich and Sheppard(2001)]%
        {Aldrich2001}
\bibfield{author}{\bibinfo{person}{Frances~K. Aldrich} {and}
  \bibinfo{person}{Linda Sheppard}.} \bibinfo{year}{2001}\natexlab{}.
\newblock \showarticletitle{Tactile graphics in school education: perspectives
  from pupils}.
\newblock \bibinfo{journal}{\emph{British Journal of Visual Impairment}}
  \bibinfo{volume}{19}, \bibinfo{number}{2} (\bibinfo{year}{2001}),
  \bibinfo{pages}{69--73}.
\newblock
\href{https://doi.org/10.1177/026461960101900204}{doi:\nolinkurl{10.1177/026461960101900204}}


\bibitem[Azenkot and Lee(2013)]%
        {Azenkot2013}
\bibfield{author}{\bibinfo{person}{Shiri Azenkot} {and}
  \bibinfo{person}{Nicole~B. Lee}.} \bibinfo{year}{2013}\natexlab{}.
\newblock \showarticletitle{Exploring the Use of Speech Input by Blind People
  on Mobile Devices}. In \bibinfo{booktitle}{\emph{Proc. ACM SIGACCESS
  Conference on Computers \& Accessibility}} (Bellevue, Washington)
  \emph{(\bibinfo{series}{ASSETS '13})}. \bibinfo{publisher}{ACM},
  \bibinfo{address}{New York, NY, USA}, Article \bibinfo{articleno}{11},
  \bibinfo{numpages}{8}~pages.
\newblock
\showISBNx{978-1-4503-2405-2}
\href{https://doi.org/10.1145/2513383.2513440}{doi:\nolinkurl{10.1145/2513383.2513440}}


\bibitem[Bartlett et~al\mbox{.}(2019)]%
        {Bartlett2019}
\bibfield{author}{\bibinfo{person}{Rachel Bartlett}, \bibinfo{person}{Yi~Xuan
  Khoo}, \bibinfo{person}{Juan~Pablo Hourcade}, {and} \bibinfo{person}{Kyle~K.
  Rector}.} \bibinfo{year}{2019}\natexlab{}.
\newblock \showarticletitle{Exploring the Opportunities for Technologies to
  Enhance Quality of Life with People who have Experienced Vision Loss}. In
  \bibinfo{booktitle}{\emph{Proc. ACM CHI Conference on Human Factors in
  Computing Systems}} (Glasgow, Scotland Uk) \emph{(\bibinfo{series}{CHI
  '19})}. \bibinfo{publisher}{ACM}, \bibinfo{address}{New York, NY, USA},
  \bibinfo{pages}{1–8}.
\newblock
\showISBNx{9781450359702}
\href{https://doi.org/10.1145/3290605.3300421}{doi:\nolinkurl{10.1145/3290605.3300421}}


\bibitem[Bartneck et~al\mbox{.}(2009)]%
        {Bartneck2009}
\bibfield{author}{\bibinfo{person}{Christoph Bartneck}, \bibinfo{person}{Dana
  Kuli{\'{c}}}, \bibinfo{person}{Elizabeth Croft}, {and}
  \bibinfo{person}{Susana Zoghbi}.} \bibinfo{year}{2009}\natexlab{}.
\newblock \showarticletitle{Measurement Instruments for the Anthropomorphism,
  Animacy, Likeability, Perceived Intelligence, and Perceived Safety of
  Robots}.
\newblock \bibinfo{journal}{\emph{International Journal of Social Robotics}}
  \bibinfo{volume}{1}, \bibinfo{number}{1} (\bibinfo{date}{01 Jan}
  \bibinfo{year}{2009}), \bibinfo{pages}{71--81}.
\newblock
\showISSN{1875-4805}
\href{https://doi.org/10.1007/s12369-008-0001-3}{doi:\nolinkurl{10.1007/s12369-008-0001-3}}


\bibitem[Bevan and Stanton~Fraser(2015)]%
        {Bevan2015}
\bibfield{author}{\bibinfo{person}{Chris Bevan} {and}
  \bibinfo{person}{Dana\"{e} Stanton~Fraser}.} \bibinfo{year}{2015}\natexlab{}.
\newblock \showarticletitle{Shaking Hands and Cooperation in Tele-present
  Human-Robot Negotiation}. In \bibinfo{booktitle}{\emph{Proc. of the Tenth
  Annual ACM/IEEE International Conference on Human-Robot Interaction}}
  (Portland, Oregon, USA) \emph{(\bibinfo{series}{HRI '15})}.
  \bibinfo{publisher}{ACM}, \bibinfo{address}{New York, NY, USA},
  \bibinfo{pages}{247–254}.
\newblock
\showISBNx{9781450328838}
\href{https://doi.org/10.1145/2696454.2696490}{doi:\nolinkurl{10.1145/2696454.2696490}}


\bibitem[Bickmore and Cassell(2001)]%
        {Bickmore2001}
\bibfield{author}{\bibinfo{person}{Timothy Bickmore} {and}
  \bibinfo{person}{Justine Cassell}.} \bibinfo{year}{2001}\natexlab{}.
\newblock \showarticletitle{Relational agents: a model and implementation of
  building user trust}. In \bibinfo{booktitle}{\emph{Proc. ACM CHI Conference
  on Human Factors in Computing Systems}} (Seattle, Washington, USA)
  \emph{(\bibinfo{series}{CHI '01})}. \bibinfo{publisher}{ACM},
  \bibinfo{address}{New York, NY, USA}, \bibinfo{pages}{396–403}.
\newblock
\showISBNx{1581133278}
\href{https://doi.org/10.1145/365024.365304}{doi:\nolinkurl{10.1145/365024.365304}}


\bibitem[Bickmore et~al\mbox{.}(2013)]%
        {Bickmore2013}
\bibfield{author}{\bibinfo{person}{Timothy~W. Bickmore},
  \bibinfo{person}{Daniel Schulman}, {and} \bibinfo{person}{Candace Sidner}.}
  \bibinfo{year}{2013}\natexlab{}.
\newblock \showarticletitle{Automated interventions for multiple health
  behaviors using conversational agents}.
\newblock \bibinfo{journal}{\emph{Patient Education and Counseling}}
  \bibinfo{volume}{92}, \bibinfo{number}{2} (\bibinfo{year}{2013}),
  \bibinfo{pages}{142 -- 148}.
\newblock
\showISSN{0738-3991}
\href{https://doi.org/10.1016/j.pec.2013.05.011}{doi:\nolinkurl{10.1016/j.pec.2013.05.011}}


\bibitem[Biocca(1999)]%
        {Biocca1999}
\bibfield{author}{\bibinfo{person}{Frank Biocca}.}
  \bibinfo{year}{1999}\natexlab{}.
\newblock \bibinfo{booktitle}{\emph{Chapter 6 The Cyborg's dilemma. Progressive
  embodiment in virtual environments} (\bibinfo{edition}{c} ed.)}.
\newblock Number~C in \bibinfo{series}{Human Factors in Information
  Technology}. \bibinfo{publisher}{Elsevier}, \bibinfo{address}{Netherlands},
  \bibinfo{pages}{113--144}.
\newblock
\href{https://doi.org/10.1016/S0923-8433(99)80011-2}{doi:\nolinkurl{10.1016/S0923-8433(99)80011-2}}


\bibitem[Bischof et~al\mbox{.}(2016)]%
        {Bischof2016}
\bibfield{author}{\bibinfo{person}{Andreas Bischof}, \bibinfo{person}{Kevin
  Lefeuvre}, \bibinfo{person}{Albrecht Kurze}, \bibinfo{person}{Michael Storz},
  \bibinfo{person}{S\"{o}ren Totzauer}, {and} \bibinfo{person}{Arne Berger}.}
  \bibinfo{year}{2016}\natexlab{}.
\newblock \showarticletitle{Exploring the Playfulness of Tools for Co-Designing
  Smart Connected Devices: A Case Study with Blind and Visually Impaired
  Students}. In \bibinfo{booktitle}{\emph{Proc. Computer-Human Interaction in
  Play Companion}} (Austin, Texas, USA) \emph{(\bibinfo{series}{CHI PLAY
  Companion '16})}. \bibinfo{publisher}{ACM}, \bibinfo{address}{New York, NY,
  USA}, \bibinfo{pages}{93–99}.
\newblock
\showISBNx{9781450344586}
\href{https://doi.org/10.1145/2968120.2987728}{doi:\nolinkurl{10.1145/2968120.2987728}}


\bibitem[Blades et~al\mbox{.}(1999)]%
        {Blades1999}
\bibfield{author}{\bibinfo{person}{Mark Blades}, \bibinfo{person}{Simon Ungar},
  {and} \bibinfo{person}{Christopher Spencer}.}
  \bibinfo{year}{1999}\natexlab{}.
\newblock \showarticletitle{Map Use by Adults with Visual Impairments}.
\newblock \bibinfo{journal}{\emph{The Professional Geographer}}
  \bibinfo{volume}{51}, \bibinfo{number}{4} (\bibinfo{year}{1999}),
  \bibinfo{pages}{539--553}.
\newblock
\href{https://doi.org/10.1111/0033-0124.00191}{doi:\nolinkurl{10.1111/0033-0124.00191}}


\bibitem[Boberg et~al\mbox{.}(2015)]%
        {Boberg2015}
\bibfield{author}{\bibinfo{person}{Marion Boberg}, \bibinfo{person}{Evangelos
  Karapanos}, \bibinfo{person}{Jussi Holopainen}, {and}
  \bibinfo{person}{Andr\'{e}s Lucero}.} \bibinfo{year}{2015}\natexlab{}.
\newblock \showarticletitle{PLEXQ: Towards a Playful Experiences
  Questionnaire}. In \bibinfo{booktitle}{\emph{Proc. of the 2015 Annual
  Symposium on Computer-Human Interaction in Play}} (London, United Kingdom)
  \emph{(\bibinfo{series}{CHI PLAY '15})}. \bibinfo{publisher}{ACM},
  \bibinfo{address}{New York, NY, USA}, \bibinfo{pages}{381–391}.
\newblock
\showISBNx{9781450334662}
\href{https://doi.org/10.1145/2793107.2793124}{doi:\nolinkurl{10.1145/2793107.2793124}}


\bibitem[Bolt(1980)]%
        {Bolt1980}
\bibfield{author}{\bibinfo{person}{Richard~A. Bolt}.}
  \bibinfo{year}{1980}\natexlab{}.
\newblock \showarticletitle{“Put-That-There”: Voice and Gesture at the
  Graphics Interface}. In \bibinfo{booktitle}{\emph{Proc. of the 7th Annual
  Conference on Computer Graphics and Interactive Techniques}} (Seattle,
  Washington, USA) \emph{(\bibinfo{series}{SIGGRAPH '80})}.
  \bibinfo{publisher}{ACM}, \bibinfo{address}{New York, NY, USA},
  \bibinfo{pages}{262–270}.
\newblock
\showISBNx{0897910214}
\href{https://doi.org/10.1145/800250.807503}{doi:\nolinkurl{10.1145/800250.807503}}


\bibitem[Borgstedt(2023)]%
        {Borgstedt2023}
\bibfield{author}{\bibinfo{person}{Jacqueline Borgstedt}.}
  \bibinfo{year}{2023}\natexlab{}.
\newblock \showarticletitle{Investigating the Potential of Life-like Haptic
  Cues for Socially Assistive Care Robots}. In
  \bibinfo{booktitle}{\emph{Companion of the 2023 ACM/IEEE International
  Conference on Human-Robot Interaction}} (Stockholm, Sweden)
  \emph{(\bibinfo{series}{HRI '23})}. \bibinfo{publisher}{ACM},
  \bibinfo{address}{New York, NY, USA}, \bibinfo{pages}{745–747}.
\newblock
\showISBNx{9781450399708}
\href{https://doi.org/10.1145/3568294.3579972}{doi:\nolinkurl{10.1145/3568294.3579972}}


\bibitem[Brown and Hurst(2012)]%
        {Brown2012}
\bibfield{author}{\bibinfo{person}{Craig Brown} {and} \bibinfo{person}{Amy
  Hurst}.} \bibinfo{year}{2012}\natexlab{}.
\newblock \showarticletitle{VizTouch: Automatically Generated Tactile
  Visualizations of Coordinate Spaces}. In \bibinfo{booktitle}{\emph{Proc.
  Tangible, Embedded, and Embodied Interaction}} (Kingston, Ontario, Canada)
  \emph{(\bibinfo{series}{TEI '12})}. \bibinfo{publisher}{ACM},
  \bibinfo{address}{New York, NY, USA}, \bibinfo{pages}{131--138}.
\newblock
\showISBNx{978-1-4503-1174-8}
\href{https://doi.org/10.1145/2148131.2148160}{doi:\nolinkurl{10.1145/2148131.2148160}}


\bibitem[Buehler et~al\mbox{.}(2016)]%
        {Buehler2016}
\bibfield{author}{\bibinfo{person}{Erin Buehler}, \bibinfo{person}{Niara
  Comrie}, \bibinfo{person}{Megan Hofmann}, \bibinfo{person}{Samantha
  McDonald}, {and} \bibinfo{person}{Amy Hurst}.}
  \bibinfo{year}{2016}\natexlab{}.
\newblock \showarticletitle{Investigating the Implications of 3D Printing in
  Special Education}.
\newblock \bibinfo{journal}{\emph{ACM Trans. Access. Comput.}}
  \bibinfo{volume}{8}, \bibinfo{number}{3}, Article \bibinfo{articleno}{11}
  (\bibinfo{date}{March} \bibinfo{year}{2016}), \bibinfo{numpages}{28}~pages.
\newblock
\showISSN{1936-7228}
\href{https://doi.org/10.1145/2870640}{doi:\nolinkurl{10.1145/2870640}}


\bibitem[Butler et~al\mbox{.}(2017)]%
        {butler2017understanding}
\bibfield{author}{\bibinfo{person}{Matthew Butler}, \bibinfo{person}{Leona
  Holloway}, \bibinfo{person}{Kim Marriott}, {and} \bibinfo{person}{Cagatay
  Goncu}.} \bibinfo{year}{2017}\natexlab{}.
\newblock \showarticletitle{Understanding the graphical challenges faced by
  vision-impaired students in Australian universities}.
\newblock \bibinfo{journal}{\emph{Higher Education Research \& Development}}
  \bibinfo{volume}{36}, \bibinfo{number}{1} (\bibinfo{year}{2017}),
  \bibinfo{pages}{59--72}.
\newblock
\href{https://doi.org/10.1080/07294360.2016.1177001}{doi:\nolinkurl{10.1080/07294360.2016.1177001}}


\bibitem[Butler et~al\mbox{.}(2021)]%
        {Butler2021}
\bibfield{author}{\bibinfo{person}{Matthew Butler}, \bibinfo{person}{Leona~M
  Holloway}, \bibinfo{person}{Samuel Reinders}, \bibinfo{person}{Cagatay
  Goncu}, {and} \bibinfo{person}{Kim Marriott}.}
  \bibinfo{year}{2021}\natexlab{}.
\newblock \showarticletitle{Technology Developments in Touch-Based Accessible
  Graphics: A Systematic Review of Research 2010-2020}. In
  \bibinfo{booktitle}{\emph{Proc. ACM CHI Conference on Human Factors in
  Computing Systems}} (Yokohama, Japan) \emph{(\bibinfo{series}{CHI '21})}.
  \bibinfo{publisher}{ACM}, \bibinfo{address}{New York, NY, USA}, Article
  \bibinfo{articleno}{278}, \bibinfo{numpages}{15}~pages.
\newblock
\showISBNx{9781450380966}
\href{https://doi.org/10.1145/3411764.3445207}{doi:\nolinkurl{10.1145/3411764.3445207}}


\bibitem[Butler et~al\mbox{.}(2023)]%
        {Butler2023}
\bibfield{author}{\bibinfo{person}{Matthew Butler}, \bibinfo{person}{Erica~J
  Tandori}, \bibinfo{person}{Vince Dziekan}, \bibinfo{person}{Kirsten Ellis},
  \bibinfo{person}{Jenna Hall}, \bibinfo{person}{Leona~M Holloway},
  \bibinfo{person}{Ruth~G Nagassa}, {and} \bibinfo{person}{Kim Marriott}.}
  \bibinfo{year}{2023}\natexlab{}.
\newblock \showarticletitle{A Gallery In My Hand: A Multi-Exhibition
  Investigation of Accessible and Inclusive Gallery Experiences for Blind and
  Low Vision Visitors}. In \bibinfo{booktitle}{\emph{Proc. ACM SIGACCESS
  Conference on Computers \& Accessibility}} (New York, NY, USA)
  \emph{(\bibinfo{series}{ASSETS '23})}. \bibinfo{publisher}{ACM},
  \bibinfo{address}{New York, NY, USA}, Article \bibinfo{articleno}{9},
  \bibinfo{numpages}{15}~pages.
\newblock
\showISBNx{9798400702204}
\href{https://doi.org/10.1145/3597638.3608391}{doi:\nolinkurl{10.1145/3597638.3608391}}


\bibitem[Carlton et~al\mbox{.}(2019)]%
        {carlton2019}
\bibfield{author}{\bibinfo{person}{Jonathan Carlton}, \bibinfo{person}{Andy
  Brown}, \bibinfo{person}{Caroline Jay}, {and} \bibinfo{person}{John Keane}.}
  \bibinfo{year}{2019}\natexlab{}.
\newblock \showarticletitle{Inferring User Engagement from Interaction Data}.
  In \bibinfo{booktitle}{\emph{Proc. ACM CHI Conference on Human Factors in
  Computing Systems}} (Glasgow, Scotland Uk) \emph{(\bibinfo{series}{CHI EA
  '19})}. \bibinfo{publisher}{ACM}, \bibinfo{address}{New York, NY, USA},
  \bibinfo{pages}{1–6}.
\newblock
\showISBNx{9781450359719}
\href{https://doi.org/10.1145/3290607.3313009}{doi:\nolinkurl{10.1145/3290607.3313009}}


\bibitem[Cassell(2000)]%
        {cassell2000more}
\bibfield{author}{\bibinfo{person}{Justine Cassell}.}
  \bibinfo{year}{2000}\natexlab{}.
\newblock \showarticletitle{More than just another pretty face: Embodied
  conversational interface agents}.
\newblock \bibinfo{journal}{\emph{Commun. ACM}} \bibinfo{volume}{43},
  \bibinfo{number}{4} (\bibinfo{year}{2000}), \bibinfo{pages}{70--78}.
\newblock


\bibitem[Cassell(2001)]%
        {Cassell2001}
\bibfield{author}{\bibinfo{person}{Justine Cassell}.}
  \bibinfo{year}{2001}\natexlab{}.
\newblock \showarticletitle{Embodied Conversational Agents: Representation and
  Intelligence in User Interfaces}.
\newblock \bibinfo{journal}{\emph{AI Magazine}} \bibinfo{volume}{22},
  \bibinfo{number}{4} (\bibinfo{date}{Dec.} \bibinfo{year}{2001}),
  \bibinfo{pages}{67}.
\newblock
\href{https://doi.org/10.1609/aimag.v22i4.1593}{doi:\nolinkurl{10.1609/aimag.v22i4.1593}}


\bibitem[Cassell et~al\mbox{.}(1999)]%
        {Cassell1999}
\bibfield{author}{\bibinfo{person}{J. Cassell}, \bibinfo{person}{T. Bickmore},
  \bibinfo{person}{M. Billinghurst}, \bibinfo{person}{L. Campbell},
  \bibinfo{person}{K. Chang}, \bibinfo{person}{H. Vilhj\'{a}lmsson}, {and}
  \bibinfo{person}{H. Yan}.} \bibinfo{year}{1999}\natexlab{}.
\newblock \showarticletitle{Embodiment in conversational interfaces: Rea}. In
  \bibinfo{booktitle}{\emph{Proc. ACM CHI Conference on Human Factors in
  Computing Systems}} (Pittsburgh, Pennsylvania, USA)
  \emph{(\bibinfo{series}{CHI '99})}. \bibinfo{publisher}{ACM},
  \bibinfo{address}{New York, NY, USA}, \bibinfo{pages}{520–527}.
\newblock
\showISBNx{0201485591}
\href{https://doi.org/10.1145/302979.303150}{doi:\nolinkurl{10.1145/302979.303150}}


\bibitem[Cassell and Thorisson(1999)]%
        {Cassell1999c}
\bibfield{author}{\bibinfo{person}{Justine Cassell} {and}
  \bibinfo{person}{Kristinn~R. Thorisson}.} \bibinfo{year}{1999}\natexlab{}.
\newblock \showarticletitle{The power of a nod and a glance: Envelope vs.
  emotional feedback in animated conversational agents}.
\newblock \bibinfo{journal}{\emph{Applied Artificial Intelligence}}
  \bibinfo{volume}{13}, \bibinfo{number}{4-5} (\bibinfo{year}{1999}),
  \bibinfo{pages}{519--538}.
\newblock
\href{https://doi.org/10.1080/088395199117360}{doi:\nolinkurl{10.1080/088395199117360}}


\bibitem[Cassell and Vilhj{\'a}lmsson(1999)]%
        {Cassell1999b}
\bibfield{author}{\bibinfo{person}{J. Cassell} {and} \bibinfo{person}{H.
  Vilhj{\'a}lmsson}.} \bibinfo{year}{1999}\natexlab{}.
\newblock \showarticletitle{Fully Embodied Conversational Avatars: Making
  Communicative Behaviors Autonomous}.
\newblock \bibinfo{journal}{\emph{Autonomous Agents and Multi-Agent Systems}}
  \bibinfo{volume}{2}, \bibinfo{number}{1} (\bibinfo{date}{01 Mar}
  \bibinfo{year}{1999}), \bibinfo{pages}{45--64}.
\newblock
\showISSN{1573-7454}
\href{https://doi.org/10.1023/A:1010027123541}{doi:\nolinkurl{10.1023/A:1010027123541}}


\bibitem[Cavazos~Quero et~al\mbox{.}(2019)]%
        {Quero2019}
\bibfield{author}{\bibinfo{person}{Luis Cavazos~Quero}, \bibinfo{person}{Jorge
  Iranzo~Bartolom\'{e}}, \bibinfo{person}{Dongmyeong Lee},
  \bibinfo{person}{Yerin Lee}, \bibinfo{person}{Sangwon Lee}, {and}
  \bibinfo{person}{Jundong Cho}.} \bibinfo{year}{2019}\natexlab{}.
\newblock \showarticletitle{Jido: A Conversational Tactile Map for Blind
  People}. In \bibinfo{booktitle}{\emph{Proc. ACM SIGACCESS Conference on
  Computers \& Accessibility}} (Pittsburgh, PA, USA)
  \emph{(\bibinfo{series}{ASSETS '19})}. \bibinfo{publisher}{ACM},
  \bibinfo{address}{New York, NY, USA}, \bibinfo{pages}{682–684}.
\newblock
\showISBNx{9781450366762}
\href{https://doi.org/10.1145/3308561.3354600}{doi:\nolinkurl{10.1145/3308561.3354600}}


\bibitem[Cavazos~Quero et~al\mbox{.}(2018)]%
        {Quero2018}
\bibfield{author}{\bibinfo{person}{Luis Cavazos~Quero}, \bibinfo{person}{Jorge
  Iranzo~Bartolom\'{e}}, \bibinfo{person}{Seonggu Lee}, \bibinfo{person}{En
  Han}, \bibinfo{person}{Sunhee Kim}, {and} \bibinfo{person}{Jundong Cho}.}
  \bibinfo{year}{2018}\natexlab{}.
\newblock \showarticletitle{An Interactive Multimodal Guide to Improve Art
  Accessibility for Blind People}. In \bibinfo{booktitle}{\emph{Proc. ACM
  SIGACCESS Conference on Computers \& Accessibility}} (Galway, Ireland)
  \emph{(\bibinfo{series}{ASSETS '18})}. \bibinfo{publisher}{ACM},
  \bibinfo{address}{New York, NY, USA}, \bibinfo{pages}{346–348}.
\newblock
\showISBNx{9781450356503}
\href{https://doi.org/10.1145/3234695.3241033}{doi:\nolinkurl{10.1145/3234695.3241033}}


\bibitem[Cavazos~Quero et~al\mbox{.}(2021)]%
        {Quero2021}
\bibfield{author}{\bibinfo{person}{Luis Cavazos~Quero}, \bibinfo{person}{Jorge
  Iranzo~Bartolomé}, {and} \bibinfo{person}{Jundong Cho}.}
  \bibinfo{year}{2021}\natexlab{}.
\newblock \showarticletitle{Accessible Visual Artworks for Blind and Visually
  Impaired People: Comparing a Multimodal Approach with Tactile Graphics}.
\newblock \bibinfo{journal}{\emph{Electronics}} \bibinfo{volume}{10},
  \bibinfo{number}{3} (\bibinfo{year}{2021}).
\newblock
\showISSN{2079-9292}
\href{https://doi.org/10.3390/electronics10030297}{doi:\nolinkurl{10.3390/electronics10030297}}


\bibitem[Cho et~al\mbox{.}(2024)]%
        {Cho2024}
\bibfield{author}{\bibinfo{person}{Haena Cho}, \bibinfo{person}{Yoonji Lee},
  \bibinfo{person}{Woohun Lee}, {and} \bibinfo{person}{Chang~Hee Lee}.}
  \bibinfo{year}{2024}\natexlab{}.
\newblock \showarticletitle{Thermo-Play: Exploring the Playful Qualities of
  Thermochromic Materials}. In \bibinfo{booktitle}{\emph{Proc. Tangible,
  Embedded, and Embodied Interaction}} (Cork, Ireland)
  \emph{(\bibinfo{series}{TEI '24})}. \bibinfo{publisher}{ACM},
  \bibinfo{address}{New York, NY, USA}, Article \bibinfo{articleno}{28},
  \bibinfo{numpages}{16}~pages.
\newblock
\showISBNx{9798400704024}
\href{https://doi.org/10.1145/3623509.3633376}{doi:\nolinkurl{10.1145/3623509.3633376}}


\bibitem[Choi et~al\mbox{.}(2020)]%
        {Choi2020}
\bibfield{author}{\bibinfo{person}{Dasom Choi}, \bibinfo{person}{Daehyun Kwak},
  \bibinfo{person}{Minji Cho}, {and} \bibinfo{person}{Sangsu Lee}.}
  \bibinfo{year}{2020}\natexlab{}.
\newblock \showarticletitle{"Nobody Speaks That Fast!" An Empirical Study of
  Speech Rate in Conversational Agents for People with Vision Impairments}. In
  \bibinfo{booktitle}{\emph{Proc. ACM CHI Conference on Human Factors in
  Computing Systems}} (Honolulu, HI, USA) \emph{(\bibinfo{series}{CHI '20})}.
  \bibinfo{publisher}{ACM}, \bibinfo{address}{New York, NY, USA},
  \bibinfo{pages}{1–13}.
\newblock
\showISBNx{9781450367080}
\href{https://doi.org/10.1145/3313831.3376569}{doi:\nolinkurl{10.1145/3313831.3376569}}


\bibitem[Coeckelbergh(2011)]%
        {Coeckelbergh2011}
\bibfield{author}{\bibinfo{person}{Mark Coeckelbergh}.}
  \bibinfo{year}{2011}\natexlab{}.
\newblock \showarticletitle{You, robot: on the linguistic construction of
  artificial others}.
\newblock \bibinfo{journal}{\emph{AI {\&} SOCIETY}} \bibinfo{volume}{26},
  \bibinfo{number}{1} (\bibinfo{date}{01 Feb} \bibinfo{year}{2011}),
  \bibinfo{pages}{61--69}.
\newblock
\showISSN{1435-5655}
\href{https://doi.org/10.1007/s00146-010-0289-z}{doi:\nolinkurl{10.1007/s00146-010-0289-z}}


\bibitem[Collins et~al\mbox{.}(2023)]%
        {Collins2023}
\bibfield{author}{\bibinfo{person}{Jazmin Collins}, \bibinfo{person}{Crescentia
  Jung}, \bibinfo{person}{Yeonju Jang}, \bibinfo{person}{Danielle Montour},
  \bibinfo{person}{Andrea~Stevenson Won}, {and} \bibinfo{person}{Shiri
  Azenkot}.} \bibinfo{year}{2023}\natexlab{}.
\newblock \showarticletitle{“The Guide Has Your Back”: Exploring How
  Sighted Guides Can Enhance Accessibility in Social Virtual Reality for Blind
  and Low Vision People}. In \bibinfo{booktitle}{\emph{Proc. ACM SIGACCESS
  Conference on Computers \& Accessibility}} (New York, NY, USA)
  \emph{(\bibinfo{series}{ASSETS '23})}. \bibinfo{publisher}{ACM},
  \bibinfo{address}{New York, NY, USA}, Article \bibinfo{articleno}{38},
  \bibinfo{numpages}{14}~pages.
\newblock
\showISBNx{9798400702204}
\href{https://doi.org/10.1145/3597638.3608386}{doi:\nolinkurl{10.1145/3597638.3608386}}


\bibitem[Cramer and Howitt(2004)]%
        {Cramer2004}
\bibfield{author}{\bibinfo{person}{Duncan Cramer} {and}
  \bibinfo{person}{Dennis. Howitt}.} \bibinfo{year}{2004}\natexlab{}.
\newblock \bibinfo{booktitle}{\emph{The Sage dictionary of statistics: a
  practical resource for students in the social sciences}}.
\newblock \bibinfo{publisher}{Sage Publications}, \bibinfo{address}{London}.
\newblock
\showISBNx{0761941371}
\newblock
\shownote{Includes bibliographical references (p. [187]-188).}.


\bibitem[Davis et~al\mbox{.}(2020)]%
        {Davis2020}
\bibfield{author}{\bibinfo{person}{Josh~Urban Davis}, \bibinfo{person}{Te-Yen
  Wu}, \bibinfo{person}{Bo Shi}, \bibinfo{person}{Hanyi Lu},
  \bibinfo{person}{Athina Panotopoulou}, \bibinfo{person}{Emily Whiting}, {and}
  \bibinfo{person}{Xing-Dong Yang}.} \bibinfo{year}{2020}\natexlab{}.
\newblock \showarticletitle{TangibleCircuits: An Interactive 3D Printed Circuit
  Education Tool for People with Visual Impairments}. In
  \bibinfo{booktitle}{\emph{Proc. ACM CHI Conference on Human Factors in
  Computing Systems}} (Honolulu, HI, USA) \emph{(\bibinfo{series}{CHI '20})}.
  \bibinfo{publisher}{ACM}, \bibinfo{address}{New York, NY, USA},
  \bibinfo{pages}{1–13}.
\newblock
\showISBNx{9781450367080}
\href{https://doi.org/10.1145/3313831.3376513}{doi:\nolinkurl{10.1145/3313831.3376513}}


\bibitem[Degachi et~al\mbox{.}(2023)]%
        {Degachi2023}
\bibfield{author}{\bibinfo{person}{Chadha Degachi},
  \bibinfo{person}{Myrthe~Lotte Tielman}, {and} \bibinfo{person}{Mohammed
  Al~Owayyed}.} \bibinfo{year}{2023}\natexlab{}.
\newblock \showarticletitle{Trust and Perceived Control in Burnout Support
  Chatbots}. In \bibinfo{booktitle}{\emph{Proc. ACM CHI Conference on Human
  Factors in Computing Systems}} (Hamburg, Germany) \emph{(\bibinfo{series}{CHI
  EA '23})}. \bibinfo{publisher}{ACM}, \bibinfo{address}{New York, NY, USA},
  Article \bibinfo{articleno}{295}, \bibinfo{numpages}{10}~pages.
\newblock
\showISBNx{9781450394222}
\href{https://doi.org/10.1145/3544549.3585780}{doi:\nolinkurl{10.1145/3544549.3585780}}


\bibitem[Doherty and Doherty(2018)]%
        {Doherty2018}
\bibfield{author}{\bibinfo{person}{Kevin Doherty} {and} \bibinfo{person}{Gavin
  Doherty}.} \bibinfo{year}{2018}\natexlab{}.
\newblock \showarticletitle{Engagement in HCI: Conception, Theory and
  Measurement}.
\newblock \bibinfo{journal}{\emph{ACM Comput. Surv.}} \bibinfo{volume}{51},
  \bibinfo{number}{5}, Article \bibinfo{articleno}{99} (\bibinfo{date}{nov}
  \bibinfo{year}{2018}), \bibinfo{numpages}{39}~pages.
\newblock
\showISSN{0360-0300}
\href{https://doi.org/10.1145/3234149}{doi:\nolinkurl{10.1145/3234149}}


\bibitem[Dourish(2001)]%
        {Dourish2001}
\bibfield{author}{\bibinfo{person}{Paul Dourish}.}
  \bibinfo{year}{2001}\natexlab{}.
\newblock \bibinfo{booktitle}{\emph{{Where the Action Is: The Foundations of
  Embodied Interaction}}}.
\newblock \bibinfo{publisher}{The MIT Press}.
\newblock
\showISBNx{9780262256056}
\href{https://doi.org/10.7551/mitpress/7221.001.0001}{doi:\nolinkurl{10.7551/mitpress/7221.001.0001}}


\bibitem[Edwards et~al\mbox{.}(2015)]%
        {Edwards2015}
\bibfield{author}{\bibinfo{person}{Alistair D.~N. Edwards},
  \bibinfo{person}{Nazatul Naquiah~Abd Hamid}, {and} \bibinfo{person}{Helen
  Petrie}.} \bibinfo{year}{2015}\natexlab{}.
\newblock \showarticletitle{Exploring Map Orientation with Interactive
  Audio-Tactile Maps}. In \bibinfo{booktitle}{\emph{Human-Computer Interaction
  -- INTERACT 2015}}, \bibfield{editor}{\bibinfo{person}{Julio Abascal},
  \bibinfo{person}{Simone Barbosa}, \bibinfo{person}{Mirko Fetter},
  \bibinfo{person}{Tom Gross}, \bibinfo{person}{Philippe Palanque}, {and}
  \bibinfo{person}{Marco Winckler}} (Eds.). \bibinfo{publisher}{Springer
  International Publishing}, \bibinfo{address}{Cham}, \bibinfo{pages}{72--79}.
\newblock
\showISBNx{978-3-319-22701-6}


\bibitem[Ghodke et~al\mbox{.}(2019)]%
        {Ghodke2019}
\bibfield{author}{\bibinfo{person}{Uttara Ghodke}, \bibinfo{person}{Lena
  Yusim}, \bibinfo{person}{Sowmya Somanath}, {and} \bibinfo{person}{Peter
  Coppin}.} \bibinfo{year}{2019}\natexlab{}.
\newblock \showarticletitle{The Cross-Sensory Globe: Participatory Design of a
  3D Audio-Tactile Globe Prototype for Blind and Low-Vision Users to Learn
  Geography}. In \bibinfo{booktitle}{\emph{Proc. ACM Designing Interactive
  Systems Conference}} (San Diego, CA, USA) \emph{(\bibinfo{series}{DIS '19})}.
  \bibinfo{publisher}{ACM}, \bibinfo{address}{New York, NY, USA},
  \bibinfo{pages}{399–412}.
\newblock
\showISBNx{9781450358507}
\href{https://doi.org/10.1145/3322276.3323686}{doi:\nolinkurl{10.1145/3322276.3323686}}


\bibitem[Giraud et~al\mbox{.}(2017)]%
        {Giraud2017}
\bibfield{author}{\bibinfo{person}{St\'{e}phanie Giraud},
  \bibinfo{person}{Anke~M Brock}, \bibinfo{person}{Marc J-M Mac\'{e}}, {and}
  \bibinfo{person}{Christophe Jouffrais}.} \bibinfo{year}{2017}\natexlab{}.
\newblock \showarticletitle{Map learning with a 3D printed interactive
  small-scale model: Improvement of space and text memorization in visually
  impaired students}.
\newblock \bibinfo{journal}{\emph{Frontiers in Psychology}}
  \bibinfo{volume}{8}, \bibinfo{number}{930} (\bibinfo{year}{2017}).
\newblock
\href{https://doi.org/10.3389/fpsyg.2017.00930}{doi:\nolinkurl{10.3389/fpsyg.2017.00930}}


\bibitem[G\"{o}tzelmann et~al\mbox{.}(2017)]%
        {Gotzelmann2017}
\bibfield{author}{\bibinfo{person}{Timo G\"{o}tzelmann}, \bibinfo{person}{Lisa
  Branz}, \bibinfo{person}{Claudia Heidenreich}, {and} \bibinfo{person}{Markus
  Otto}.} \bibinfo{year}{2017}\natexlab{}.
\newblock \showarticletitle{A Personal Computer-based Approach for 3D Printing
  Accessible to Blind People}. In \bibinfo{booktitle}{\emph{Proc. PErvasive
  Technologies Related to Assistive Environments}} (Island of Rhodes, Greece)
  \emph{(\bibinfo{series}{PETRA '17})}. \bibinfo{publisher}{ACM},
  \bibinfo{address}{New York, NY, USA}, \bibinfo{pages}{1--4}.
\newblock
\showISBNx{978-1-4503-5227-7}
\href{https://doi.org/10.1145/3056540.3064954}{doi:\nolinkurl{10.1145/3056540.3064954}}


\bibitem[Gual et~al\mbox{.}(2012)]%
        {gual2012visual}
\bibfield{author}{\bibinfo{person}{Jaume Gual}, \bibinfo{person}{Marina
  Puyuelo}, \bibinfo{person}{Joaquim Llover{\'a}s}, {and} \bibinfo{person}{Lola
  Merino}.} \bibinfo{year}{2012}\natexlab{}.
\newblock \showarticletitle{Visual Impairment and urban orientation. Pilot
  study with tactile maps produced through 3D Printing}.
\newblock \bibinfo{journal}{\emph{Psyecology}} \bibinfo{volume}{3},
  \bibinfo{number}{2} (\bibinfo{year}{2012}), \bibinfo{pages}{239--250}.
\newblock


\bibitem[Gulati et~al\mbox{.}(2018)]%
        {Gulati2018}
\bibfield{author}{\bibinfo{person}{Siddharth Gulati}, \bibinfo{person}{Sonia
  Sousa}, {and} \bibinfo{person}{David Lamas}.}
  \bibinfo{year}{2018}\natexlab{}.
\newblock \showarticletitle{Modelling trust in human-like technologies}. In
  \bibinfo{booktitle}{\emph{Proc. Indian Conference on Human-Computer
  Interaction}} (Bangalore, India) \emph{(\bibinfo{series}{IndiaHCI '18})}.
  \bibinfo{publisher}{ACM}, \bibinfo{address}{New York, NY, USA},
  \bibinfo{pages}{1–10}.
\newblock
\showISBNx{9781450362146}
\href{https://doi.org/10.1145/3297121.3297124}{doi:\nolinkurl{10.1145/3297121.3297124}}


\bibitem[Guo et~al\mbox{.}(2022)]%
        {Guo2022}
\bibfield{author}{\bibinfo{person}{Lijie Guo}, \bibinfo{person}{Elizabeth~M.
  Daly}, \bibinfo{person}{Oznur Alkan}, \bibinfo{person}{Massimiliano
  Mattetti}, \bibinfo{person}{Owen Cornec}, {and} \bibinfo{person}{Bart
  Knijnenburg}.} \bibinfo{year}{2022}\natexlab{}.
\newblock \showarticletitle{Building Trust in Interactive Machine Learning via
  User Contributed Interpretable Rules}. In \bibinfo{booktitle}{\emph{Proc.
  International Conference on Intelligent User Interfaces}} (Helsinki, Finland)
  \emph{(\bibinfo{series}{IUI '22})}. \bibinfo{publisher}{ACM},
  \bibinfo{address}{New York, NY, USA}, \bibinfo{pages}{537–548}.
\newblock
\showISBNx{9781450391443}
\href{https://doi.org/10.1145/3490099.3511111}{doi:\nolinkurl{10.1145/3490099.3511111}}


\bibitem[Hasper et~al\mbox{.}(2015)]%
        {Hasper2015}
\bibfield{author}{\bibinfo{person}{Eric Hasper}, \bibinfo{person}{Rogier~A.
  Windhorst}, \bibinfo{person}{Terri Hedgpeth}, \bibinfo{person}{Leanne~Van
  Tuyl}, \bibinfo{person}{Ashleigh Gonzales}, \bibinfo{person}{Britta
  Martinez}, \bibinfo{person}{Hongyu Yu}, \bibinfo{person}{Zoltan Farkas},
  {and} \bibinfo{person}{Debra~P. Baluch}.} \bibinfo{year}{2015}\natexlab{}.
\newblock \showarticletitle{Methods for Creating and Evaluating 3D Tactile
  Images to Teach STEM Courses to the Visually Impaired}.
\newblock \bibinfo{journal}{\emph{Journal of College Science Teaching}}
  \bibinfo{volume}{44}, \bibinfo{number}{6} (\bibinfo{year}{2015}),
  \bibinfo{pages}{92--99}.
\newblock
\showISSN{0047231X, 19434898}
\urldef\tempurl%
\url{http://www.jstor.org/stable/43632001}
\showURL{%
\tempurl}


\bibitem[Heuwinkel(2013)]%
        {Heuwinkel2013}
\bibfield{author}{\bibinfo{person}{Kerstin Heuwinkel}.}
  \bibinfo{year}{2013}\natexlab{}.
\newblock \bibinfo{booktitle}{\emph{Framing the Invisible -- The Social
  Background of Trust}}.
\newblock \bibinfo{publisher}{Springer Berlin Heidelberg},
  \bibinfo{address}{Berlin, Heidelberg}, \bibinfo{pages}{16--26}.
\newblock
\showISBNx{978-3-642-37346-6}
\href{https://doi.org/10.1007/978-3-642-37346-6_3}{doi:\nolinkurl{10.1007/978-3-642-37346-6_3}}


\bibitem[Holloway et~al\mbox{.}(2022)]%
        {Holloway2022}
\bibfield{author}{\bibinfo{person}{Leona Holloway}, \bibinfo{person}{Swamy
  Ananthanarayan}, \bibinfo{person}{Matthew Butler},
  \bibinfo{person}{Madhuka~Thisuri De~Silva}, \bibinfo{person}{Kirsten Ellis},
  \bibinfo{person}{Cagatay Goncu}, \bibinfo{person}{Kate Stephens}, {and}
  \bibinfo{person}{Kim Marriott}.} \bibinfo{year}{2022}\natexlab{}.
\newblock \showarticletitle{Animations at Your Fingertips: Using a Refreshable
  Tactile Display to Convey Motion Graphics for People Who Are Blind or Have
  Low Vision}. In \bibinfo{booktitle}{\emph{Proc. ACM SIGACCESS Conference on
  Computers \& Accessibility}} (Athens, Greece) \emph{(\bibinfo{series}{ASSETS
  '22})}. \bibinfo{publisher}{ACM}, \bibinfo{address}{New York, NY, USA},
  Article \bibinfo{articleno}{32}, \bibinfo{numpages}{16}~pages.
\newblock
\showISBNx{9781450392587}
\href{https://doi.org/10.1145/3517428.3544797}{doi:\nolinkurl{10.1145/3517428.3544797}}


\bibitem[Holloway et~al\mbox{.}(2018)]%
        {Holloway2018}
\bibfield{author}{\bibinfo{person}{Leona Holloway}, \bibinfo{person}{Matthew
  Butler}, {and} \bibinfo{person}{Kim Marriott}.}
  \bibinfo{year}{2018}\natexlab{}.
\newblock \showarticletitle{Accessible maps for the blind: Comparing 3D printed
  models with tactile graphics}. In \bibinfo{booktitle}{\emph{Proc. ACM CHI
  Conference on Human Factors in Computing Systems}} (Montréal, QC, Canada)
  \emph{(\bibinfo{series}{CHI '18})}. \bibinfo{publisher}{ACM}.
\newblock
\href{https://doi.org/10.1145/3173574.3173772}{doi:\nolinkurl{10.1145/3173574.3173772}}


\bibitem[Holloway et~al\mbox{.}(2019a)]%
        {Holloway2019}
\bibfield{author}{\bibinfo{person}{Leona Holloway}, \bibinfo{person}{Kim
  Marriott}, \bibinfo{person}{Matthew Butler}, {and} \bibinfo{person}{Alan
  Borning}.} \bibinfo{year}{2019}\natexlab{a}.
\newblock \showarticletitle{Making Sense of Art: Access for Gallery Visitors
  with Vision Impairments}. In \bibinfo{booktitle}{\emph{Proc. ACM CHI
  Conference on Human Factors in Computing Systems}} (Glasgow, Scotland Uk)
  \emph{(\bibinfo{series}{CHI '19})}. \bibinfo{publisher}{ACM},
  \bibinfo{address}{New York, NY, USA}, \bibinfo{pages}{1–12}.
\newblock
\showISBNx{9781450359702}
\href{https://doi.org/10.1145/3290605.3300250}{doi:\nolinkurl{10.1145/3290605.3300250}}


\bibitem[Holloway et~al\mbox{.}(2019b)]%
        {Holloway2019b}
\bibfield{author}{\bibinfo{person}{Leona Holloway}, \bibinfo{person}{Kim
  Marriott}, \bibinfo{person}{Matthew Butler}, {and} \bibinfo{person}{Samuel
  Reinders}.} \bibinfo{year}{2019}\natexlab{b}.
\newblock \showarticletitle{3D Printed Maps and Icons for Inclusion: Testing in
  the Wild by People Who Are Blind or Have Low Vision}. In
  \bibinfo{booktitle}{\emph{Proc. ACM SIGACCESS Conference on Computers \&
  Accessibility}} (Pittsburgh, PA, USA) \emph{(\bibinfo{series}{ASSETS '19})}.
  \bibinfo{publisher}{ACM}, \bibinfo{address}{New York, NY, USA},
  \bibinfo{pages}{183–195}.
\newblock
\showISBNx{9781450366762}
\href{https://doi.org/10.1145/3308561.3353790}{doi:\nolinkurl{10.1145/3308561.3353790}}


\bibitem[Hu(2015)]%
        {Hu2015}
\bibfield{author}{\bibinfo{person}{Michele Hu}.}
  \bibinfo{year}{2015}\natexlab{}.
\newblock \showarticletitle{Exploring New Paradigms for Accessible 3D Printed
  Graphs}. In \bibinfo{booktitle}{\emph{Proc. ACM SIGACCESS Conference on
  Computers \& Accessibility}} (Lisbon, Portugal)
  \emph{(\bibinfo{series}{ASSETS '15})}. \bibinfo{publisher}{ACM},
  \bibinfo{address}{New York, NY, USA}, \bibinfo{pages}{365--366}.
\newblock
\showISBNx{978-1-4503-3400-6}
\href{https://doi.org/10.1145/2700648.2811330}{doi:\nolinkurl{10.1145/2700648.2811330}}


\bibitem[Inc({[n.\,d.]})]%
        {TTT}
\bibfield{author}{\bibinfo{person}{Touch~Graphics Inc}.}
  \bibinfo{year}{[n.\,d.]}\natexlab{}.
\newblock \bibinfo{title}{T3 Tactile Tablet}.  (\bibinfo{year}{[n.\,d.]}).
\newblock
\newblock
\shownote{Available from \url{https://www.touchgraphics.com/education/t3}}.


\bibitem[Iranzo~Bartolome et~al\mbox{.}(2019)]%
        {Bartolome2019}
\bibfield{author}{\bibinfo{person}{Jorge Iranzo~Bartolome},
  \bibinfo{person}{Luis Cavazos~Quero}, \bibinfo{person}{Sunhee Kim},
  \bibinfo{person}{Myung-Yong Um}, {and} \bibinfo{person}{Jundong Cho}.}
  \bibinfo{year}{2019}\natexlab{}.
\newblock \showarticletitle{Exploring Art with a Voice Controlled Multimodal
  Guide for Blind People}. In \bibinfo{booktitle}{\emph{Proc. Tangible,
  Embedded, and Embodied Interaction}} (Tempe, Arizona, USA)
  \emph{(\bibinfo{series}{TEI '19})}. \bibinfo{publisher}{ACM},
  \bibinfo{address}{New York, NY, USA}, \bibinfo{pages}{383–390}.
\newblock
\showISBNx{9781450361965}
\href{https://doi.org/10.1145/3294109.3300994}{doi:\nolinkurl{10.1145/3294109.3300994}}


\bibitem[Iravantchi et~al\mbox{.}(2020)]%
        {Iravantchi2020}
\bibfield{author}{\bibinfo{person}{Yasha Iravantchi}, \bibinfo{person}{Mayank
  Goel}, {and} \bibinfo{person}{Chris Harrison}.}
  \bibinfo{year}{2020}\natexlab{}.
\newblock \showarticletitle{Digital Ventriloquism: Giving Voice to Everyday
  Objects}. In \bibinfo{booktitle}{\emph{Proc. ACM CHI Conference on Human
  Factors in Computing Systems}} (Honolulu, HI, USA)
  \emph{(\bibinfo{series}{CHI '20})}. \bibinfo{publisher}{ACM},
  \bibinfo{address}{New York, NY, USA}, \bibinfo{pages}{1–10}.
\newblock
\showISBNx{9781450367080}
\href{https://doi.org/10.1145/3313831.3376503}{doi:\nolinkurl{10.1145/3313831.3376503}}


\bibitem[Kane and Bigham(2014)]%
        {kane2014}
\bibfield{author}{\bibinfo{person}{Shaun~K. Kane} {and}
  \bibinfo{person}{Jeffrey~P. Bigham}.} \bibinfo{year}{2014}\natexlab{}.
\newblock \showarticletitle{Tracking@ stemxcomet: teaching programming to blind
  students via 3D printing, crisis management, and twitter}. In
  \bibinfo{booktitle}{\emph{Proc. ACM Computer Science Education}}. ACM,
  \bibinfo{pages}{247--252}.
\newblock


\bibitem[Karaduman et~al\mbox{.}(2022)]%
        {karaduman2022beyond}
\bibfield{author}{\bibinfo{person}{H{\i}d{\i}r Karaduman},
  \bibinfo{person}{{\"U}mran Alan}, {and} \bibinfo{person}{E~{\"O}zlem
  Yi{\u{g}}it}.} \bibinfo{year}{2022}\natexlab{}.
\newblock \showarticletitle{Beyond “do not touch”: the experience of a
  three-dimensional printed artifacts museum as an alternative to traditional
  museums for visitors who are blind and partially sighted}.
\newblock \bibinfo{journal}{\emph{Universal Access in the Information Society}}
  (\bibinfo{year}{2022}), \bibinfo{pages}{1--14}.
\newblock


\bibitem[Karim et~al\mbox{.}(2023)]%
        {Karim2023}
\bibfield{author}{\bibinfo{person}{Saman Karim}, \bibinfo{person}{Jin Kang},
  {and} \bibinfo{person}{Audrey Girouard}.} \bibinfo{year}{2023}\natexlab{}.
\newblock \showarticletitle{Exploring Rulebook Accessibility and Companionship
  in Board Games via Voiced-based Conversational Agent Alexa}. In
  \bibinfo{booktitle}{\emph{Proc. ACM Designing Interactive Systems
  Conference}} (Pittsburgh, PA, USA) \emph{(\bibinfo{series}{DIS '23})}.
  \bibinfo{publisher}{ACM}, \bibinfo{address}{New York, NY, USA},
  \bibinfo{pages}{2221–2232}.
\newblock
\showISBNx{9781450398930}
\href{https://doi.org/10.1145/3563657.3595970}{doi:\nolinkurl{10.1145/3563657.3595970}}


\bibitem[Keeffe(2005)]%
        {Keeffe2005}
\bibfield{author}{\bibinfo{person}{Jill Keeffe}.}
  \bibinfo{year}{2005}\natexlab{}.
\newblock \showarticletitle{Psychosocial Impact of Vision Impairment}.
\newblock \bibinfo{journal}{\emph{International Congress Series}}
  \bibinfo{volume}{1282} (\bibinfo{year}{2005}), \bibinfo{pages}{167--173}.
\newblock


\bibitem[Kidd and Breazeal(2004)]%
        {Kidd2004}
\bibfield{author}{\bibinfo{person}{C.D. Kidd} {and} \bibinfo{person}{C.
  Breazeal}.} \bibinfo{year}{2004}\natexlab{}.
\newblock \showarticletitle{Effect of a robot on user perceptions}. In
  \bibinfo{booktitle}{\emph{Proc. IEEE/RSJ Intelligent Robots and Systems}},
  Vol.~\bibinfo{volume}{4}. \bibinfo{pages}{3559--3564 vol.4}.
\newblock
\href{https://doi.org/10.1109/IROS.2004.1389967}{doi:\nolinkurl{10.1109/IROS.2004.1389967}}


\bibitem[Kim and Yeh(2015)]%
        {kim2015}
\bibfield{author}{\bibinfo{person}{Jeeeun Kim} {and} \bibinfo{person}{Tom
  Yeh}.} \bibinfo{year}{2015}\natexlab{}.
\newblock \showarticletitle{Toward {3D}-printed movable tactile pictures for
  children with visual impairments}. In \bibinfo{booktitle}{\emph{Proc. ACM CHI
  Conference on Human Factors in Computing Systems}}. ACM,
  \bibinfo{pages}{2815--2824}.
\newblock


\bibitem[Kontogiorgos et~al\mbox{.}(2020)]%
        {Kontogiorgos2020}
\bibfield{author}{\bibinfo{person}{Dimosthenis Kontogiorgos},
  \bibinfo{person}{Sanne van Waveren}, \bibinfo{person}{Olle Wallberg},
  \bibinfo{person}{Andre Pereira}, \bibinfo{person}{Iolanda Leite}, {and}
  \bibinfo{person}{Joakim Gustafson}.} \bibinfo{year}{2020}\natexlab{}.
\newblock \showarticletitle{Embodiment Effects in Interactions with Failing
  Robots}. In \bibinfo{booktitle}{\emph{Proc. ACM CHI Conference on Human
  Factors in Computing Systems}} (Honolulu, HI, USA)
  \emph{(\bibinfo{series}{CHI '20})}. \bibinfo{publisher}{ACM},
  \bibinfo{address}{New York, NY, USA}, \bibinfo{pages}{1–14}.
\newblock
\showISBNx{9781450367080}
\href{https://doi.org/10.1145/3313831.3376372}{doi:\nolinkurl{10.1145/3313831.3376372}}


\bibitem[Landau(2009)]%
        {Landau2009}
\bibfield{author}{\bibinfo{person}{Steven Landau}.}
  \bibinfo{year}{2009}\natexlab{}.
\newblock \showarticletitle{An Interactive Talking Campus Model at Carroll
  Center for the Blind}.
\newblock  (\bibinfo{year}{2009}).
\newblock
\urldef\tempurl%
\url{http://www.touchgraphics.com/downloads/carrollcentertalkingcampusmodelfinalreportlow.pdf}
\showURL{%
\tempurl}


\bibitem[Lankton and Tripp(2015)]%
        {Lankton2015}
\bibfield{author}{\bibinfo{person}{D.~Harrison; Lankton, Nancy K.;~McKnight}
  {and} \bibinfo{person}{John Tripp}.} \bibinfo{year}{2015}\natexlab{}.
\newblock \showarticletitle{Technology, Humanness, and Trust: Rethinking Trust
  in Technology}.
\newblock \bibinfo{journal}{\emph{Journal of the Association for Information
  Systems}} \bibinfo{volume}{16}, \bibinfo{number}{10} (\bibinfo{year}{2015}).
\newblock
\href{https://doi.org/10.17705/1jais.00411}{doi:\nolinkurl{10.17705/1jais.00411}}


\bibitem[Lester et~al\mbox{.}(1997)]%
        {Lester1997}
\bibfield{author}{\bibinfo{person}{James~C. Lester},
  \bibinfo{person}{Sharolyn~A. Converse}, \bibinfo{person}{Susan~E. Kahler},
  \bibinfo{person}{S.~Todd Barlow}, \bibinfo{person}{Brian~A. Stone}, {and}
  \bibinfo{person}{Ravinder~S. Bhogal}.} \bibinfo{year}{1997}\natexlab{}.
\newblock \showarticletitle{The persona effect: affective impact of animated
  pedagogical agents}. In \bibinfo{booktitle}{\emph{Proc. ACM CHI Conference on
  Human Factors in Computing Systems}} (Atlanta, Georgia, USA)
  \emph{(\bibinfo{series}{CHI '97})}. \bibinfo{publisher}{ACM},
  \bibinfo{address}{New York, NY, USA}, \bibinfo{pages}{359–366}.
\newblock
\showISBNx{0897918029}
\href{https://doi.org/10.1145/258549.258797}{doi:\nolinkurl{10.1145/258549.258797}}


\bibitem[Lester et~al\mbox{.}(1999)]%
        {Lester1999}
\bibfield{author}{\bibinfo{person}{James~C. Lester}, \bibinfo{person}{Brian~A.
  Stone}, {and} \bibinfo{person}{Gary~D. Stelling}.}
  \bibinfo{year}{1999}\natexlab{}.
\newblock \showarticletitle{Lifelike Pedagogical Agents for Mixed-initiative
  Problem Solving in Constructivist Learning Environments}.
\newblock \bibinfo{journal}{\emph{User Modeling and User-Adapted Interaction}}
  \bibinfo{volume}{9}, \bibinfo{number}{1} (\bibinfo{date}{01 Apr}
  \bibinfo{year}{1999}), \bibinfo{pages}{1--44}.
\newblock
\showISSN{1573-1391}
\href{https://doi.org/10.1023/A:1008374607830}{doi:\nolinkurl{10.1023/A:1008374607830}}


\bibitem[Liao et~al\mbox{.}(2018)]%
        {Liao2018}
\bibfield{author}{\bibinfo{person}{Q.~Vera Liao}, \bibinfo{person}{Muhammed
  Mas-ud Hussain}, \bibinfo{person}{Praveen Chandar}, \bibinfo{person}{Matthew
  Davis}, \bibinfo{person}{Yasaman Khazaeni}, \bibinfo{person}{Marco~Patricio
  Crasso}, \bibinfo{person}{Dakuo Wang}, \bibinfo{person}{Michael Muller},
  \bibinfo{person}{N.~Sadat Shami}, {and} \bibinfo{person}{Werner Geyer}.}
  \bibinfo{year}{2018}\natexlab{}.
\newblock \showarticletitle{All Work and No Play?}. In
  \bibinfo{booktitle}{\emph{Proc. ACM CHI Conference on Human Factors in
  Computing Systems}} (Montreal QC, Canada) \emph{(\bibinfo{series}{CHI '18})}.
  \bibinfo{publisher}{ACM}, \bibinfo{address}{New York, NY, USA},
  \bibinfo{pages}{1–13}.
\newblock
\showISBNx{9781450356206}
\href{https://doi.org/10.1145/3173574.3173577}{doi:\nolinkurl{10.1145/3173574.3173577}}


\bibitem[Luger and Sellen(2016)]%
        {Luger2016}
\bibfield{author}{\bibinfo{person}{Ewa Luger} {and} \bibinfo{person}{Abigail
  Sellen}.} \bibinfo{year}{2016}\natexlab{}.
\newblock \showarticletitle{"Like Having a Really Bad PA": The Gulf between
  User Expectation and Experience of Conversational Agents}. In
  \bibinfo{booktitle}{\emph{Proc. ACM CHI Conference on Human Factors in
  Computing Systems}} (San Jose, California, USA) \emph{(\bibinfo{series}{CHI
  '16})}. \bibinfo{publisher}{ACM}, \bibinfo{address}{New York, NY, USA},
  \bibinfo{pages}{5286–5297}.
\newblock
\showISBNx{9781450333627}
\href{https://doi.org/10.1145/2858036.2858288}{doi:\nolinkurl{10.1145/2858036.2858288}}


\bibitem[Luria et~al\mbox{.}(2017)]%
        {Luria2017}
\bibfield{author}{\bibinfo{person}{Michal Luria}, \bibinfo{person}{Guy
  Hoffman}, {and} \bibinfo{person}{Oren Zuckerman}.}
  \bibinfo{year}{2017}\natexlab{}.
\newblock \showarticletitle{Comparing Social Robot, Screen and Voice Interfaces
  for Smart-Home Control}. In \bibinfo{booktitle}{\emph{Proc. ACM CHI
  Conference on Human Factors in Computing Systems}} (Denver, Colorado, USA)
  \emph{(\bibinfo{series}{CHI '17})}. \bibinfo{publisher}{ACM},
  \bibinfo{address}{New York, NY, USA}, \bibinfo{pages}{580–628}.
\newblock
\showISBNx{9781450346559}
\href{https://doi.org/10.1145/3025453.3025786}{doi:\nolinkurl{10.1145/3025453.3025786}}


\bibitem[Luria et~al\mbox{.}(2019)]%
        {Luria2019}
\bibfield{author}{\bibinfo{person}{Michal Luria}, \bibinfo{person}{Samantha
  Reig}, \bibinfo{person}{Xiang~Zhi Tan}, \bibinfo{person}{Aaron Steinfeld},
  \bibinfo{person}{Jodi Forlizzi}, {and} \bibinfo{person}{John Zimmerman}.}
  \bibinfo{year}{2019}\natexlab{}.
\newblock \showarticletitle{Re-Embodiment and Co-Embodiment: Exploration of
  Social Presence for Robots and Conversational Agents}. In
  \bibinfo{booktitle}{\emph{Proc. ACM Designing Interactive Systems
  Conference}} (San Diego, CA, USA) \emph{(\bibinfo{series}{DIS '19})}.
  \bibinfo{publisher}{ACM}, \bibinfo{address}{New York, NY, USA},
  \bibinfo{pages}{633–644}.
\newblock
\showISBNx{9781450358507}
\href{https://doi.org/10.1145/3322276.3322340}{doi:\nolinkurl{10.1145/3322276.3322340}}


\bibitem[McDaniel et~al\mbox{.}(2018)]%
        {McDaniel2018}
\bibfield{author}{\bibinfo{person}{Troy McDaniel}, \bibinfo{person}{Samjhana
  Devkota}, \bibinfo{person}{Ramin Tadayon}, \bibinfo{person}{Bryan Duarte},
  \bibinfo{person}{Bijan Fakhri}, {and} \bibinfo{person}{Sethuraman
  Panchanathan}.} \bibinfo{year}{2018}\natexlab{}.
\newblock \showarticletitle{Tactile Facial Action Units Toward Enriching Social
  Interactions for Individuals Who Are Blind}. In
  \bibinfo{booktitle}{\emph{Smart Multimedia}},
  \bibfield{editor}{\bibinfo{person}{Anup Basu} {and} \bibinfo{person}{Stefano
  Berretti}} (Eds.). \bibinfo{publisher}{Springer International Publishing},
  \bibinfo{address}{Cham}, \bibinfo{pages}{3--14}.
\newblock
\showISBNx{978-3-030-04375-9}


\bibitem[McDonald et~al\mbox{.}(2014)]%
        {McDonald2014}
\bibfield{author}{\bibinfo{person}{Samantha McDonald}, \bibinfo{person}{Joshua
  Dutterer}, \bibinfo{person}{Ali Abdolrahmani}, \bibinfo{person}{Shaun~K.
  Kane}, {and} \bibinfo{person}{Amy Hurst}.} \bibinfo{year}{2014}\natexlab{}.
\newblock \showarticletitle{Tactile Aids for Visually Impaired Graphical Design
  Education}. In \bibinfo{booktitle}{\emph{Proc. ACM SIGACCESS Conference on
  Computers \& Accessibility}} (Rochester, New York, USA)
  \emph{(\bibinfo{series}{ASSETS '14})}. \bibinfo{publisher}{ACM},
  \bibinfo{address}{New York, NY, USA}, \bibinfo{pages}{275--276}.
\newblock
\showISBNx{978-1-4503-2720-6}
\href{https://doi.org/10.1145/2661334.2661392}{doi:\nolinkurl{10.1145/2661334.2661392}}


\bibitem[Miele et~al\mbox{.}(2006)]%
        {Miele2006}
\bibfield{author}{\bibinfo{person}{Joshua~A. Miele}, \bibinfo{person}{Steven
  Landau}, {and} \bibinfo{person}{Deborah Gilden}.}
  \bibinfo{year}{2006}\natexlab{}.
\newblock \showarticletitle{Talking TMAP: Automated generation of audio-tactile
  maps using Smith-Kettlewell's TMAP software}.
\newblock \bibinfo{journal}{\emph{British Journal of Visual Impairment}}
  \bibinfo{volume}{24}, \bibinfo{number}{2} (\bibinfo{year}{2006}),
  \bibinfo{pages}{93--100}.
\newblock
\href{https://doi.org/10.1177/0264619606064436}{doi:\nolinkurl{10.1177/0264619606064436}}


\bibitem[Minjin~Rheu and Huh-Yoo(2021)]%
        {Rheu2021}
\bibfield{author}{\bibinfo{person}{Wei~Peng Minjin~Rheu, Ji Youn~Shin} {and}
  \bibinfo{person}{Jina Huh-Yoo}.} \bibinfo{year}{2021}\natexlab{}.
\newblock \showarticletitle{Systematic Review: Trust-Building Factors and
  Implications for Conversational Agent Design}.
\newblock \bibinfo{journal}{\emph{International Journal of Human–Computer
  Interaction}} \bibinfo{volume}{37}, \bibinfo{number}{1}
  (\bibinfo{year}{2021}), \bibinfo{pages}{81--96}.
\newblock
\href{https://doi.org/10.1080/10447318.2020.1807710}{doi:\nolinkurl{10.1080/10447318.2020.1807710}}


\bibitem[Moreno et~al\mbox{.}(2001)]%
        {Moreno2001}
\bibfield{author}{\bibinfo{person}{Roxana Moreno}, \bibinfo{person}{Richard~E.
  Mayer}, \bibinfo{person}{Hiller~A. Spires}, {and} \bibinfo{person}{James~C.
  Lester}.} \bibinfo{year}{2001}\natexlab{}.
\newblock \showarticletitle{The Case for Social Agency in Computer-Based
  Teaching: Do Students Learn More Deeply When They Interact With Animated
  Pedagogical Agents?}
\newblock \bibinfo{journal}{\emph{Cognition and Instruction}}
  \bibinfo{volume}{19}, \bibinfo{number}{2} (\bibinfo{year}{2001}),
  \bibinfo{pages}{177--213}.
\newblock
\href{https://doi.org/10.1207/S1532690XCI1902\_02}{doi:\nolinkurl{10.1207/S1532690XCI1902\_02}}


\bibitem[Nagassa et~al\mbox{.}(2023)]%
        {Nagassa2023}
\bibfield{author}{\bibinfo{person}{Ruth~G Nagassa}, \bibinfo{person}{Matthew
  Butler}, \bibinfo{person}{Leona Holloway}, \bibinfo{person}{Cagatay Goncu},
  {and} \bibinfo{person}{Kim Marriott}.} \bibinfo{year}{2023}\natexlab{}.
\newblock \showarticletitle{3D Building Plans: Supporting Navigation by People
  who are Blind or have Low Vision in Multi-Storey Buildings}. In
  \bibinfo{booktitle}{\emph{Proc. ACM CHI Conference on Human Factors in
  Computing Systems}} (Hamburg, Germany) \emph{(\bibinfo{series}{CHI '23})}.
  \bibinfo{publisher}{ACM}, \bibinfo{address}{New York, NY, USA}, Article
  \bibinfo{articleno}{539}, \bibinfo{numpages}{19}~pages.
\newblock
\showISBNx{9781450394215}
\href{https://doi.org/10.1145/3544548.3581389}{doi:\nolinkurl{10.1145/3544548.3581389}}


\bibitem[Nie et~al\mbox{.}(2012)]%
        {Nie2012}
\bibfield{author}{\bibinfo{person}{Jiaqi Nie}, \bibinfo{person}{Michelle Pak},
  \bibinfo{person}{Angie~Lorena Marin}, {and} \bibinfo{person}{S.~Shyam
  Sundar}.} \bibinfo{year}{2012}\natexlab{}.
\newblock \showarticletitle{Can you hold my hand? physical warmth in
  human-robot interaction}. In \bibinfo{booktitle}{\emph{Proc. ACM/IEEE
  Human-Robot Interaction}} (Boston, Massachusetts, USA)
  \emph{(\bibinfo{series}{HRI '12})}. \bibinfo{publisher}{ACM},
  \bibinfo{address}{New York, NY, USA}, \bibinfo{pages}{201–202}.
\newblock
\showISBNx{9781450310635}
\href{https://doi.org/10.1145/2157689.2157755}{doi:\nolinkurl{10.1145/2157689.2157755}}


\bibitem[Nowak and Biocca(2003)]%
        {Nowak2003}
\bibfield{author}{\bibinfo{person}{Kristine~L. Nowak} {and}
  \bibinfo{person}{Frank Biocca}.} \bibinfo{year}{2003}\natexlab{}.
\newblock \showarticletitle{The Effect of the Agency and Anthropomorphism on
  Users' Sense of Telepresence, Copresence, and Social Presence in Virtual
  Environments}.
\newblock \bibinfo{journal}{\emph{Presence}} \bibinfo{volume}{12},
  \bibinfo{number}{5} (\bibinfo{year}{2003}), \bibinfo{pages}{481--494}.
\newblock
\href{https://doi.org/10.1162/105474603322761289}{doi:\nolinkurl{10.1162/105474603322761289}}


\bibitem[O'Brien(2016)]%
        {OBrien2016}
\bibfield{author}{\bibinfo{person}{Heather O'Brien}.}
  \bibinfo{year}{2016}\natexlab{}.
\newblock \bibinfo{booktitle}{\emph{Theoretical Perspectives on User
  Engagement}}.
\newblock \bibinfo{publisher}{Springer International Publishing},
  \bibinfo{address}{Cham}, \bibinfo{pages}{1--26}.
\newblock
\showISBNx{978-3-319-27446-1}
\href{https://doi.org/10.1007/978-3-319-27446-1_1}{doi:\nolinkurl{10.1007/978-3-319-27446-1_1}}


\bibitem[O'Brien and Lebow({[n.\,d.]})]%
        {OBrien2013}
\bibfield{author}{\bibinfo{person}{Heather~L. O'Brien} {and}
  \bibinfo{person}{Mahria Lebow}.} \bibinfo{year}{[n.\,d.]}\natexlab{}.
\newblock \showarticletitle{Mixed-methods approach to measuring user experience
  in online news interactions}.
\newblock \bibinfo{journal}{\emph{Journal of the American Society for
  Information Science and Technology}} \bibinfo{volume}{64},
  \bibinfo{number}{8} (\bibinfo{year}{[n.\,d.]}), \bibinfo{pages}{1543--1556}.
\newblock
\href{https://doi.org/10.1002/asi.22871}{doi:\nolinkurl{10.1002/asi.22871}}


\bibitem[of~the Blind(2009)]%
        {NFIB2009}
\bibfield{author}{\bibinfo{person}{National~Federation of~the Blind}.}
  \bibinfo{year}{2009}\natexlab{}.
\newblock \bibinfo{title}{The Braille Literacy Crisis in America: Facing the
  Truth, Reversing the Trend, Empowering the Blind}.  (\bibinfo{year}{2009}).
\newblock
\newblock
\shownote{Available from
  \url{https://nfb.org/images/nfb/documents/pdf/braille_literacy_report_web.pdf}}.


\bibitem[O’Brien et~al\mbox{.}(2018)]%
        {Obrien2018}
\bibfield{author}{\bibinfo{person}{Heather~L. O’Brien}, \bibinfo{person}{Paul
  Cairns}, {and} \bibinfo{person}{Mark Hall}.} \bibinfo{year}{2018}\natexlab{}.
\newblock \showarticletitle{A practical approach to measuring user engagement
  with the refined user engagement scale (UES) and new UES short form}.
\newblock \bibinfo{journal}{\emph{International Journal of Human-Computer
  Studies}}  \bibinfo{volume}{112} (\bibinfo{year}{2018}),
  \bibinfo{pages}{28--39}.
\newblock
\showISSN{1071-5819}
\href{https://doi.org/10.1016/j.ijhcs.2018.01.004}{doi:\nolinkurl{10.1016/j.ijhcs.2018.01.004}}


\bibitem[Parkes(1994)]%
        {NOMAD}
\bibfield{author}{\bibinfo{person}{Don Parkes}.}
  \bibinfo{year}{1994}\natexlab{}.
\newblock \showarticletitle{Audio tactile systems for designing and learning
  complex environments as a vision impaired person: static and dynamic spatial
  information access}.
\newblock \bibinfo{journal}{\emph{Learning Environment Technology: Selected
  Papers from LETA}}  \bibinfo{volume}{94} (\bibinfo{year}{1994}),
  \bibinfo{pages}{219--223}.
\newblock


\bibitem[Phillips and Zhao(1993)]%
        {Betsy1993}
\bibfield{author}{\bibinfo{person}{Betsy Phillips} {and}
  \bibinfo{person}{Hongxin Zhao}.} \bibinfo{year}{1993}\natexlab{}.
\newblock \showarticletitle{Predictors of Assistive Technology Abandonment}.
\newblock \bibinfo{journal}{\emph{Assistive Technology}} \bibinfo{volume}{5},
  \bibinfo{number}{1} (\bibinfo{year}{1993}), \bibinfo{pages}{36--45}.
\newblock
\href{https://doi.org/10.1080/10400435.1993.10132205}{doi:\nolinkurl{10.1080/10400435.1993.10132205}}
\newblock
\shownote{PMID: 10171664}.


\bibitem[Pradhan et~al\mbox{.}(2019)]%
        {Pradhan2019}
\bibfield{author}{\bibinfo{person}{Alisha Pradhan}, \bibinfo{person}{Leah
  Findlater}, {and} \bibinfo{person}{Amanda Lazar}.}
  \bibinfo{year}{2019}\natexlab{}.
\newblock \showarticletitle{"Phantom Friend" or "Just a Box with Information":
  Personification and Ontological Categorization of Smart Speaker-based Voice
  Assistants by Older Adults}.
\newblock \bibinfo{journal}{\emph{Proc. ACM Hum.-Comput. Interact.}}
  \bibinfo{volume}{3}, \bibinfo{number}{CSCW}, Article \bibinfo{articleno}{214}
  (\bibinfo{date}{nov} \bibinfo{year}{2019}), \bibinfo{numpages}{21}~pages.
\newblock
\href{https://doi.org/10.1145/3359316}{doi:\nolinkurl{10.1145/3359316}}


\bibitem[Pradhan et~al\mbox{.}(2018)]%
        {Pradhan2018}
\bibfield{author}{\bibinfo{person}{Alisha Pradhan}, \bibinfo{person}{Kanika
  Mehta}, {and} \bibinfo{person}{Leah Findlater}.}
  \bibinfo{year}{2018}\natexlab{}.
\newblock \showarticletitle{"Accessibility Came by Accident": Use of
  Voice-Controlled Intelligent Personal Assistants by People with
  Disabilities}. In \bibinfo{booktitle}{\emph{Proc. ACM CHI Conference on Human
  Factors in Computing Systems}} (Montreal QC, Canada)
  \emph{(\bibinfo{series}{CHI '18})}. \bibinfo{publisher}{ACM},
  \bibinfo{address}{New York, NY, USA}, Article \bibinfo{articleno}{459},
  \bibinfo{numpages}{13}~pages.
\newblock
\showISBNx{978-1-4503-5620-6}
\href{https://doi.org/10.1145/3173574.3174033}{doi:\nolinkurl{10.1145/3173574.3174033}}


\bibitem[Purington et~al\mbox{.}(2017)]%
        {Purington2017}
\bibfield{author}{\bibinfo{person}{Amanda Purington},
  \bibinfo{person}{Jessie~G. Taft}, \bibinfo{person}{Shruti Sannon},
  \bibinfo{person}{Natalya~N. Bazarova}, {and} \bibinfo{person}{Samuel~Hardman
  Taylor}.} \bibinfo{year}{2017}\natexlab{}.
\newblock \showarticletitle{"Alexa is my new BFF": Social Roles, User
  Satisfaction, and Personification of the Amazon Echo}. In
  \bibinfo{booktitle}{\emph{Proc. ACM CHI Conference on Human Factors in
  Computing Systems}} (Denver, Colorado, USA) \emph{(\bibinfo{series}{CHI EA
  '17})}. \bibinfo{publisher}{ACM}, \bibinfo{address}{New York, NY, USA},
  \bibinfo{pages}{2853–2859}.
\newblock
\showISBNx{9781450346566}
\href{https://doi.org/10.1145/3027063.3053246}{doi:\nolinkurl{10.1145/3027063.3053246}}


\bibitem[Qiu et~al\mbox{.}(2020)]%
        {Qiu2020}
\bibfield{author}{\bibinfo{person}{Shi Qiu}, \bibinfo{person}{Pengcheng An},
  \bibinfo{person}{Jun Hu}, \bibinfo{person}{Ting Han}, {and}
  \bibinfo{person}{Matthias Rauterberg}.} \bibinfo{year}{2020}\natexlab{}.
\newblock \showarticletitle{Understanding visually impaired people's
  experiences of social signal perception in face-to-face communication}.
\newblock \bibinfo{journal}{\emph{Universal Access in the Information Society}}
  \bibinfo{volume}{19}, \bibinfo{number}{4} (\bibinfo{date}{01 Nov}
  \bibinfo{year}{2020}), \bibinfo{pages}{873--890}.
\newblock
\showISSN{1615-5297}
\href{https://doi.org/10.1007/s10209-019-00698-3}{doi:\nolinkurl{10.1007/s10209-019-00698-3}}


\bibitem[Qiu et~al\mbox{.}(2016)]%
        {Qiu2016}
\bibfield{author}{\bibinfo{person}{Shi Qiu}, \bibinfo{person}{Siti~Aisyah
  Anas}, \bibinfo{person}{Hirotaka Osawa}, \bibinfo{person}{Matthias
  Rauterberg}, {and} \bibinfo{person}{Jun Hu}.}
  \bibinfo{year}{2016}\natexlab{}.
\newblock \showarticletitle{E-Gaze Glasses: Simulating Natural Gazes for Blind
  People}. In \bibinfo{booktitle}{\emph{Proc. Tangible, Embedded, and Embodied
  Interaction}} (Eindhoven, Netherlands) \emph{(\bibinfo{series}{TEI '16})}.
  \bibinfo{publisher}{ACM}, \bibinfo{address}{New York, NY, USA},
  \bibinfo{pages}{563–569}.
\newblock
\showISBNx{9781450335829}
\href{https://doi.org/10.1145/2839462.2856518}{doi:\nolinkurl{10.1145/2839462.2856518}}


\bibitem[Rader et~al\mbox{.}(2014)]%
        {rader2014}
\bibfield{author}{\bibinfo{person}{Joshua Rader}, \bibinfo{person}{Troy
  McDaniel}, \bibinfo{person}{Artemio Ramirez}, \bibinfo{person}{Shantanu
  Bala}, {and} \bibinfo{person}{Sethuraman Panchanathan}.}
  \bibinfo{year}{2014}\natexlab{}.
\newblock \showarticletitle{A Wizard of Oz Study Exploring How
  Agreement/Disagreement Nonverbal Cues Enhance Social Interactions for
  Individuals Who Are Blind}. In \bibinfo{booktitle}{\emph{HCI International
  2014 - Posters' Extended Abstracts}},
  \bibfield{editor}{\bibinfo{person}{Constantine Stephanidis}} (Ed.).
  \bibinfo{publisher}{Springer International Publishing},
  \bibinfo{address}{Cham}, \bibinfo{pages}{243--248}.
\newblock
\showISBNx{978-3-319-07854-0}


\bibitem[Reeves et~al\mbox{.}(2004)]%
        {Reeves2004}
\bibfield{author}{\bibinfo{person}{Leah~M. Reeves}, \bibinfo{person}{Jennifer
  Lai}, \bibinfo{person}{James~A. Larson}, \bibinfo{person}{Sharon Oviatt},
  \bibinfo{person}{T.~S. Balaji}, \bibinfo{person}{St{\'e}phanie Buisine},
  \bibinfo{person}{Penny Collings}, \bibinfo{person}{Phil Cohen},
  \bibinfo{person}{Ben Kraal}, \bibinfo{person}{Jean-Claude Martin},
  \bibinfo{person}{Michael McTear}, \bibinfo{person}{TV Raman},
  \bibinfo{person}{Kay~M. Stanney}, \bibinfo{person}{Hui Su}, {and}
  \bibinfo{person}{Qian~Ying Wang}.} \bibinfo{year}{2004}\natexlab{}.
\newblock \showarticletitle{Guidelines for Multimodal User Interface Design}.
\newblock \bibinfo{journal}{\emph{Commun. ACM}} \bibinfo{volume}{47},
  \bibinfo{number}{1} (\bibinfo{date}{Jan.} \bibinfo{year}{2004}),
  \bibinfo{pages}{57--59}.
\newblock
\showISSN{0001-0782}
\href{https://doi.org/10.1145/962081.962106}{doi:\nolinkurl{10.1145/962081.962106}}


\bibitem[Reichinger et~al\mbox{.}(2016)]%
        {Reichinger2016}
\bibfield{author}{\bibinfo{person}{Andreas Reichinger}, \bibinfo{person}{Anton
  Fuhrmann}, \bibinfo{person}{Stefan Maierhofer}, {and} \bibinfo{person}{Werner
  Purgathofer}.} \bibinfo{year}{2016}\natexlab{}.
\newblock \showarticletitle{Gesture-Based Interactive Audio Guide on Tactile
  Reliefs}. In \bibinfo{booktitle}{\emph{Proc. ACM SIGACCESS Conference on
  Computers \& Accessibility}} (Reno, Nevada, USA)
  \emph{(\bibinfo{series}{ASSETS '16})}. \bibinfo{publisher}{ACM},
  \bibinfo{address}{New York, NY, USA}, \bibinfo{pages}{91--100}.
\newblock
\showISBNx{978-1-4503-4124-0}
\href{https://doi.org/10.1145/2982142.2982176}{doi:\nolinkurl{10.1145/2982142.2982176}}


\bibitem[Reinders et~al\mbox{.}(2023)]%
        {Reinders2023}
\bibfield{author}{\bibinfo{person}{Samuel Reinders}, \bibinfo{person}{Swamy
  Ananthanarayan}, \bibinfo{person}{Matthew Butler}, {and} \bibinfo{person}{Kim
  Marriott}.} \bibinfo{year}{2023}\natexlab{}.
\newblock \showarticletitle{Designing Conversational Multimodal 3D Printed
  Models with People who are Blind}. In \bibinfo{booktitle}{\emph{Proc. ACM
  Designing Interactive Systems Conference}} (Pittsburgh, PA, USA)
  \emph{(\bibinfo{series}{DIS '23})}. \bibinfo{publisher}{ACM},
  \bibinfo{address}{New York, NY, USA}, \bibinfo{pages}{2172–2188}.
\newblock
\showISBNx{9781450398930}
\href{https://doi.org/10.1145/3563657.3595989}{doi:\nolinkurl{10.1145/3563657.3595989}}


\bibitem[Reinders et~al\mbox{.}(2020)]%
        {Reinders2020}
\bibfield{author}{\bibinfo{person}{Samuel Reinders}, \bibinfo{person}{Matthew
  Butler}, {and} \bibinfo{person}{Kim Marriott}.}
  \bibinfo{year}{2020}\natexlab{}.
\newblock \showarticletitle{"Hey Model!" - Natural User Interactions and Agency
  in Accessible Interactive 3D Models}. In \bibinfo{booktitle}{\emph{Proc. ACM
  CHI Conference on Human Factors in Computing Systems}} (Honolulu, HI, USA)
  \emph{(\bibinfo{series}{CHI '20})}. \bibinfo{publisher}{ACM},
  \bibinfo{address}{New York, NY, USA}, \bibinfo{pages}{1–13}.
\newblock
\showISBNx{9781450367080}
\href{https://doi.org/10.1145/3313831.3376145}{doi:\nolinkurl{10.1145/3313831.3376145}}


\bibitem[Rosenblum and Herzberg(2015)]%
        {Rosenblum2015}
\bibfield{author}{\bibinfo{person}{L.~Penny Rosenblum} {and}
  \bibinfo{person}{Tina~S. Herzberg}.} \bibinfo{year}{2015}\natexlab{}.
\newblock \showarticletitle{Braille and Tactile Graphics: Youths with Visual
  Impairments Share Their Experiences}.
\newblock \bibinfo{journal}{\emph{Journal of Visual Impairment \& Blindness}}
  \bibinfo{volume}{109}, \bibinfo{number}{3} (\bibinfo{year}{2015}),
  \bibinfo{pages}{173--184}.
\newblock
\href{https://doi.org/10.1177/0145482X1510900302}{doi:\nolinkurl{10.1177/0145482X1510900302}}


\bibitem[Rowell and Ungar(2005)]%
        {Rowell2005}
\bibfield{author}{\bibinfo{person}{Jonathan Rowell} {and}
  \bibinfo{person}{Simon Ungar}.} \bibinfo{year}{2005}\natexlab{}.
\newblock \showarticletitle{Feeling our way: tactile map user requirements –
  a survey}. In \bibinfo{booktitle}{\emph{Proc. International Cartographic
  Conference}}.
\newblock


\bibitem[Salah et~al\mbox{.}(2023)]%
        {Salah2023}
\bibfield{author}{\bibinfo{person}{Mohammed Salah}, \bibinfo{person}{Hussam
  Alhalbusi}, \bibinfo{person}{Maria~Mohd Ismail}, {and} \bibinfo{person}{Fadi
  Abdelfattah}.} \bibinfo{year}{2023}\natexlab{}.
\newblock \showarticletitle{Chatting with ChatGPT: decoding the mind of Chatbot
  users and unveiling the intricate connections between user perception, trust
  and stereotype perception on self-esteem and psychological well-being}.
\newblock \bibinfo{journal}{\emph{Current Psychology}} (\bibinfo{date}{20 Jul}
  \bibinfo{year}{2023}).
\newblock
\showISSN{1936-4733}
\href{https://doi.org/10.1007/s12144-023-04989-0}{doi:\nolinkurl{10.1007/s12144-023-04989-0}}


\bibitem[Schroeder et~al\mbox{.}(2013)]%
        {Schroeder2013}
\bibfield{author}{\bibinfo{person}{Noah~L. Schroeder},
  \bibinfo{person}{Olusola~O. Adesope}, {and} \bibinfo{person}{Rachel~Barouch
  Gilbert}.} \bibinfo{year}{2013}\natexlab{}.
\newblock \showarticletitle{How Effective are Pedagogical Agents for Learning?
  A Meta-Analytic Review}.
\newblock \bibinfo{journal}{\emph{Journal of Educational Computing Research}}
  \bibinfo{volume}{49}, \bibinfo{number}{1} (\bibinfo{year}{2013}),
  \bibinfo{pages}{1--39}.
\newblock
\href{https://doi.org/10.2190/EC.49.1.a}{doi:\nolinkurl{10.2190/EC.49.1.a}}


\bibitem[Shamekhi et~al\mbox{.}(2018)]%
        {Shamekhi2018}
\bibfield{author}{\bibinfo{person}{Ameneh Shamekhi}, \bibinfo{person}{Q.~Vera
  Liao}, \bibinfo{person}{Dakuo Wang}, \bibinfo{person}{Rachel K.~E. Bellamy},
  {and} \bibinfo{person}{Thomas Erickson}.} \bibinfo{year}{2018}\natexlab{}.
\newblock \showarticletitle{Face Value? Exploring the Effects of Embodiment for
  a Group Facilitation Agent}. In \bibinfo{booktitle}{\emph{Proc. ACM CHI
  Conference on Human Factors in Computing Systems}} (Montreal QC, Canada)
  \emph{(\bibinfo{series}{CHI '18})}. \bibinfo{publisher}{ACM},
  \bibinfo{address}{New York, NY, USA}, \bibinfo{pages}{1–13}.
\newblock
\showISBNx{9781450356206}
\href{https://doi.org/10.1145/3173574.3173965}{doi:\nolinkurl{10.1145/3173574.3173965}}


\bibitem[Sheffield(2016)]%
        {Sheffield2016}
\bibfield{author}{\bibinfo{person}{Rebecca Sheffield}.}
  \bibinfo{year}{2016}\natexlab{}.
\newblock \bibinfo{title}{International Approaches to Rehabilitation Programs
  for Adults who are Blind or Visually Impaired: Delivery Models, Services,
  Challenges and Trends}.
\newblock


\bibitem[Shi et~al\mbox{.}(2019)]%
        {Shi2019}
\bibfield{author}{\bibinfo{person}{Lei Shi}, \bibinfo{person}{Holly Lawson},
  \bibinfo{person}{Zhuohao Zhang}, {and} \bibinfo{person}{Shiri Azenkot}.}
  \bibinfo{year}{2019}\natexlab{}.
\newblock \showarticletitle{Designing Interactive 3D Printed Models with
  Teachers of the Visually Impaired}. In \bibinfo{booktitle}{\emph{Proc. ACM
  CHI Conference on Human Factors in Computing Systems}} (Glasgow, Scotland Uk)
  \emph{(\bibinfo{series}{CHI '19})}. \bibinfo{publisher}{ACM},
  \bibinfo{address}{New York, NY, USA}, Article \bibinfo{articleno}{197},
  \bibinfo{numpages}{14}~pages.
\newblock
\showISBNx{978-1-4503-5970-2}
\href{https://doi.org/10.1145/3290605.3300427}{doi:\nolinkurl{10.1145/3290605.3300427}}


\bibitem[Shi et~al\mbox{.}(2016)]%
        {Shi2016}
\bibfield{author}{\bibinfo{person}{Lei Shi}, \bibinfo{person}{Idan Zelzer},
  \bibinfo{person}{Catherine Feng}, {and} \bibinfo{person}{Shiri Azenkot}.}
  \bibinfo{year}{2016}\natexlab{}.
\newblock \showarticletitle{Tickers and Talker: An accessible labeling toolkit
  for 3D printed models}. In \bibinfo{booktitle}{\emph{Proc. of the 34rd Annual
  ACM Conference on Human Factors in Computing Systems (CHI'16)}}.
\newblock
\href{https://doi.org/10.1145/2858036.2858507}{doi:\nolinkurl{10.1145/2858036.2858507}}


\bibitem[Shi et~al\mbox{.}(2017)]%
        {Shi2017b}
\bibfield{author}{\bibinfo{person}{Lei Shi}, \bibinfo{person}{Yuhang Zhao},
  {and} \bibinfo{person}{Shiri Azenkot}.} \bibinfo{year}{2017}\natexlab{}.
\newblock \showarticletitle{Designing Interactions for 3D Printed Models with
  Blind People}. In \bibinfo{booktitle}{\emph{Proc. ACM SIGACCESS Conference on
  Computers \& Accessibility}} (Baltimore, Maryland, USA)
  \emph{(\bibinfo{series}{ASSETS '17})}. \bibinfo{publisher}{ACM},
  \bibinfo{address}{New York, NY, USA}, \bibinfo{pages}{200--209}.
\newblock
\showISBNx{978-1-4503-4926-0}
\href{https://doi.org/10.1145/3132525.3132549}{doi:\nolinkurl{10.1145/3132525.3132549}}


\bibitem[Shi et~al\mbox{.}(2020)]%
        {Shi2020}
\bibfield{author}{\bibinfo{person}{Lei Shi}, \bibinfo{person}{Yuhang Zhao},
  \bibinfo{person}{Ricardo Gonzalez~Penuela}, \bibinfo{person}{Elizabeth
  Kupferstein}, {and} \bibinfo{person}{Shiri Azenkot}.}
  \bibinfo{year}{2020}\natexlab{}.
\newblock \showarticletitle{Molder: An Accessible Design Tool for Tactile
  Maps}. In \bibinfo{booktitle}{\emph{Proc. ACM CHI Conference on Human Factors
  in Computing Systems}} (Honolulu, HI, USA) \emph{(\bibinfo{series}{CHI
  '20})}. \bibinfo{publisher}{ACM}, \bibinfo{address}{New York, NY, USA},
  \bibinfo{pages}{1–14}.
\newblock
\showISBNx{9781450367080}
\href{https://doi.org/10.1145/3313831.3376431}{doi:\nolinkurl{10.1145/3313831.3376431}}


\bibitem[Siddharth~Gulati and Lamas(2019)]%
        {Gulati2019}
\bibfield{author}{\bibinfo{person}{Sonia~Sousa Siddharth~Gulati} {and}
  \bibinfo{person}{David Lamas}.} \bibinfo{year}{2019}\natexlab{}.
\newblock \showarticletitle{Design, development and evaluation of a
  human-computer trust scale}.
\newblock \bibinfo{journal}{\emph{Behaviour \& Information Technology}}
  \bibinfo{volume}{38}, \bibinfo{number}{10} (\bibinfo{year}{2019}),
  \bibinfo{pages}{1004--1015}.
\newblock
\href{https://doi.org/10.1080/0144929X.2019.1656779}{doi:\nolinkurl{10.1080/0144929X.2019.1656779}}


\bibitem[Sidner et~al\mbox{.}(2018)]%
        {Sidner2018}
\bibfield{author}{\bibinfo{person}{Candace~L. Sidner}, \bibinfo{person}{Timothy
  Bickmore}, \bibinfo{person}{Bahador Nooraie}, \bibinfo{person}{Charles Rich},
  \bibinfo{person}{Lazlo Ring}, \bibinfo{person}{Mahni Shayganfar}, {and}
  \bibinfo{person}{Laura Vardoulakis}.} \bibinfo{year}{2018}\natexlab{}.
\newblock \showarticletitle{Creating New Technologies for Companionable Agents
  to Support Isolated Older Adults}.
\newblock \bibinfo{journal}{\emph{ACM Trans. Interact. Intell. Syst.}}
  \bibinfo{volume}{8}, \bibinfo{number}{3}, Article \bibinfo{articleno}{17}
  (\bibinfo{date}{July} \bibinfo{year}{2018}), \bibinfo{numpages}{27}~pages.
\newblock
\showISSN{2160-6455}
\href{https://doi.org/10.1145/3213050}{doi:\nolinkurl{10.1145/3213050}}


\bibitem[Stangl et~al\mbox{.}(2015)]%
        {Stangl2015}
\bibfield{author}{\bibinfo{person}{Abigale Stangl}, \bibinfo{person}{Chia-Lo
  Hsu}, {and} \bibinfo{person}{Tom Yeh}.} \bibinfo{year}{2015}\natexlab{}.
\newblock \showarticletitle{Transcribing Across the Senses: Community Efforts
  to Create 3D Printable Accessible Tactile Pictures for Young Children with
  Visual Impairments}. In \bibinfo{booktitle}{\emph{Proc. ACM SIGACCESS
  Conference on Computers \& Accessibility}} (Lisbon, Portugal)
  \emph{(\bibinfo{series}{ASSETS '15})}. \bibinfo{publisher}{ACM},
  \bibinfo{address}{New York, NY, USA}, \bibinfo{pages}{127--137}.
\newblock
\showISBNx{978-1-4503-3400-6}
\href{https://doi.org/10.1145/2700648.2809854}{doi:\nolinkurl{10.1145/2700648.2809854}}


\bibitem[Taylor et~al\mbox{.}(2015)]%
        {Taylor2015}
\bibfield{author}{\bibinfo{person}{Brandon~T. Taylor},
  \bibinfo{person}{Anind~K. Dey}, \bibinfo{person}{Dan~P. Siewiorek}, {and}
  \bibinfo{person}{Asim Smailagic}.} \bibinfo{year}{2015}\natexlab{}.
\newblock \showarticletitle{TactileMaps.net: A web interface for generating
  customized 3D-printable tactile maps}. In \bibinfo{booktitle}{\emph{Proc. ACM
  SIGACCESS Conference on Computers \& Accessibility}}. ACM,
  \bibinfo{pages}{427--428}.
\newblock
\href{https://doi.org/10.1145/2700648.2811336}{doi:\nolinkurl{10.1145/2700648.2811336}}


\bibitem[ViewPlus({[n.\,d.]})]%
        {IVEO}
\bibfield{author}{\bibinfo{person}{ViewPlus}.}
  \bibinfo{year}{[n.\,d.]}\natexlab{}.
\newblock \bibinfo{title}{IVEO 3 Hands-On Learning System}.
  (\bibinfo{year}{[n.\,d.]}).
\newblock
\newblock
\shownote{Available from
  \url{https://viewplus.com/product/iveo-3-hands-on-learning-system/}}.


\bibitem[Wedler et~al\mbox{.}(2012)]%
        {wedler2012applied}
\bibfield{author}{\bibinfo{person}{Henry~B Wedler}, \bibinfo{person}{Sarah~R
  Cohen}, \bibinfo{person}{Rebecca~L Davis}, \bibinfo{person}{Jason~G
  Harrison}, \bibinfo{person}{Matthew~R Siebert}, \bibinfo{person}{Dan
  Willenbring}, \bibinfo{person}{Christian~S Hamann}, \bibinfo{person}{Jared~T
  Shaw}, {and} \bibinfo{person}{Dean~J Tantillo}.}
  \bibinfo{year}{2012}\natexlab{}.
\newblock \showarticletitle{Applied computational chemistry for the blind and
  visually impaired}.
\newblock \bibinfo{journal}{\emph{Journal of Chemical Education}}
  \bibinfo{volume}{89}, \bibinfo{number}{11} (\bibinfo{year}{2012}),
  \bibinfo{pages}{1400--1404}.
\newblock


\bibitem[Wiebe et~al\mbox{.}(2014)]%
        {wiebe2014}
\bibfield{author}{\bibinfo{person}{Eric~N. Wiebe}, \bibinfo{person}{Allison
  Lamb}, \bibinfo{person}{Megan Hardy}, {and} \bibinfo{person}{David Sharek}.}
  \bibinfo{year}{2014}\natexlab{}.
\newblock \showarticletitle{Measuring engagement in video game-based
  environments: Investigation of the User Engagement Scale}.
\newblock \bibinfo{journal}{\emph{Computers in Human Behavior}}
  \bibinfo{volume}{32} (\bibinfo{year}{2014}), \bibinfo{pages}{123--132}.
\newblock
\showISSN{0747-5632}
\href{https://doi.org/10.1016/j.chb.2013.12.001}{doi:\nolinkurl{10.1016/j.chb.2013.12.001}}


\bibitem[Wu et~al\mbox{.}(2017)]%
        {Wu2017}
\bibfield{author}{\bibinfo{person}{Shaomei Wu}, \bibinfo{person}{Jeffrey
  Wieland}, \bibinfo{person}{Omid Farivar}, {and} \bibinfo{person}{Julie
  Schiller}.} \bibinfo{year}{2017}\natexlab{}.
\newblock \showarticletitle{Automatic Alt-Text: Computer-Generated Image
  Descriptions for Blind Users on a Social Network Service}. In
  \bibinfo{booktitle}{\emph{Proc. ACM Computer Supported Cooperative Work and
  Social Computing}} (Portland, Oregon, USA) \emph{(\bibinfo{series}{CSCW
  '17})}. \bibinfo{publisher}{ACM}, \bibinfo{address}{New York, NY, USA},
  \bibinfo{pages}{1180–1192}.
\newblock
\showISBNx{9781450343350}
\href{https://doi.org/10.1145/2998181.2998364}{doi:\nolinkurl{10.1145/2998181.2998364}}


\end{thebibliography}
\newpage
\appendix
\label{sec:Appendix}

\section{Godspeed Questionnaire (GSQ)}\label{sec:GSQ}
\small
GSQ items were mixed to mask intention and used a 5-point semantic scale.

\subsection*{\small Anthropomorphism}
\begin{itemize}[labelsep=0.3cm, leftmargin=0.7cm, itemsep=0.1em]
    \item Machine-like \hfill \makebox[4.5cm]{\hfill} \hfill Human-like
    \item Unconscious \hfill \makebox[4.5cm]{\hfill} \hfill Conscious
\end{itemize}

\subsection*{\small Animacy}
\begin{itemize}[labelsep=0.3cm, leftmargin=0.7cm, itemsep=0.1em]
    \item Artificial \hfill \makebox[4.5cm]{\hfill} \hfill Lifelike
    \item Inert \hfill \makebox[4.5cm]{\hfill} \hfill Interactive
\end{itemize}

\subsection*{\small Likeability}
\begin{itemize}[labelsep=0.3cm, leftmargin=0.7cm, itemsep=0.1em]
    \item Dislike \hfill \makebox[4.5cm]{\hfill} \hfill Like
    \item Unfriendly \hfill \makebox[4.5cm]{\hfill} \hfill Friendly
    \item Unpleasant \hfill \makebox[4.5cm]{\hfill} \hfill Pleasant
\end{itemize}

\subsection*{\small Perceived Intelligence}
\begin{itemize}[labelsep=0.3cm, leftmargin=0.7cm, itemsep=0.1em]
    \item Ignorant \hfill \makebox[4.5cm]{\hfill} \hfill Knowledgeable
    \item Unintelligent \hfill \makebox[4.5cm]{\hfill} \hfill Intelligent
\end{itemize}

\section{User Engagement Scale Short Form (UES-SF)}\label{sec:UES}
\small
UES-SF items were mixed to mask intention and used a 5-point Likert scale.

\subsection*{\small Focused Attention}
\begin{itemize}[labelsep=0.3cm, leftmargin=0.7cm, itemsep=0.1em]
    \item I lost myself in this experience.
    \item The time spent using the model just slipped away.
    \item I was absorbed in this experience.
\end{itemize}

\subsection*{\small Perceived Usability}
\begin{itemize}[labelsep=0.3cm, leftmargin=0.7cm, itemsep=0.1em]
    \item I felt frustrated while using the model.
    \item I found the model confusing to use.
    \item Using the model was taxing.
\end{itemize}

\subsection*{\small Aesthetic Appeal}
\begin{itemize}[labelsep=0.3cm, leftmargin=0.7cm, itemsep=0.1em]
    \item This model was attractive to my senses.
    \item This model was aesthetically pleasing to my senses.
    \item This model appealed to my senses.
\end{itemize}

\subsection*{\small Reward}
\begin{itemize}[labelsep=0.3cm, leftmargin=0.7cm, itemsep=0.1em]
    \item Using the model was worthwhile.
    \item This experience was rewarding.
    \item I felt interested in this experience.
\end{itemize}

\section{Playful Experiences Questionnaire (PLEXQ)}\label{sec:PLEXQ}
\small
PLEXQ items were mixed to mask intention and used a 5-point Likert scale.

\subsection*{\small Captivation}
\begin{itemize}[labelsep=0.3cm, leftmargin=0.7cm, itemsep=0.1em]
    \item I forgot my surroundings.
    \item I felt completely absorbed.
    \item I lost track of space and time.
\end{itemize}

\subsection*{\small Challenge}
\begin{itemize}[labelsep=0.3cm, leftmargin=0.7cm, itemsep=0.1em]
    \item It stimulated me to learn new things.
    \item It was a true learning experience.
    \item I enjoyed learning new things.
\end{itemize}

\subsection*{\small Control}
\begin{itemize}[labelsep=0.3cm, leftmargin=0.7cm, itemsep=0.1em]
    \item I had the capability to influence what was happening.
    \item I felt powerful.
    \item I enjoyed being in control.
\end{itemize}

\subsection*{\small Discovery}
\begin{itemize}[labelsep=0.3cm, leftmargin=0.7cm, itemsep=0.1em]
    \item I enjoyed discovering new things.
    \item I enjoyed finding useful new ways of using it.
    \item I enjoyed finding something unexpected.
\end{itemize}

\subsection*{\small Exploration}
\begin{itemize}[labelsep=0.3cm, leftmargin=0.7cm, itemsep=0.1em]
    \item I felt curious.
    \item I enjoyed experimenting.
    \item I enjoyed trying out new things.
\end{itemize}

\subsection*{\small Humor}
\begin{itemize}[labelsep=0.3cm, leftmargin=0.7cm, itemsep=0.1em]
    \item It made me laugh.
    \item I had fun.
    \item I experienced funny situations.
\end{itemize}

\subsection*{\small Relaxation}
\begin{itemize}[labelsep=0.3cm, leftmargin=0.7cm, itemsep=0.1em]
    \item I felt relaxed.
    \item I enjoyed passing time with it.
    \item I felt relieved from stress.
\end{itemize}

\subsection*{\small Sensation}
\begin{itemize}[labelsep=0.3cm, leftmargin=0.7cm, itemsep=0.1em]
    \item I felt pleased by its aesthetics to my senses.
    \item I enjoyed the aesthetics to my senses.
    \item I felt pleased by the quality of it.
\end{itemize}

\section{Human-Computer Trust Model (HCTM)}\label{sec:HCTM}
\small
HCTM items were mixed to mask intention and used a 5-point Likert scale.

\subsection*{\small Perceived Risk}
\begin{itemize}[labelsep=0.3cm, leftmargin=0.7cm, itemsep=0.1em]
    \item I believe that there could be negative consequences using the model.
    \item I feel I must be cautious when using the model.
    \item It is risky to interact with the model.
\end{itemize}

\subsection*{\small Benevolence}
\begin{itemize}[labelsep=0.3cm, leftmargin=0.7cm, itemsep=0.1em]
    \item I believe that the model will act in my best interest.
    \item I believe that the model will do its best to help me if I need help.
    \item I believe that the model is interested in understanding my needs and preferences.
\end{itemize}

\subsection*{\small Competence}
\begin{itemize}[labelsep=0.3cm, leftmargin=0.7cm, itemsep=0.1em]
    \item I think the model is competent and effective in facilitating my learning.
    \item I think the model performs its role as a learning material very well.
    \item I believe that the model has all the functionalities I would expect from a learning material.
\end{itemize}

\subsection*{\small Reciprocity}
\begin{itemize}[labelsep=0.3cm, leftmargin=0.7cm, itemsep=0.1em]
    \item If I use the model, I think I would be able to depend on it completely.
    \item I can always rely on the model for facilitating my learning.
    \item I can trust the information presented to me by the model.
\end{itemize}

%TC:endignore
\end{document}